%% file: main.tex
\documentclass{ws-rv9x6}
\usepackage{subfigure}   
\usepackage{ws-rv-thm}   
\usepackage{ws-rv-van}   
\makeindex
\usepackage[linewidth=1pt]{mdframed}
\usepackage[colorlinks=true,citecolor=blue,urlcolor=blue,linkcolor=blue,pdfstartview=FitH,bookmarksopen]{hyperref}
\usepackage{caption}
\usepackage[normalem]{ulem}

\newcommand{\bfrac}[2]{\left(\frac{#1}{#2}\right)}

\newcommand{\infrac}[2]{{#1}/{#2}}

\newcommand{\derr}[2]{\frac{d{#1}}{d{#2}}}

\newcommand{\inderr}[2]{d{#1}/d{#2}}

\newcommand{\parr}[2]{\frac{\partial{#1}}{\partial{#2}}}
\newcommand{\inparr}[2]{\partial{#1}/\partial{#2}}

\newcommand{\snn} {\sqrt{s_{_{\rm NN}}}}

\newcommand{\ekin} {E_{\rm{kin}}}

\newcommand{\commentout}[1]  {}

\usepackage{bm}

\newcommand{\fluid}{\mathrm{fluid}}
\newcommand{\prt}{\mathrm{part}}
\newcommand{\source}{\mathrm{source}}

\begin{document}

\title{Quark-Gluon Plasma 6}

\chapter[The QCD phase diagram and Beam Energy Scan physics]{The QCD phase diagram and Beam Energy Scan physics:\\
  a theory overview}
\label{ch:BES_review}

\author[L.~Du, A.~Sorensen, M.~Stephanov]{
Lipei Du,$^{1,2}$\footnote{lipei.du@mail.mcgill.ca} 
Agnieszka Sorensen,$^{3}$\footnote{amsorens@uw.edu} 
and Mikhail Stephanov$^{4,5}$\footnote{misha@uic.edu}}

\address{ 
$^1$Department of Physics, McGill University, Montreal, Quebec H3A 2T8, Canada\\
$^2$Nuclear Science Division, Lawrence Berkeley National Laboratory, Berkeley, CA 94270, USA\\
$^3$Institute for Nuclear Theory, University of Washington, Seattle, WA 98195, USA\\
$^4$Department of Physics and Laboratory for Quantum Theory at the Extremes, University of Illinois, Chicago, IL 60607,
USA\\
$^5$Kadanoff  Center  for  Theoretical  Physics,  University  of  Chicago,  Chicago,  Illinois  60637,  USA
}

\authormark{L.~Du, A.~Sorensen, and M.~Stephanov}


\begin{abstract}
 We review recent theoretical developments relevant to heavy-ion experiments carried out within the Beam Energy Scan program at the Relativistic Heavy Ion Collider.
 Our main focus is on the description of the dynamics of systems created in heavy-ion collisions and establishing the necessary connection between the experimental observables and the QCD
 phase diagram.
\end{abstract}



\body
\setcounter{tocdepth}{2} 
\tableofcontents


\section{Introduction}
\label{sec:intro}
\input{intro.tex}

\section{Multistage description of bulk dynamics}
\label{sec:hydro}
\input{hydro}

\section{Microscopic transport description of dense nuclear matter dynamics}
\label{sec:microscopic_transport}
\input{microscopic_transport.tex}

\section{Fluctuations: critical point, hydrodynamics, freeze-out}
\label{sec:fluctuations}
\input{fluctuations}

\label{sec:summary}
\input{summary}

\section*{Acknowledgments}
This work was partly supported by the Natural Sciences and Engineering
Research Council of Canada (L.D.), and partly by the U.S. Department of
Energy, Office of Science, Office of Nuclear Physics, under Grants
No. DE-FG02-00ER41132 (A.S.) and No. DE-FG0201ER41195 (M.S.). 
L.D.\ acknowledges C.~Chattopadhyay, C.~Gale, U.~Heinz, S.~Jaiswal, S.~Jeon, B.~Schenke, and C.~Shen for useful comments on the manuscript.
A.S.\ wants to thank Jan Steinheimer for helpful discussions, and Manjunath Omana Kuttan for sharing data tables.

\bibliographystyle{ws-rv-van}
\bibliography{fluctuations,hydro,microscopic_transport,non_inspire_QGP6}


\end{document}

%% file: intro.tex
\begin{figure}[h]
  \centering
  \includegraphics[width=0.99\textwidth]{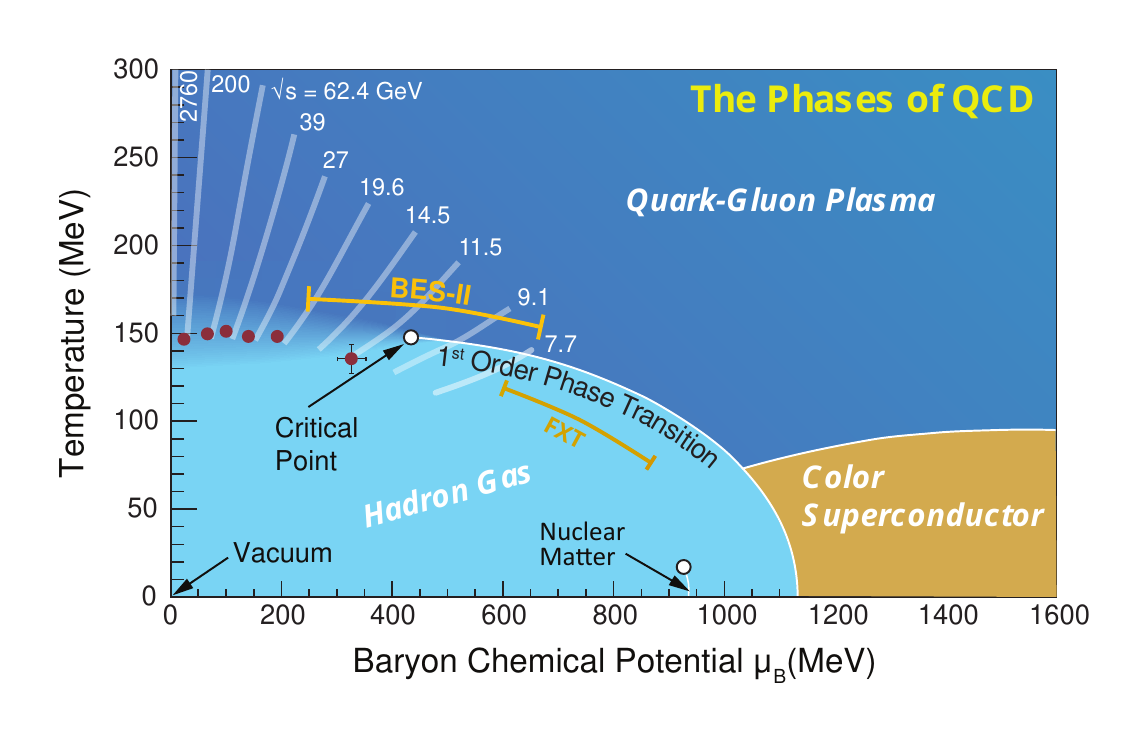}     \caption{The QCD phase diagram as conjectured based on the current
    understanding. Red circles mark the location of freeze-out
    points determined experimentally. Trajectories followed by
    hydrodynamically expanding matter in the fireball are indicated
    schematically by white lines. These trajectories are labeled by the
    initial collision energies $\snn$ in GeV, demonstrating how a
    beam-energy scan translates into the scan of the QCD phase
    diagram. The range covered by the second phase of the Beam Energy
    Scan program is marked as ``BES-II'' (collider mode) and ``FXT''
    (fixed target). Figure modified from Ref.~\cite{Bzdak:2019pkr}.
	}
  \label{fig:phase-diagram}
\end{figure}

Quantum Chromodynamics~(QCD) is a non-abelian gauge theory describing
a vast range of the strong interaction phenomena using a tight set of
fundamental principles. 
However, extracting theoretical predictions from QCD
is remarkably difficult due to the non-perturbative nature of the 
dominant
fundamental strong interaction phenomena: quark and gluon
confinement and spontaneous chiral symmetry breaking. 
One of the most challenging and still open questions is the full structure of the QCD phase diagram at finite temperature and finite baryon density, a tentative sketch of which is shown in Fig.~\ref{fig:phase-diagram}.
The exploration of the QCD phase diagram proceeds along several intertwined
directions.

On the purely theoretical side, the most prominent approach is based
on first-principle lattice calculations. While the equation of
state~(EOS) at finite temperature and {\em zero} chemical potential
can be reliably calculated, the notorious sign problem prevents
lattice simulations from extending this first-principle method to {\em
  finite} baryon density. The recent advances and challenges in this
area of the QCD phase diagram research, covered in reviews such as
Ref.~\cite{Ratti:2018ksb}, are not within the scope of this
chapter. Among other theoretical attempts to shed light on the QCD
phase diagram are microscopic approaches such as the Functional
Renormalization Group (FRG)\cite{Fu:2022gou} as well as
various calculations based on models as simple as the
Random Matrix Model~(RMM)\cite{Halasz:1998qr} or Nambu--Jona-Lasinio~(NJL)
models\cite{Berges:1998rc,Stephanov:2004wx} and as sophisticated as models based on applications of
gauge-gravity (or AdS/CFT) duality to QCD\cite{Hippert:2023bel}.

Experiments involving heavy-ion collisions can be utilized to explore the QCD EOS and phase diagram.
By colliding heavy nuclei at center-of-mass energies per nucleon pair, $\snn$, ranging from a few GeV to a few TeV, one can scan and explore the range of temperatures and baryon chemical potentials where the most theoretically challenging phenomena associated with the transition between the two major QCD phases --- the
Quark-Gluon Plasma~(QGP) and Hadron Resonance Gas~(HRG) --- occur (see Fig.~\ref{fig:phase-diagram}). 
Another avenue for exploration of the QCD phase diagram has been recently opened by gravitational wave observations of neutron star mergers \cite{Radice:2020ddv,Dexheimer:2020zzs}. 
While, in contrast to laboratory experiments, these natural phenomena cannot be planned or controlled, they have the advantage of probing the QCD phase diagram in the regime complementary to that explored by heavy-ion collisions, that is at high baryon densities and low temperatures as well as at substantial isospin fractions.

The ultimate goal of research centered on the QCD phase diagram is to determine the QCD EOS quantitatively. In using heavy-ion collision experiments to explore the QCD phase diagram, the challenge for theory is predicting the experimental signatures of the phenomena associated with the QCD phase structure and interpreting experimental observations in terms of the QCD EOS. 
The task of connecting theory and experiment is the main focus of this review.

One question central to our understanding of the QCD phase diagram is the existence and the location of the QCD critical point, see Fig.~\ref{fig:phase-diagram}. 
This point anchors the expected first-order phase transition separating the QGP and HRG phases of QCD. 
Discovering this QCD first-order phase transition has been a key objective of heavy-ion collision research since its inception.
It is now understood, largely due to lattice calculations,
that such a discontinuous transition does not take place at zero baryon chemical potential, \mbox{$\mu_B=0$}. Instead, at $\mu_B=0$, the
transition between the two phases happens gradually, or smoothly,
\textit{via} a crossover at a temperature of approximately
\mbox{150--160 MeV}, as shown in Fig.~\ref{fig:phase-diagram}. 
While
lattice calculations are unfortunately impeded by the sign problem,
most theoretical approaches --- from RMM to FRG to AdS/CFT ---
consistently indicate that the transition becomes discontinuous above a
certain critical baryon chemical potential, where, by definition, the
QCD critical point is located.

Heavy-ion collisions have the potential
to answer the question of the existence of the QCD critical point and
to determine its location by scanning the QCD phase diagram
\textit{via} varying the crucial experimental control parameter: the
beam energy, or the center-of-mass collision energy per nucleon pair $\snn$, as shown in Fig.~\ref{fig:phase-diagram}. This is the major motivation behind the Beam Energy Scan (BES) program at the Relativistic Heavy Ion Collider (RHIC).

The basic theoretical strategy for
translating the experimental measurements into the knowledge of the
QCD phase diagram is as follows~\cite{An:2021wof}. Using a putative QCD EOS one can
describe the evolution of the hot and dense system created in
heavy-ion collisions --- the fireball --- which expands and cools
hydrodynamically and then breaks up, or freezes out, into observed
particles.  The predicted multiplicities as well as their fluctuations and
correlations depend on the EOS.  Establishing this dependence
theoretically allows one to connect experimental observations to the
QCD EOS.  This is not a simple task due to the highly dynamic nature
of the collisions, 
and this chapter will cover several key
ingredients needed to complete it. 
We briefly outline these ingredients below.

For heavy-ion collisions at intermediate and high energies, the
system's evolution is captured by multistage hydrodynamic simulations.
This includes modeling various stages such as the initial collision, hydrodynamic expansion of the fireball, the transition from thermodynamic degrees of freedom to particles, and the inclusion of hadronic rescatterings. 
These aspects will be discussed in Section~\ref{sec:hydro}.

At low collision energies, microscopic degrees of freedom play a prominent role in the dynamics which proceeds out-of-equilibrium for substantial fractions of the evolution time. 
Modeling heavy-ion collisions in this regime, which probes the highest density regions of the QCD phase diagram available in laboratory experiments, is the subject of Section~\ref{sec:microscopic_transport}.

Finally, many experimental signatures of the QCD critical point are based on
fluctuation and correlation phenomena. The description of fluctuations in a dynamical environment characteristic of
relativistic heavy-ion collisions and, importantly, the translation of
hydrodynamic fluctuations into fluctuations of observables 
has been developed only recently, as discussed in Section~\ref{sec:fluctuations}. 

We aim to cover recent and ongoing developments in these areas and hope this chapter will serve as a comprehensive yet concise guide for researchers engaged or planning to engage in the subject.

%% file: hydro.tex
Heavy-ion collisions generate highly dynamic systems undergoing various phases. Multistage descriptions, incorporating diverse physics, have become the ``standard model'' for characterizing these collisions \cite{Bass:2000ib,Teaney:2001av,Hirano:2005xf,Nonaka:2006yn,Petersen:2008dd,Werner:2010aa,Song:2010mg,Karpenko:2012yf}. A crucial aspect is the quantitative understanding of bulk dynamics, representing the final soft hadrons with transverse momenta below 3 GeV/c (constituting more than 99\% of all particles produced in the collisions). This understanding forms the foundation for exploring other facets of heavy-ion physics, including critical phenomena \cite{Rajagopal:2019xwg,Du:2020bxp}, hard probes \cite{Cao:2024pxc}, and electromagnetic probes \cite{Churchill:2023vpt,Churchill:2023zkk,Churchill:2023hog,Gale:2018vuh,Shen:2023aeg}. Previous comprehensive reviews on multistage descriptions with hydrodynamics as a core can be found in Refs.~\cite{Hirano:2012kj,Gale:2013da,Petersen:2014yqa}. Here, we focus on recent theoretical developments related to Beam Energy Scan (BES) physics for collisions at $\snn\gtrsim 7$ GeV, with particular attention to considerations of finite charge densities. It is important to note that there are also microscopic descriptions, such as \texttt{JAM} \cite{Nara:1999dz}, \texttt{PHSD} \cite{Cassing:2008sv,Bleicher:2022kcu}, \texttt{EPOS} \cite{Werner:2008zza}, \texttt{AMPT} \cite{Lin:2004en}, \texttt{SMASH} \cite{SMASH:2016zqf}, \texttt{UrQMD} \cite{Bass:1998ca,Bleicher:1999xi}, and \texttt{PHQMD} \cite{Aichelin:2019tnk}, which are not covered in this section. Among these, the pure hadronic descriptions hold particular significance for low-energy collisions at $\snn\lesssim 7$ GeV, as discussed in Sec.~\ref{sec:microscopic_transport}.

\subsection{Prehydrodynamic stage}\label{sec:prehydro}

During the initial phase in a heavy-ion collision, the two colliding nuclei first penetrate each other, and, subsequently, the produced system undergoes hydrodynamization—a phase referred to as the prehydrodynamic stage. During this stage, the nucleons or partons initially at beam rapidity ($y_\mathrm{b}$) undergo collisions that result in the loss of some of their initial energy and longitudinal momentum. This lost energy and momentum contribute to the formation of the quark-gluon plasma. As these particles collide and lose energy and longitudinal momentum, their rapidities shift away from the beam rapidity and distribute continuously within the range from beam rapidity to midrapidity. Additionally, the charges, including baryon number, strangeness, and electric charge, carried by the partons also undergo redistribution processes during this phase.\footnote{%
Baryon, strangeness, and electric charges are key observables in heavy-ion collisions owing to their conservation laws and their relevance to the study of strong and electromagnetic interactions and the properties of the QGP.
}
A deep understanding of the energy loss and charge stopping mechanisms plays a pivotal role in determining the initial distributions of quantities such as energy density, flow velocity, and baryon density. These initial conditions serve as the foundation for the entire subsequent evolution of the collision, making this stage the fundamental basis for comprehending the overall dynamic evolution of a heavy-ion collision \cite{Luzum:2013yya}.

In ultra-relativistic collisions, the energy deposition is often modeled using the Glauber model \cite{Miller:2007ri}. The model is based on the assumption that each nucleon or parton can either participate in a collision or remain a spectator. The determination of whether nucleons or partons interact is governed by probabilistic considerations based on collision geometry and cross sections. In ultra-relativistic scenarios, the colliding nuclei are significantly Lorentz-contracted in the beam direction. Consequently, it is reasonable to assume that all collisions happen nearly instantaneously. The subsequent expansion in the longitudinal direction is assumed to be boost-invariant and undergo the so-called Bjorken expansion \cite{Bjorken:1982qr}. In this scenario, it is common to model a (2+1)-dimensional system and analyze the observables measured at midrapidity. In practice, when building the initial transverse energy or entropy distribution, an overall normalization factor is often adjusted to match the measured charged particle multiplicity around midrapidity. Moreover, the baryon density is routinely assumed to be negligible in this region, as it is predominantly occupied by partons with small $x$-values, primarily composed of gluons.

\begin{figure}[t]
    \centering
    \includegraphics[width=0.46\linewidth]{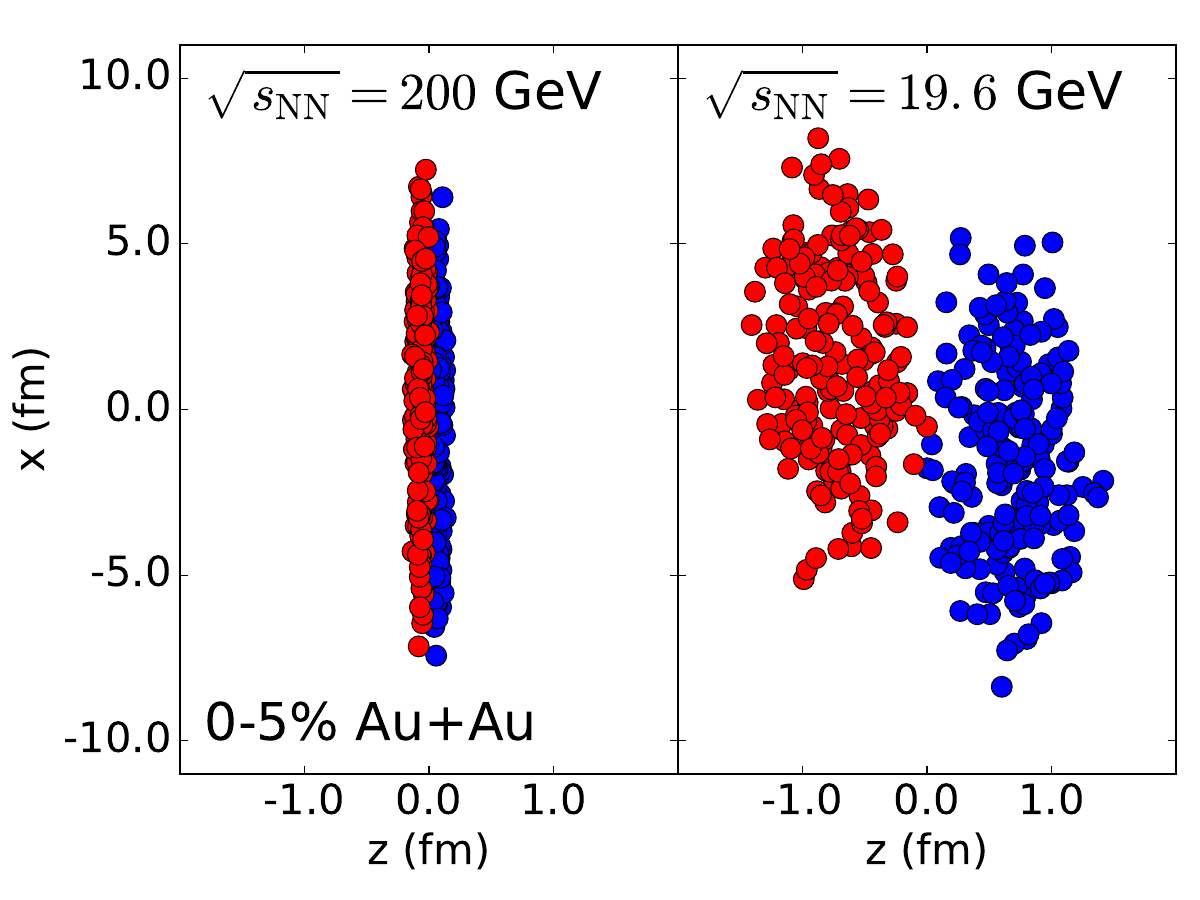}
    \hspace{4mm}
    \includegraphics[width=0.45\linewidth]{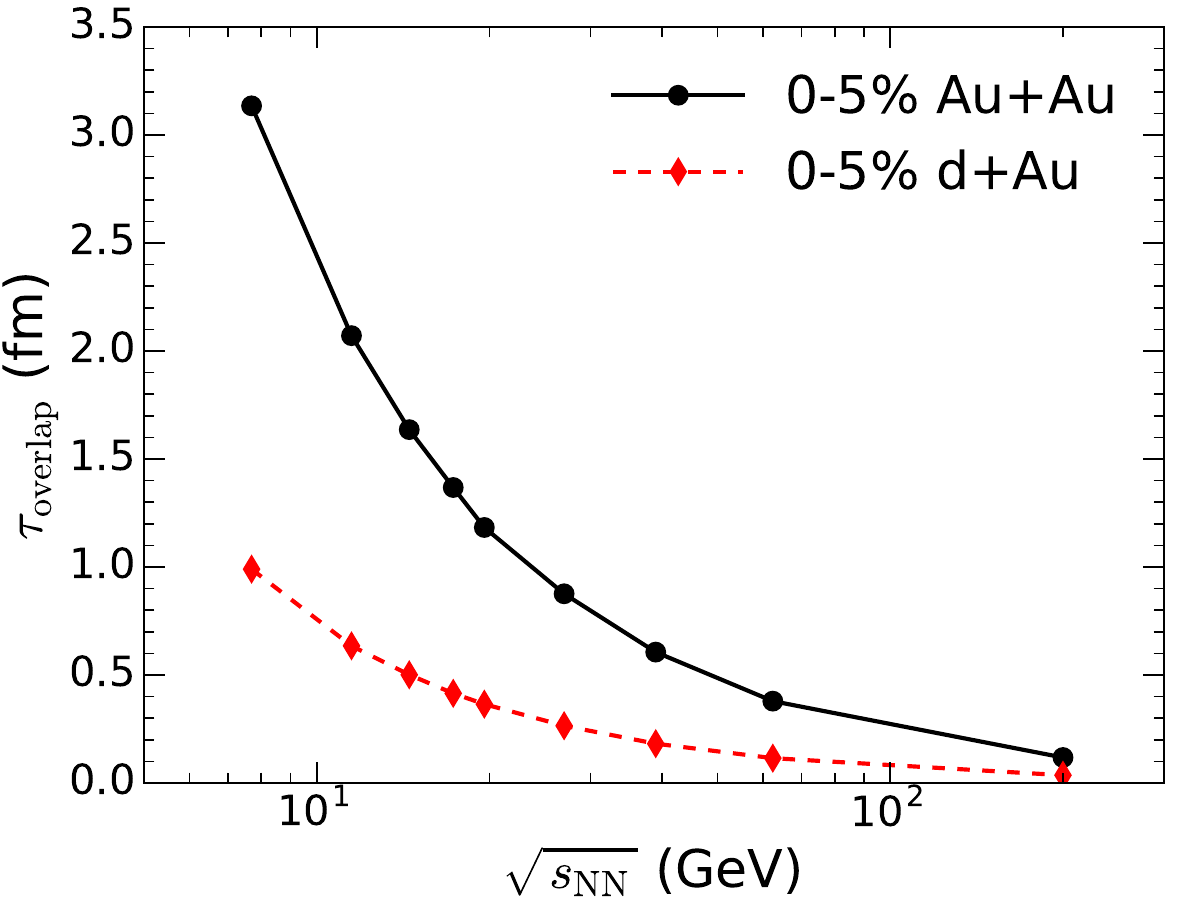}
    \caption{\textit{Left:} Nucleon positions plotted against transverse ($x$) and longitudinal ($z$) coordinates for two distinct collision energies. Note that the thickness in $z$ at 19.6 GeV is visually exaggerated due to the aspect ratio of the plot. \textit{Right:} The nuclear overlapping time for 0-5\% central d+Au and Au+Au collisions depicted as a function of collision energy. Figures from Ref.~\cite{Shen:2017bsr}.}
    \label{fig:overlap}
\end{figure}

However, when we consider heavy-ion collisions at the BES and lower energies, the dynamics of the initial stage become considerably more intricate. This complexity becomes evident when we examine the overlap time \cite{Shen:2017bsr} between the two colliding nuclei in the center-of-mass frame, which is determined by the formula:
\begin{equation}
    t_\mathrm{overlap}=\frac{2R}{\gamma v_z}=\frac{2R}{\sinh(y_\mathrm{b})}\,.
\end{equation}
Here, $R$ is the radius of the colliding nuclei, $\gamma=1/\sqrt{1-v_z^2}$  represents the Lorentz contraction factor, and $y_\mathrm{b}$ is the beam rapidity, calculated as 
$y_\mathrm{b}=\textrm{arccosh}(\snn/(2m_p))$. As collision energy decreases, the overlap time significantly extends and may even become comparable to the duration of the QGP stage (see Fig.~\ref{fig:overlap}). Consequently, in such scenarios, the collision locations in space and time vary considerably, necessitating the construction of dynamic, (3+1)-dimensional initial conditions. Moreover, it's important to note that the net baryon number is non-zero and highly non-uniform, as evidenced by the experimentally measured varied distribution of net proton yields in rapidity \cite{BRAHMS:2003wwg}. Therefore, it becomes imperative to model the initial distributions of baryon charge as an integral aspect of the initial conditions. At present, the primary challenge in constructing (3+1)-dimensional dynamic initial conditions lies in the lack of a comprehensive understanding of the mechanisms governing baryon stopping and energy deposition during this prehydrodynamic phase. This, in turn, makes the extraction of hydrodynamic transport coefficients from experimental measurements an extremely difficult task.

\subsubsection{Parametric initial conditions}\label{sec:para_init}

A straightforward yet effective approach to address the previously mentioned challenges is the use of parametric initial conditions. This method often involves supplementing transverse distributions with longitudinally parametrized profiles. The transverse distributions of energy or entropy and baryon densities are typically derived from nuclear thickness functions through Glauber-like models, while the longitudinal profiles are parameterized to account for specific observables sensitive to longitudinal bulk dynamics, such as longitudinal decorrelation and rapidity-dependent identified particle yields. These initial conditions are employed in subsequent hydrodynamic evolution at a specific initial time ($\tau_0$), and the initial longitudinal flow is initiated as Bjorken flow. These models become less reliable at lower beam energies, where the overlap time is long and the Bjorken boost-invariant approximation becomes less appropriate. However, they retain their strength in a different way by enabling the identification of favored longitudinal profiles from experimental data more cleanly, compared to dynamical models with stochastic fluctuations. These extracted profiles can provide valuable insights into the mechanisms of energy deposition and baryon stopping.

For symmetric collision systems, the initial energy or entropy density in spacetime rapidity is often parametrized using a plateau-like function. This function is flat around midrapidity and smoothly transitions to half of a Gaussian function in the forward and backward spacetime rapidity ($\eta_s$) regions \cite{Morita:1999vj,Hirano:2001yi,Hirano:2001eu,Morita:2002av,Hirano:2002ds}. These initial profiles are tuned to match the measured charged particle multiplicity in pseudorapidity ($\eta$), while the resulting profile can be influenced by various factors in the subsequent hydrodynamic evolution, such as the equation of state, initial time, and shear viscosity \cite{Satarov:2006iw,Bozek:2009ty,Bozek:2007qt}.

It was later recognized that the presence of a nonzero directed flow of charged particles, $v_1^{\rm ch}(\eta)$, indicates the need to break symmetry in the reaction plane. This can be achieved by introducing asymmetry in either the initial flow or the initial density, or in both \cite{Bozek:2010bi}. The underlying reason for this is that the left- and right-going participants within the projectile and target at a given point in the transverse plane possess an imbalance in momentum, resulting in a nonzero total longitudinal momentum. To account for this total longitudinal momentum, a shift in the longitudinal density profile, as seen in the shifted initial densities, was introduced \cite{Hirano:2002ds}. However, this shifted initial profile was found to yield a $v_1^{\rm ch}(\eta)$ with an incorrect dependence on rapidity \cite{Bozek:2010bi}. Nevertheless, a recent study that imposed local energy-momentum conservation by introducing a similar shift can generate correct rapidity-dependent $v_1$ for pions \cite{Shen:2020jwv}. Furthermore, the authors also introduced an initial longitudinal flow velocity in addition to the shift in initial energy density to ensure local energy-momentum conservation. Their findings indicated that a concurrent explanation of both the global polarization of the $\Lambda$ hyperon and the slope of the pion's directed flow $v_1^{\pi}(y)$ can substantially constrain the longitudinal flow at the initial stages of hydrodynamic evolution \cite{Ryu:2021lnx,Alzhrani:2022dpi}.

\begin{figure}[t]
    \centering
    \includegraphics[width=0.38\linewidth]{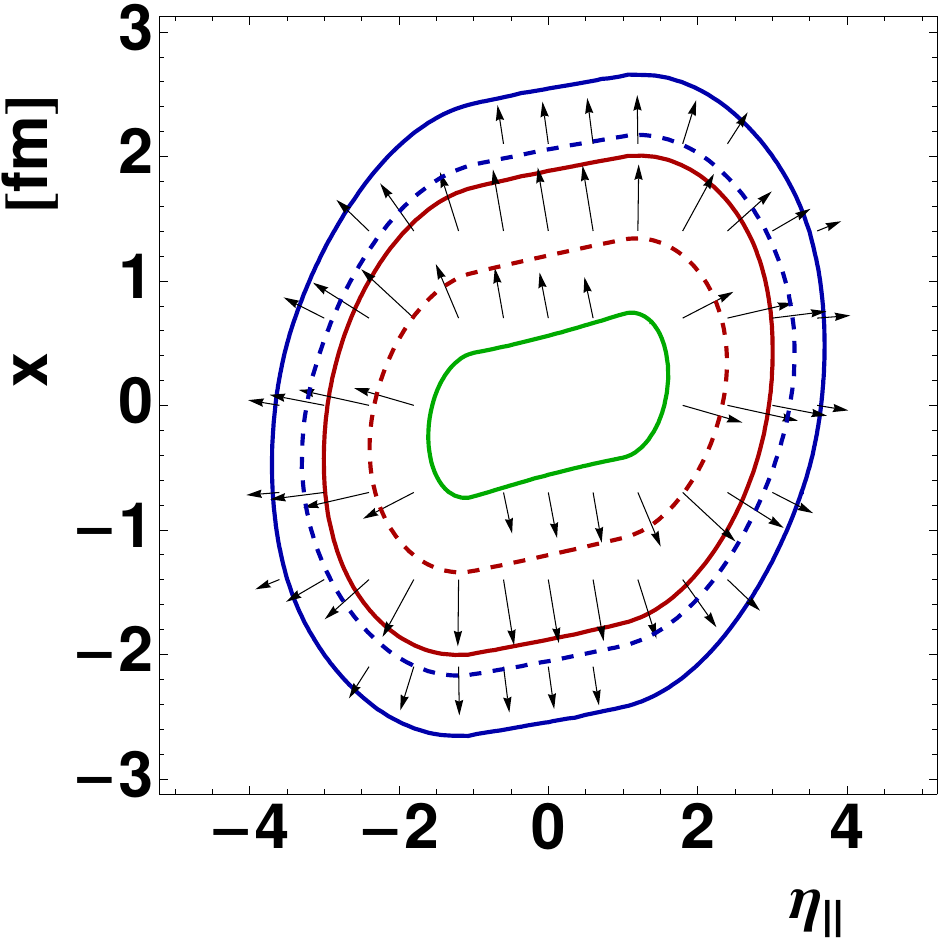}\qquad
	\hspace{4mm}
    \includegraphics[width=0.47\linewidth]{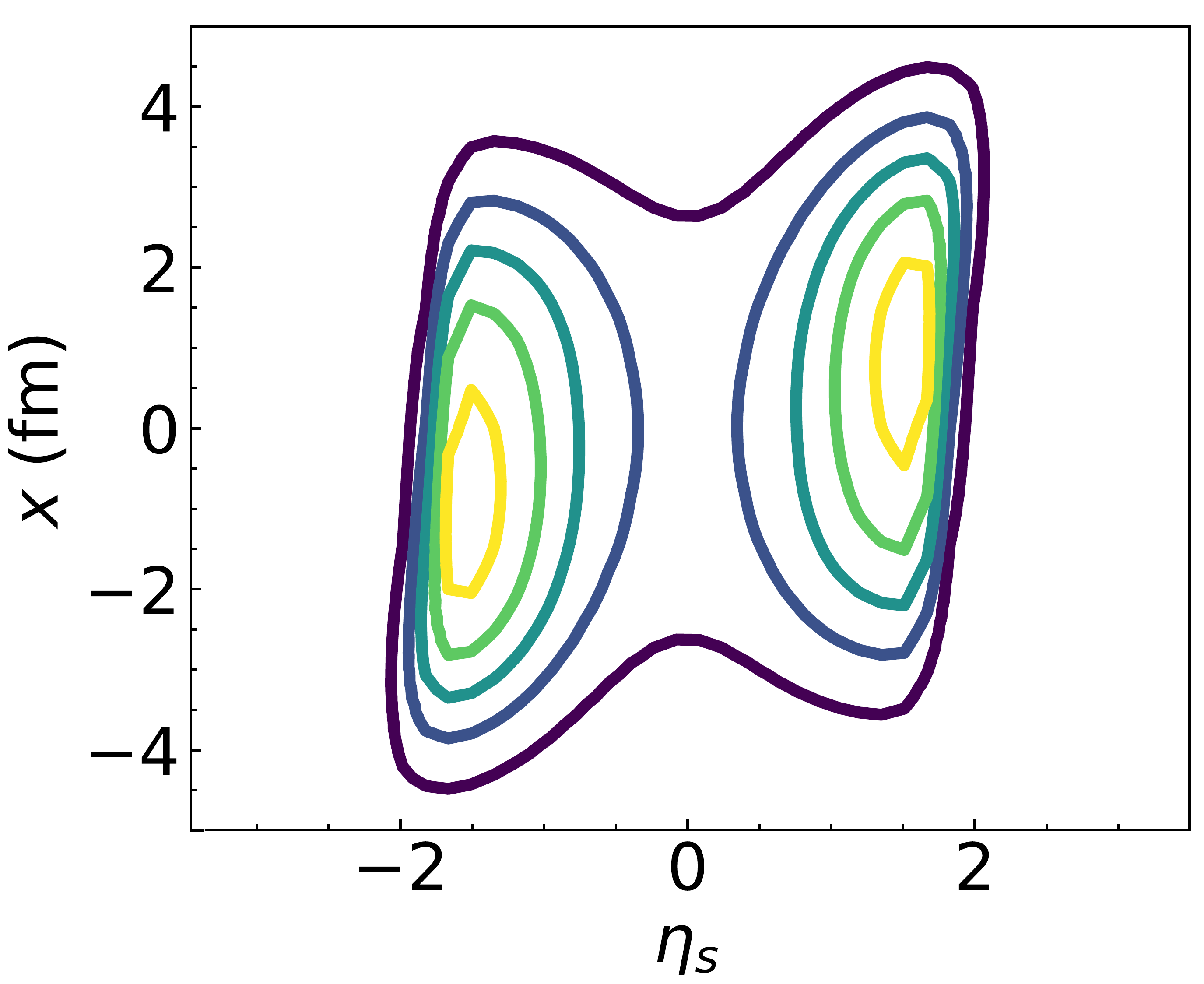}
    \caption{Parametric initial distributions for \textit{left:} energy density \cite{Bozek:2010bi} and \textit{right:} baryon density \cite{Du:2022yok} in the reaction plane.}
    \label{fig:para_init}
\end{figure}

An alternative category of initial conditions considers a preference for gluon emission near the rapidity of the participant nucleon \cite{Bozek:2010bi}. For example, in this framework, nucleons from the projectile with positive rapidity emit more gluons in the forward rapidity regions than in the backward regions \cite{Bialas:2004su,Bzdak:2009xq,Gazdzicki:2005rr,Bzdak:2009dr}. The initial energy density is constructed as a sum of contributions from both forward- and backward-moving participant nucleons, and possibly binary collisions that are assumed to contribute symmetrically. This discrepancy in forward and backward emission results in a tilt of the source within the reaction plane (see the left panel of the Fig.~\ref{fig:para_init}), breaking symmetry in the longitudinal direction and generating directed flow $v_1(y)$ with a negative slope around midrapidity for charged particles and mesons. The asymmetric deposition of entropy also results in a rapidity-dependent orientation for the event plane angle \cite{Bozek:2010vz}, making it useful for exploring longitudinal decorrelation \cite{Bozek:2015bha,Bozek:2015bna}. Such tilted initial conditions continue to be widely used in current studies on directed flows \cite{Chatterjee:2017ahy,Bozek:2022svy,Du:2022yok}. Initial conditions that aim for realistic event-by-event simulations, like the \texttt{TRENTo} 3D model \cite{Ke:2016jrd,Soeder:2023vdn}, build upon similar ideas of extending two-dimensional transverse profiles around midrapidity \cite{Moreland:2014oya} with the inclusion of longitudinal profiles.

The initial profile of baryon density has received considerably less attention in the community. This could be attributed to several factors, with two notable ones as follows: From an experimental standpoint, investigating baryon charge necessitates the measurement of baryons' excess over antibaryons (typically, proton and antiproton), which requires particle identification. This, in turn, leads to a limited number of available measurements for studying the longitudinal baryon profile. On the theoretical side, investigating the evolution of the baryon distribution significantly adds to the complexity of simulations. The initial longitudinal baryon profile is parametrized as the sum of two Gaussian functions with peaks symmetrically centered around midrapidity  \cite{Ishii:1992xi,Hirano:2001yi,Morita:2002av}. This parametrization is motivated by the double-humped structure observed in the net-proton yields in rapidity. In some cases, experimentalists employ a similar approach by fitting the net proton yields with the sum of two Gaussian functions, enabling them to reconstruct the entire distribution (see, e.g., Ref.~\cite{E917:2000spt}). In recent studies, similar initial baryon profiles are still commonly used, while the Gaussian function associated with the projectile or target has been generalized into an asymmetric Gaussian function \cite{Denicol:2018wdp,Shen:2020jwv,Du:2021zqz,Ryu:2021lnx,Alzhrani:2022dpi,Du:2023gnv}. This function features a peak with half-Gaussian functions on each side, and it allows for asymmetric widths and thereby enhances flexibility.

The baryon profile described above can easily reproduce the double-humped net proton yields in rapidity. However, it tends to yield a $v_1(y)$ with a notably positive slope, and its magnitude significantly exceeds experimental measurements \cite{Shen:2020jwv}. This occurs because the directed flow of baryons is primarily due to the initial asymmetric baryon distribution with respect to the beam axis driven by the transverse expansion \cite{Du:2022yok}. To tackle this problem, Ref.~\cite{Du:2022yok} introduced an additional rapidity-independent plateau component within the baryon profile (see the right panel of Fig.~\ref{fig:para_init}) \cite{Du:2022yok}. This component symmetrically contributes to the baryon density with respect to the beam axis in the reaction plane, resulting in a substantial reduction in $v_1(y)$ magnitude for baryons while effectively reproducing the net proton yields. With this component, the authors found that the primary characteristic features of $v_1(y)$ are naturally explained, including the sign change in the slope of $v_1(y)$ around midrapidity. The inclusion of a plateau component in the baryon profile implies a new baryon stopping mechanism, which can possibly be attributed to the string junction conjecture \cite{Kharzeev:1996sq,Sjostrand:2002ip}. Other studies also exist intending to understand the splitting of proton-antiproton directed flow by adjusting the initial baryon profile \cite{Bozek:2022svy,Parida:2022ppj}.

\subsubsection{Dynamical initialization}\label{sec:dynamical_init}

As previously discussed and evident in Fig.~\ref{fig:overlap}, collisions occurring at center-of-mass energies in the range of tens of GeV and below exhibit long overlapping time due to reduced Lorentz contraction of the colliding nuclei and, consequently, spacetime-dependent interactions among nucleons or partons. This phenomenon results in different regions of the fireball ``hydrodynamizing'' at various times. Thus, transitioning to a hydrodynamic description inherently necessitates a (3+1)-dimensional dynamical initial condition. The application of this approach was initially explored during the SPS era \cite{Kajantie:1982jt,Kajantie:1982nh}. Recently, there has been a resurgence in the development of more intricate models aimed at achieving quantitative simulations for collisions at BES \cite{Shen:2017bsr,Shen:2022oyg,Du:2018mpf,Akamatsu:2018olk,De:2022yxq}.

\begin{figure}[t]
    \centering
    \includegraphics[width=0.45\linewidth]{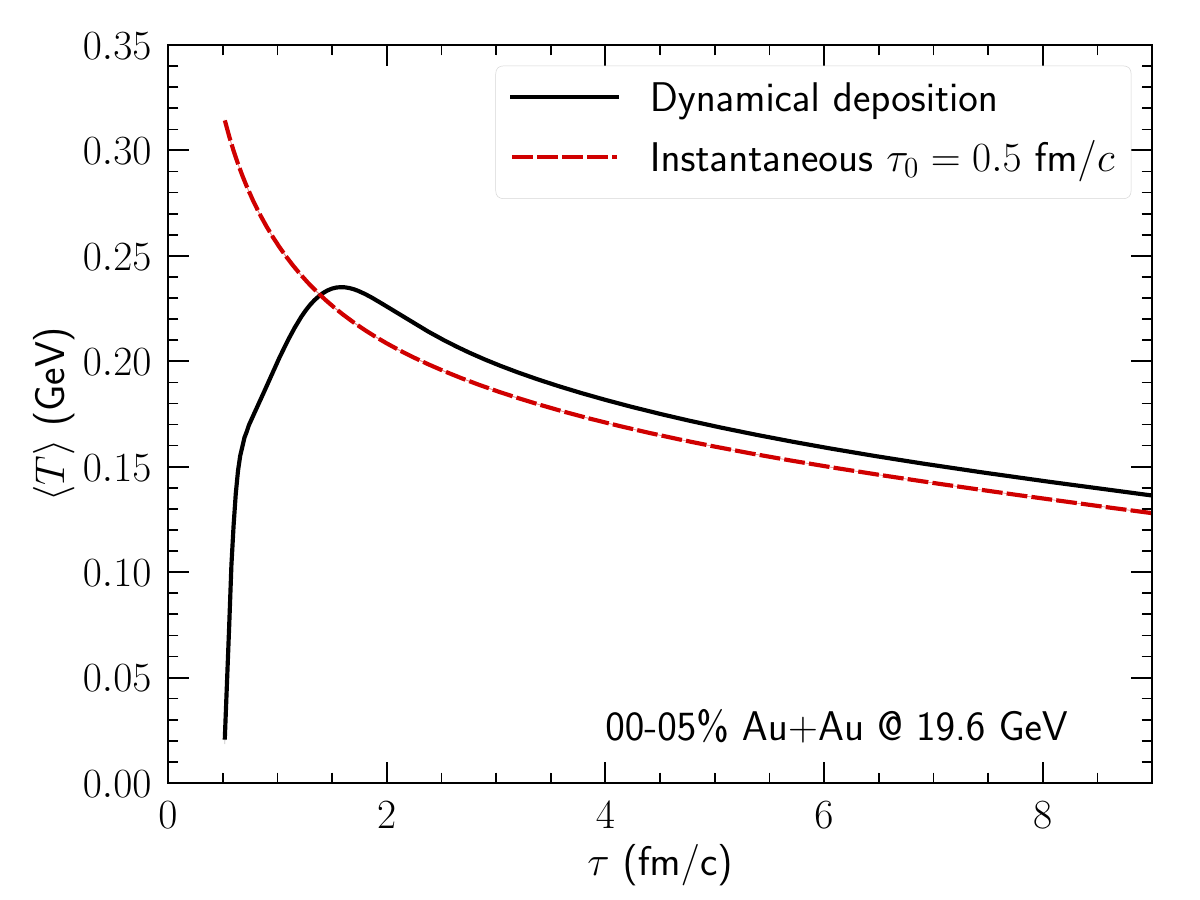}
	\hspace{4mm}
    \includegraphics[width=0.45\linewidth]{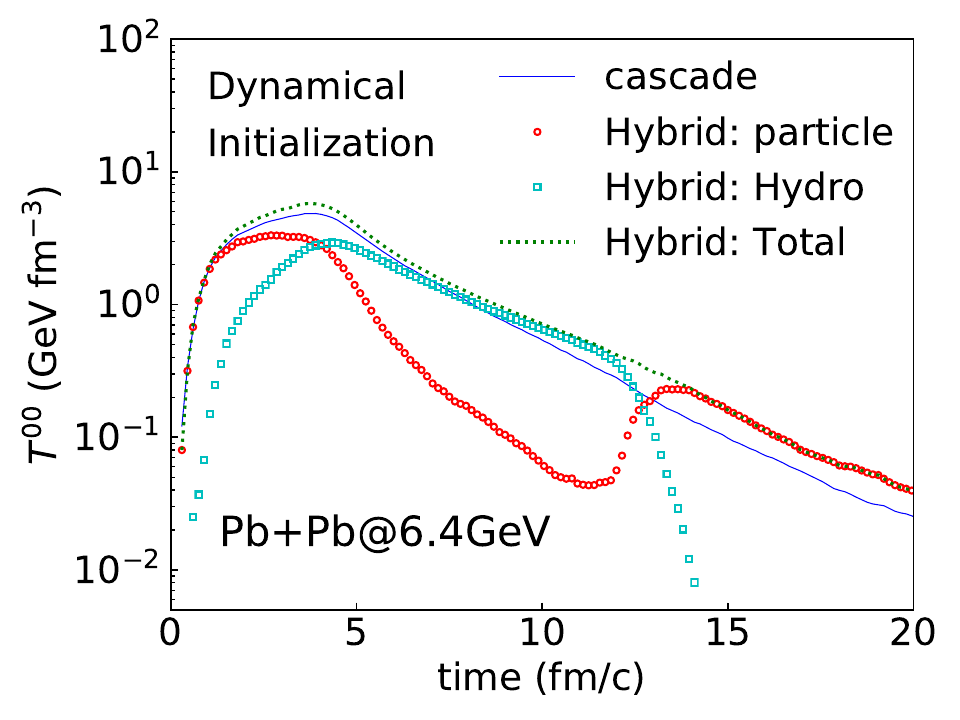}
    \caption{\textit{Left:} Time evolution of the averaged temperature in Au+Au collisions at $\snn=19.6$ GeV for both dynamical initialization and instantaneous hydrodynamization scenarios \cite{Shen:2023aeg}. \textit{Right:} Time evolution of the energy density from \texttt{JAM} hybrid simulations in central Pb+Pb collisions at $\snn=6.4$ GeV \cite{Akamatsu:2018olk}. Circles represent the contribution from particles, while squares represent the fluid contribution to the energy density.}
    \label{fig:dynamical_init}
\end{figure}

In this dynamical initialization approach, a spacetime-dependent transition occurs from a microscopic partonic description to a macroscopic hydrodynamic one once certain criteria are met, such as reaching a specified energy density. Throughout this transition, conservation laws persist for energy-momentum and charges accounting for contributions from both the partons and the subsequent fluid.
Expressed through the local conservation laws for energy-momentum and conserved charges, the following dynamical initialization equations can be obtained:
\begin{eqnarray}
\partial_\mu T^{\mu\nu}_{\fluid}(x) &=&J_{\source}^\nu(x) \equiv-\partial_\mu T^{\mu\nu}_{\prt}(x) \;,\label{eq-js}\\
\partial_\mu N^\mu_{\fluid}(x) &=&\rho_{\source}(x) \equiv-\partial_\mu N^\mu_{\prt}(x) \;,\label{eq-rhob}
\end{eqnarray}
where $J_{\text{source}}^\nu$ and $\rho_{\text{source}}$ represent the \textit{dynamical sources} contributed by the partons to the fluid.\footnote{%
This resembles modeling how the medium responds when energetic partons traverse through it in some studies focused on jet-medium interactions.
}
These equations illustrate that, as collisions occur among partons, additional energy-momentum and charges (represented by $J_{\text{source}}^\nu$ and $\rho_{\text{source}}$) are deposited into the fluid. This leads to a progressive increase in the densities of the fluid component within the collision fireball during the prehydrodynamic stage. On average, these densities peak once the collisions conclude, marking the transition to a purely hydrodynamic description (see Fig.~\ref{fig:dynamical_init}).

Implementing Eqs.~(\ref{eq-js},\,\ref{eq-rhob}) practically involves constructing $J_{\source}^\nu(x)$ and $\rho_{\source}(x)$ or continuous $T^{\mu\nu}_{\prt}(x)$ and $N^\mu_{\prt}(x)$ from discrete partons. This construction requires modeling collision dynamics, which encompasses energy or rapidity loss, typically described by stochastic processes during collisions. The energy loss of colliding partons carrying baryon charges might contribute to initial baryon stopping, thereby correlating energy deposition with baryon stopping. Understanding these mechanisms involves phenomenological studies, as these processes are not clear from first principles. Modeling these mechanisms often draws insights from longitudinal profiles preferred by experimental measurements used in parametric initial conditions discussed in Sec.~\ref{sec:para_init}. These mechanisms are significant in grasping baryon fluctuations during the initial stage, which contribute to measured proton cumulants and thus are pivotal for their interpretation \cite{Shen:2017bsr}. 

Recent advances in dynamical initialization have encompassed diverse approaches like color string dynamics \cite{Shen:2017bsr,Shen:2022oyg} and transport models such as \texttt{UrQMD} \cite{Du:2018mpf} and \texttt{JAM} \cite{Akamatsu:2018olk}. We note that these approaches are unable to elucidate the mechanisms of thermalization \cite{Berges:2020fwq,Schlichting:2019abc} or even hydrodynamization. The criteria for the time-dependent transition between initial conditions and hydrodynamics are somewhat ad hoc. Nevertheless, these methods are crucial for comprehending boost-noninvariant dynamics and establishing correlations between rapidities in both coordinate and momentum spaces during the prehydrodynamic stage \cite{Monnai:2015sca,Shen:2017bsr,Du:2021fyr,Shen:2022oyg}. 
Such correlations are often oversimplified in parametric initial conditions that assume Bjorken flow, where the rapidity ($y$) is equated with the spacetime rapidity ($\eta_s$). The non-Bjorken prehydrodynamic flows can significantly impact the subsequent hydrodynamic evolution and thus warrant a more thorough investigation. Additionally, the nontrivial impact of initial dissipative effects on the system's evolution through the phase diagram should also be investigated (see Sec.~\ref{sec:dissipation}).

There is another category of initial conditions that incorporate longitudinal dependence, yet the transition to hydrodynamics occurs at a fixed time. For instance, some studies utilize initial conditions derived from transport models, including \texttt{UrQMD} \cite{Petersen:2008dd,Karpenko:2015xea}, \texttt{SMASH} \cite{Schafer:2021csj}, and \texttt{AMPT} \cite{Pang:2012he}. The exploration of baryon or energy stopping has also been conducted using Color Glass Condensate approaches \cite{Li:2016wzh,McLerran:2018avb,Li:2018ini}, holographic approaches \cite{Casalderrey-Solana:2016xfq,vanderSchee:2015rta,Attems:2018gou,Kolbe:2020hem}, and the parton-based Gribov-Regge model \texttt{NEXUS} \cite{Andrade:2008xh,Drescher:2000ec}. These models offer more intricate longitudinal distributions than parametric ones while maintaining a fixed-time transition to hydrodynamics, thereby avoiding the complexity of a dynamical initialization scheme.

Hydrodynamic descriptions with dynamical sources find various applications on diverse topics. The three-fluid hydrodynamic model \cite{Ivanov:2013wha,Ivanov:2013yqa,Ivanov:2013yla,Cimerman:2023hjw} characterizes interactions among the fluids of the target, projectile, and fireball via friction terms, which serve a comparable role to the dynamical source terms. The utilization of Eqs.~(\ref{eq-js},\,\ref{eq-rhob}) extends beyond modeling the prehydrodynamic stage in nuclear collisions at lower beam energies. They have been utilized in various contexts, including modeling core-corona interactions \cite{Kanakubo:2019ogh,Kanakubo:2021qcw} and investigating the impact of mini-jets on the bulk dynamics \cite{Okai:2017ofp,Pablos:2022piv}.

Expanding initial conditions from 2-dimensional transverse distributions at a fixed time to a (3+1)-dimensional framework encompassing both time and rapidity dependence presents considerable challenges, yet ongoing advancements show promise. This extension is crucial for laying the foundational groundwork to understand the thermodynamic and transport properties of QCD matter at finite chemical potentials created in Beam Energy Scan collisions. Comprehensive understanding hinges on rapidity-dependent measurements, including crucial observables like identified particle yields, anisotropic flow coefficients, event plane decorrelation, and Hanbury-Brown--Twiss (HBT) interferometry. Additionally, 3-dimensional jet tomography serves as a means to scrutinize the longitudinal structure of initial conditions \cite{Adil:2005qn,Adil:2005bb}. A systematic exploration across various measurements in rapidity, covering a range of collision centralities, beam energies, and system sizes, becomes imperative. Such an analysis is critical to evaluate which description of the initial state yields a more coherent and comprehensive explanation for various available rapidity data.

\subsection{Hydrodynamics with multiple conserved charges}
\begin{figure}[t]
    \centering
    \includegraphics[width=\linewidth]{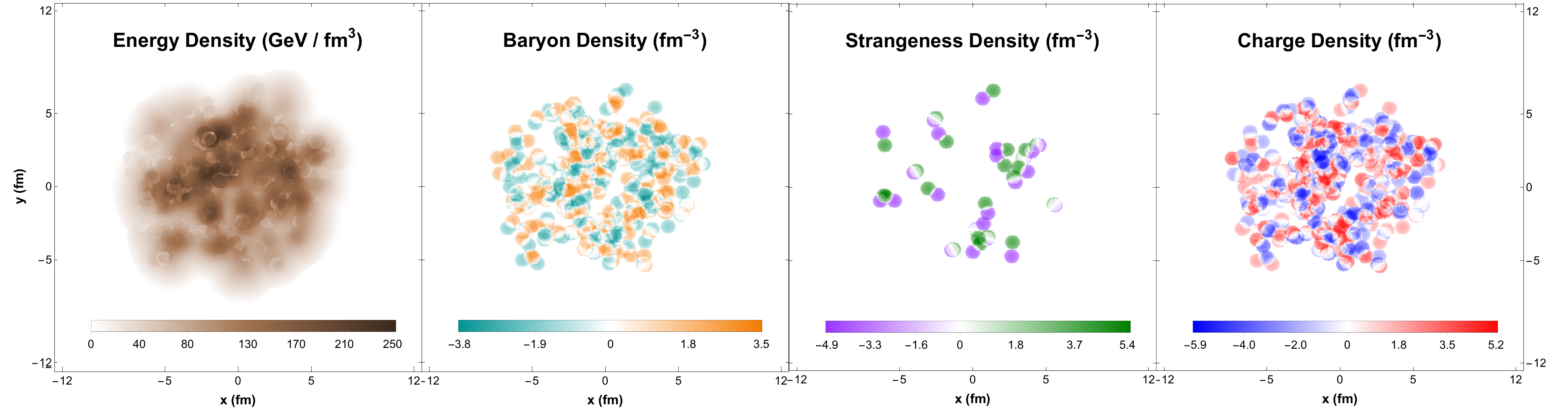}
    \caption{Energy and charge density distributions for an ICCING event with Green's function evolution of energy and charge perturbations from gluon to quark-antiquark splittings \cite{Carzon:2019qja,Carzon:2023zfp}.}
    \label{fig:multi_charge_init}
\end{figure}

As mentioned previously, in the context of ultra-relativistic nuclear collisions, the primary focus often centers on the midrapidity charge-neutral region that is formed through the interaction of low-$x$ gluons, which do not carry conserved charges. However, at BES and lower energies,  there is a growing need to investigate the longitudinal dynamics across both forward and backward rapidity regions, where the baryon density can be notably higher, and, in fact, a significant fraction of the incoming baryon charge can be stopped even near midrapidity. 
The measurements of hadron species with different charges indicate that the fluid created in intermediate- and low-energy collisions carries various quantum numbers, where we must consider the conservation of baryon number ($B$), strangeness ($S$), and electric charge ($Q$), which are conserved by the strong interaction and all hold significant relevance in the phenomenology of heavy-ion collisions (see Fig.~\ref{fig:multi_charge_init}). Consequently, it becomes imperative to develop a hydrodynamic framework that accounts for the dynamic evolution of multiple conserved charges.

\subsubsection{Conservation equations}
For the systems involving $N_q$ types of charges, hydrodynamics serves as a macroscopic theory that describes the spacetime evolution of the energy-momentum tensor $T^{\mu\nu}$ and the charge currents $N_q^\mu$, encompassing a total of $10{\,+\,}4N_q$ independent components (10 from $T^{\mu\nu}$ and 4 from each charge current). A set of evolution equations that govern these components can be derived from the conservation laws for energy, momentum, and charges:
\begin{equation}
    \partial_\mu T^{\mu\nu}=0,\,\quad\partial_\mu N_q^\mu=0\,,\label{eq:conslaws}
\end{equation}
where $q{\,\in\,}\{B,\,Q,\,S\}$, and $\partial_\mu$ represents the covariant derivative in a general curvilinear coordinates. In the current context involving $BQS$ charges, we shall take $N_q{\,=\,}3$ from now on.

The independent components of $T^{\mu\nu}$ and $N_q^\mu$ have a more physical representation in terms of the hydrodynamic decomposition of these tensors, given by:
\begin{eqnarray}
T^{\mu\nu} &=& \epsilon u^{\mu}u^{\nu}-(p+\Pi)\Delta^{\mu\nu}+\pi^{\mu\nu}\;, \label{eq-dec-T}\\
N_q^{\mu} &=& n_q u^{\mu}+n_q^{\mu}\;,\quad q{\,\in\,}\{B,\,Q,\,S\}. \label{eq-dec-N}
\end{eqnarray}
Here, the flow four-velocity $u^\mu(x)$, normalized with $u^\mu u_\mu=1$, is defined as the time-like eigenvector of the energy-momentum tensor:
\begin{equation}\label{eq:flow_landau}
T^{\mu\nu}u_\nu=\epsilon u^\mu\;,
\end{equation}
and it specifies the local rest frame (LRF) of the fluid at point $x$; this definition employs the so-called ``Landau frame.''\footnote{%
In the Landau frame, the total energy-momentum diffusion current (or heat flow vector) is zero, i.e., $W^\mu\equiv u_\nu\delta T^{\mu\nu}=0$. Alternatively, one might opt for the Eckart frame to define the local rest frame by requiring that the net charge diffusion current is zero \cite{Monnai:2019jkc}. However, in scenarios where a charge-neutral fluid is produced in ultra-relativistic collisions or in a fluid with multiple conserved charges generated in lower-energy collisions, the definition of such a frame is less suitable or even ill-defined \cite{Fotakis:2022usk}. One can also define the theory in a general hydrodynamic frame \cite{Rocha:2023ilf}.
} 
The tensors $u^{\mu}u^{\nu}$ and $\Delta^{\mu\nu} \equiv g^{\mu\nu} - u^{\mu}u^{\nu}$ are projectors on the temporal and spatial directions in the LRF. $\epsilon$ and $n_q$ represent the energy and net charge densities in the LRF, and they can be obtained as the following projections of $T^{\mu\nu}$ and $N^\mu$:
\begin{equation}
\epsilon = u_\mu T^{\mu\nu} u_\nu\;,\qquad n_q = u_\mu N_q^{\mu}\;.\label{eq-landau}
\end{equation}
From these quantities, the local equilibrium pressure $p$ is determined using the equation of state, $p=p(\epsilon,n_B, n_Q, n_S)$, at finite charge densities (see Sec.~\ref{sec:eos} for more discussion). The shear stress tensor $\pi^{\mu\nu}$, bulk viscous pressure $\Pi$, and net charge diffusion currents $n_q^\mu$ describe dissipative flows that account for deviations from local equilibrium. The shear stress tensor $\pi^{\mu\nu}$ is subject to the conditions of tracelessness, $\pi^\mu_\mu=0$, and orthogonality, $\pi^{\mu\nu}u_\mu=0$; thus, it has 5 unknowns. The diffusion currents satisfy $u_\mu n_q^{\mu}=0$ and represent the flow of charges in the local rest frame; thus, they have $3N_q$ unknowns. The bulk viscous pressure $\Pi$ is another unknown. Therefore, we have $6+3N_q$ unknown dissipative components.

Using the decomposition (\ref{eq-dec-T},\ref{eq-dec-N}) the conservation laws (\ref{eq:conslaws}) can be brought into the physically intuitive form \cite{Jeon:2015dfa} 
\begin{eqnarray}
 D\epsilon &=& -(\epsilon{+}p{+}\Pi)\theta + \pi_{\mu\nu}\sigma^{\mu\nu}\;,
\label{eq:vhydro-E}\\
  (\epsilon{+}p{+}\Pi)\, Du^\mu &=& \nabla^\mu(p{+}\Pi)
    - \Delta^{\mu\nu} \nabla^\sigma\pi_{\nu\sigma} + \pi^{\mu\nu} Du_\nu\;,
\label{eq:vhydro-u}\\
 Dn_q &=& -n_q\theta -\nabla_\mu n_q^\mu\;.
\label{eq:vhydro-N}
\end{eqnarray}
Here $D=u_\mu \partial^\mu$ denotes the time derivative in the LRF, $\theta=\partial_\mu u^\mu$ is the scalar expansion rate, $\nabla^\mu=\partial^{\langle\mu\rangle}$ (where generally $A^{\langle\mu\rangle} \equiv \Delta^{\mu\nu}A_\nu$) denotes the spatial gradient in the LRF, and $\sigma^{\mu\nu} = \nabla^{\langle\mu}u^{\nu\rangle}$ (where generally $B^{\langle\mu\nu\rangle} \equiv \Delta^{\mu\nu}_{\alpha\beta} B^{\alpha\beta}$, with the traceless spatial projector $\Delta^{\mu\nu}_{\alpha\beta} \equiv \frac{1}{2} (\Delta^\mu_\alpha \Delta^\nu_\beta + \Delta^\nu_\alpha \Delta^\mu_\beta) - \frac{1}{3} \Delta^{\mu\nu} \Delta_{\alpha\beta}$) is the shear flow tensor. Note that there are $1+3+N_q=7$ independent conservation equations in Eqs.~(\ref{eq:vhydro-E}-\ref{eq:vhydro-N}), and the pressure $p$ is determined from the equation of state. These equations provide a clear insight into the underlying physics, especially when considered in the local rest frame. 

Notably, the bulk viscous pressure, denoted as $\Pi$, is always added to the equilibrium pressure $p$, making $p{+}\Pi$ the effective pressure. Eq.~\eqref{eq:vhydro-E} demonstrates that, in the ideal scenario, the energy density $\epsilon$ decreases within expanding fluids, as indicated by the fact that $D\epsilon$ is negative when the expansion rate $\theta$ is positive. The presence of the shear viscous term, $\pi_{\mu\nu}\sigma^{\mu\nu}$, mitigates the decrease in energy density when it is positive, a phenomenon known as ``viscous heating.'' However, in the case of far-off-equilibrium fluids, this term can turn negative, indicating what is referred to as ``viscous cooling'' \cite{Chattopadhyay:2022sxk}. Eq.~\eqref{eq:vhydro-N} conveys a similar message regarding charge densities $n_q$ as Eq.~\eqref{eq:vhydro-E} does for energy density. A diffusion current $n_q^\mu$ can either reduce or increase the charge density by carrying away or bringing in charges to a particular location via the divergence term. Eq.~\eqref{eq:vhydro-u} bears a resemblance to Newton's second law: the acceleration of the flow velocity $Du^\mu$ is driven by the gradient of the effective pressure $\nabla^\mu(p{+}\Pi)$, while the effective enthalpy density $(\epsilon{+}p{+}\Pi)$ serves as the inertia. The bulk viscous pressure $\Pi$ tends to reduce the driving force of the radial flow, if it is negative. Moreover, the presence of the shear viscous term, $\pi^{\mu\nu} Du_\nu$, couples the flow evolution in different directions. 

As previously mentioned, once the equation of state (EOS) is given, there remain $10+4N_q$ independent unknowns in $T^{\mu\nu}$ and $N_q^\mu$, and $4+N_q$ evolution equations arising from the conservation laws \eqref{eq:vhydro-E}-\eqref{eq:vhydro-N}. To complete the equation system, we require additional $6+3N_q$ evolution equations for the dissipative components. The current state-of-the-art in relativistic dissipative hydrodynamic formalism is built upon the pioneering work of M\"uller \cite{Muller:1967zza} and Israel-Stewart \cite{Israel:1979wp} (for comprehensive overviews, see Refs.~\cite{Denicol:2014loa,Rocha:2023ilf}). For instance, the Denicol-Niemi-Moln\'ar-Rischke (DNMR) approach \cite{Denicol:2012cn} derives the equations of motion from the Boltzmann equation using the method of moments, applying a systematic power-counting scheme in Knudsen and inverse Reynolds numbers. This method is applied to a fluid with a single conserved charge. Using the same methodology, Ref.~\cite{Fotakis:2022usk} has derived multicomponent relativistic second-order dissipative hydrodynamics for a reactive mixture of various particle species with $N_q$ conserved charges. The dynamic equations for the $BQS$ diffusion currents were obtained (see also Ref.~\cite{Hu:2022vph}). Additionally, the stability conditions associated with these equations were examined, as discussed in Refs.~\cite{Brito:2020nou,Almaalol:2022pjc}. Earlier efforts to derive these equations encompassed the pioneering work in Refs.~\cite{Prakash:1993bt,Monnai:2010qp,Monnai:2012jc,Kikuchi:2015swa}. Together with the $4+N_q$ evolution equations arising from the conservation laws, the resulting $6+3N_q$ second-order equations of motion for the dissipative quantities can close the system. 

\subsubsection{Hydrodynamic quantities from kinetic theory}

The macroscopic hydrodynamic formalism can be derived from the underlying microscopic kinetic theory governed by the Boltzmann equation \cite{Denicol:2014loa}. In the case of a mixture consisting of various particle species, each species $i$ is characterized by its single-particle distribution function, $f_{i, \mathbf{p}}\equiv f(x,p_i)$. The spacetime evolution of these distribution functions is governed by the relativistic Boltzmann equation, given as:
\begin{equation}
p^\mu_i\partial_\mu f_{i, \mathbf{p}}=C[f_{i, \mathbf{p}}]\,,\label{boltzeq}
\end{equation}
where external forces are neglected, and $C[f_{i, \mathbf{p}}]$ represents the collision term, which depends on all the distribution functions. As indicated by the Boltzmann $H$-theorem, collisions on the right-hand side drive the system towards the local equilibrium state $f^0_{i, \mathbf{p}}$, where entropy is maximized. Meanwhile, the derivatives on the left-hand side push the distribution $f_{i, \mathbf{p}}$ away from the local equilibrium state $f^0_{i, \mathbf{p}}$ with a deviation $\delta f_{i, \mathbf{p}}$, denoted as
\begin{equation}\label{eq:fk_decomp}
    f_{i, \mathbf{p}}\equiv f^0_{i, \mathbf{p}}+\delta f_{i, \mathbf{p}}\,.
\end{equation}
Here, collisions and derivatives compete in the evolution towards local equilibrium. When the collision terms exhibit large cross sections, and/or when the thermodynamic and flow gradients are relatively small, the influence of collisions becomes dominant in the competition. In such cases, the distribution $f_{i, \mathbf{p}}$ tends to approach the equilibrium state $f^0_{i, \mathbf{p}}$, leading to a small out-of-equilibrium part $\delta f_{i, \mathbf{p}}$.

The equilibrium state $f^0_{i, \mathbf{p}}$ is given by the J\"uttner distribution function
\begin{equation}\label{eq:fk_eq}
    f^0_{i, \mathbf{p}}= \left[\exp\left(\frac{u_\mu(x) p^\mu_i- \mu_i(x)}{T(x)}\right) +a_i\right]^{-1}.
\end{equation}
Here $a_i=\pm1$ accounts for the fermionic ($+$) or bosonic ($-$) quantum statistics of the particle species $i$, and $u^\mu(x)$ is the fluid flow velocity, $T(x)$ the local temperature and $\mu_i(x)$ the local chemical potential at point $x$. In local equilibrium, the chemical potential $\mu_i(x)$ of a given chemically equilibrated species $i$ can be expressed as
\begin{equation}\label{eq:hadron_chem}
    \mu_i = B_i \mu_B +Q_i\mu_Q + S_i\mu_S\,,
\end{equation}
where $(B_i, Q_i, S_i)$ are the baryon number, electric charge, and strangeness carried by particle species $i$, and $(\mu_B, \mu_Q, \mu_S)$ are the associated baryon, electric and strangeness chemical potentials. In a formal sense, solving Eq.~\eqref{boltzeq} provides the spacetime evolution of $f_{i, \mathbf{p}}$ and, consequently, the evolution of $\delta f_{i, \mathbf{p}}$. This further serves as a basis for determining the evolution of the macroscopic dissipative variable.

To establish a connection between the microscopic phase-space distribution and macroscopic hydrodynamic fields, we introduce the notation for the following momentum integrals:
\begin{equation}
    \langle\cdots\rangle_{i,0}\equiv\int dP_i(\cdots)_i f^0_{i, \mathbf{p}}\,,\quad \langle\cdots\rangle_{i,\delta}\equiv\int dP_i(\cdots)_i \delta f_{i, \mathbf{p}}\,,
\end{equation}
where $dP_i\equiv d^3 p_i/[(2\pi)^3p_i^0]$ is the Lorentz-invariant integration measure. The total fluid net charge currents and energy-momentum tensor of the mixture of particle species can be expressed as follows:\footnote{
Here, no degeneracy factor is explicitly introduced, but the spin degrees of freedom should be summed over; similarly in Eq.~\eqref{eq:P_had} and Eqs.~\eqref{eq:CF_Nq}-\eqref{eq:CF_Tmunu}.
}
\begin{eqnarray}
    T^{\mu\nu}&\equiv&\sum_i \int dP_i p_i^\mu p_i^\nu(f^0_{i, \mathbf{p}}+\delta f_{i, \mathbf{p}})=\sum_i \big(\langle p^\mu p^\nu\rangle_{i,0}+\langle p^\mu p^\nu\rangle_{i,\delta}\big)\,,\label{eq:fk_Tmunu}\\
    N_q^\mu&\equiv&\sum_i q_i\int dP_i p_i^\mu(f^0_{i, \mathbf{p}}+\delta f_{i, \mathbf{p}})=\sum_i q_i\big(\langle p^\mu\rangle_{i,0}+\langle p^\mu\rangle_{i,\delta}\big)\,,\label{eq:fk_Nq}
\end{eqnarray}
where the summation is taken over all particle species in the mixture, and $q\in\{B,\,Q,\,S\}$ denotes the three types of charges. Using these two expressions, we effectively treat the mixture, consisting of multiple particle species carrying various charges, as a single fluid with multiple charge components.

It's important to note that the decomposition \eqref{eq:fk_decomp} of $f_{i, \mathbf{p}}$  into its equilibrium and off-equilibrium parts lacks uniqueness until the local equilibrium parameters  $u^\mu(x)$, $T(x)$ and $\mu(x)$ in $f^0_{i, \mathbf{p}}$ as given in Eq.~\eqref{eq:fk_eq} are accurately defined. Additionally, the link between the decompositions in Eqs.~(\ref{eq:fk_Tmunu},\ref{eq:fk_Nq}) and those in Eqs.~(\ref{eq-dec-T},\ref{eq-dec-N}) has not yet been established. In fact, the flow velocity $u^\mu(x)$  has been established in Eq.~\eqref{eq:flow_landau} by selecting the Landau frame, consequently fixing the local energy density $\epsilon(x)$. To further determine $T(x)$ and $\mu(x)$, we require that $\delta f_{i, \mathbf{p}}$ contributes nothing to the local energy density and baryon density. This condition is commonly referred to as the Landau matching condition. By combining Eqs.~\eqref{eq-landau} with Eqs.~(\ref{eq:fk_Tmunu},\ref{eq:fk_Nq}), we can derive the following expressions for the local energy and baryon densities:
\begin{eqnarray}
    \epsilon = u_\mu T^{\mu\nu} u_\nu\equiv\sum_i \langle (u_\mu p^\mu)^2\rangle_{i,0}\,,\quad    n_q = u_\mu N_q^{\mu}\equiv \sum_i q_i \langle u_\mu p^\mu\rangle_{i,0}\,.\label{eq:enlandau}
\end{eqnarray}
Meanwhile, the pressure can also be obtained, as given by
\begin{equation}
    p = -\frac{1}{3} \sum_i \langle \Delta_{\mu\nu}p^\mu p^\nu\rangle_{i,0}\,.\label{eq:p_landau}
\end{equation}
In Eqs.~\eqref{eq:enlandau}, the conditions $\sum_i\langle (u_\mu p^\mu)^2\rangle_{i,\delta}{=}0$ and $\sum_i\langle u_\mu p^\mu\rangle_{i,\delta}{=}0$ are imposed in accordance with the matching condition. 
Equations \eqref{eq:enlandau} define $T(x)$ and $\mu(x)$ and, in combination with the defined $u^\mu(x)$, determine $f^0_{i, \mathbf{p}}$. 

Projecting out the additional components, as introduced in Eqs.~(\ref{eq-dec-T},\ref{eq-dec-N}), from Eqs.~(\ref{eq:fk_Tmunu},\ref{eq:fk_Nq}), we can deduce the dissipative components:
\begin{eqnarray}
    \pi^{\mu\nu}\equiv \Delta^{\mu\nu}_{\alpha\beta}T^{\alpha\beta}&=&\sum_i \langle p^{\langle\mu}p^{\nu\rangle}\rangle_{i,\delta}\,,\label{eq:fk_pimunu}\\
    \Pi \equiv -\frac{1}{3}T^{\mu\nu}\Delta_{\mu\nu}-p&=&-\frac{1}{3} \sum_i \langle \Delta_{\mu\nu}p^\mu p^\nu\rangle_{i,\delta}\,,\label{eq:fk_Pi}\\
    n_q^\mu\equiv \Delta^\mu_\nu N_q^\nu&=&\sum_i q_i \langle p^{\langle\mu\rangle}\rangle_{i,\delta}\,.\label{eq:fk_nqmu}
\end{eqnarray}
Equations (\ref{eq:fk_pimunu}-\ref{eq:fk_nqmu}) indicate that, under the matching condition, the dissipative components only receive contributions from the off-equilibrium $\delta f_{i, \mathbf{p}}$. If we introduce the irreducible moments of tensor-rank $\ell$ and energy-rank $r$ of $\delta f_{i, \mathbf{p}}$ for a given particle species $i$, 
\begin{equation}
    \rho_{i,r}^{\mu_1\cdots\mu_\ell}\equiv\Delta_{\nu_1\cdots\nu_\ell}^{\mu_1\cdots\mu_\ell}\int dP_i E^r_{i, \mathbf{p}} p_i^{\mu_1}\cdots p_i^{\mu_\ell} \delta f_{i, \mathbf{p}}=\langle E^r_{\mathbf{p}} p^{\langle\mu_1}\cdots p^{\mu_\ell\rangle}\rangle_{i,\delta}\,,
\end{equation}
in terms of irreducible tensors $p^{\langle\mu_1}\cdots p^{\mu_\ell\rangle}$ that are orthogonal to the flow velocity, it's straightforward to verify that Eqs.~(\ref{eq:fk_pimunu}-\ref{eq:fk_nqmu}) yield
\begin{equation}
    \pi^{\mu\nu}=\sum_i \rho_{i,0}^{\mu\nu}\,,\quad \Pi=-\frac{1}{3} \sum_i (m_i^2 \rho_{i,0}-\rho_{i,2})\,,\quad n_q^\mu=\sum_i q_i\rho_{i,0}^\mu\,,\label{eq:diss_rho}
\end{equation}
where $E_{i, \mathbf{p}}=p_i^\mu u_\mu$ and $m_i^2=p_i^\mu k_{i,\mu}$ denotes the energy and mass of the given particle species $i$, respectively. 

\subsubsection{Dissipative equations of motion}\label{sec:dissipation}

In principle, one can choose to solve the Boltzmann equation (\ref{boltzeq}) to obtain $f_{i, \mathbf{p}}$. Once $f_{i, \mathbf{p}}$ is obtained, it can be used to calculate the evolving dissipative variables given by Eqs.~\eqref{eq:fk_pimunu}-\eqref{eq:fk_nqmu}. However, this is generally a very complicated task. Alternatively, one can re-formulate the Boltzmann equation into an infinite system of coupled evolution equations for the moments $\rho_{i,r}^{\mu_1\cdots\mu_\ell}$. The solution of these equations governs the evolution of the dissipative variables through Eq.~\eqref{eq:diss_rho}. Employing this approach, the DNMR theory \cite{Denicol:2012cn} derived these dissipative evolution equations for a single-component fluid. In Refs.~\cite{Jaiswal:2015mxa,Fotakis:2022usk,Chattopadhyay:2022sxk}, these equations were studied for multicomponent fluids.

More specifically, to derive the equations of motion for the irreducible moments, the equation for the comoving derivative $\dot\rho_{i,r}^{\langle\mu_1\cdots\mu_\ell\rangle}\equiv \Delta_{\nu_1\cdots\nu_\ell}^{\mu_1\cdots\mu_\ell} D \rho_{i,r}^{\nu_1\cdots\nu_\ell}$, one can proceed by multiplying the Boltzmann equation (\ref{boltzeq}) in the following form:
\begin{equation}
    \delta\dot f_{i, \mathbf{p}}=-\dot f^0_{i, \mathbf{p}}-E^{-1}_{i, \mathbf{p}}p_{i,\nu}\nabla^\nu f^0_{i, \mathbf{p}}-E^{-1}_{i, \mathbf{p}}p_{i,\nu}\nabla^\nu \delta f_{i, \mathbf{p}}+E^{-1}_{i, \mathbf{p}} C[f_{i, \mathbf{p}}]
\end{equation}
by $E^r_{i, \mathbf{p}} p_i^{\langle\mu_1}\cdots p_i^{\mu_\ell\rangle}$ and subsequently integrating over momentum space. After linearizing the collision integral with respect to the quantities $\phi_{i, \mathbf{p}}\equiv \delta f_{i, \mathbf{p}}/[f^0_{i, \mathbf{p}}(1-a_if^0_{i, \mathbf{p}})]$, which implies that
\begin{equation}\label{eq:df_expansion}
    \delta f_{i, \mathbf{p}}\equiv [f^0_{i, \mathbf{p}}(1-a_if^0_{i, \mathbf{p}})]\phi_{i, \mathbf{p}}\,,
\end{equation}
the irreducible moments of the collision integral $C[f_{i, \mathbf{p}}]$ can be expressed as a linear combination of the irreducible moments $\rho_{i,r}^{\mu_1\cdots\mu_\ell}$. A closed set of equations of motion for the dissipative quantities $\Pi, n_q^\mu, \text{and } \pi^{\mu\nu}$ is consequently obtained by truncating the infinite set of moment equations in the $(10+4N_q)$-moment approximation \cite{Fotakis:2022usk}.

The dissipative equations of motion are of relaxation type. For the evolution equations of the bulk viscous pressure and shear stress tensor, they are given by
\begin{eqnarray}
    \tau_{\Pi }\dot{\Pi}+\Pi &=& -\zeta \theta +(\textrm{higher-order terms})\,,\label{eq-relax-Pi}\\
    \tau _{\pi }\dot\pi^{\left\langle \mu \nu \right\rangle }+\pi ^{\mu \nu }
    &=& 2\eta \sigma ^{\mu \nu }+(\textrm{higher-order terms})\,,\label{eq-relax-pi}
\end{eqnarray}
where $\dot \Pi\equiv D\Pi$ and $\dot\pi^{\langle\mu\nu\rangle} \equiv \Delta^{\mu\nu}_{\alpha\beta} D\pi^{\alpha\beta}$, ensuring that all terms are purely spatial in the LRF and, if applicable, traceless. $\tau_{\Pi}$ and $\tau_{\pi}$ represent the relaxation times for $\Pi$ and $\pi^{\mu\nu}$, respectively. They determine the rate at which the dissipative flows relax to their Navier-Stokes limits, characterized by $\Pi_{\mathrm{NS}} \equiv -\zeta \theta$, and $\pi^{\mu\nu }_{\mathrm{NS}} \equiv 2\eta \sigma^{\mu\nu}$. Here, $\zeta$ and $\eta$ denote the bulk viscosity and shear viscosity, respectively. These parameters describe the first-order response of the dissipative flows to their driving forces --- the negative of the scalar expansion rate $\theta$ and the shear flow tensor $\sigma ^{\mu\nu}$ --- leading the system away from local equilibrium. The determination of these parameters will be further discussed in Sec.~\ref{sec:coeff}.

Similarly, the equation for the diffusion current of charge $q$ is given by:
\begin{equation}\label{eq:charges_evo}
    \sum_{q'}\tau_{qq'} \dot n_{q'}^{\langle\mu\rangle}+n_{q}^\mu=\sum_{q'}\kappa_{qq'}\nabla^\mu \left(\frac{\mu_{q'}}{T}\right)+(\textrm{higher-order terms})\,,
\end{equation}
where the sum is taken over $BQS$ charges. Importantly, due to the presence of off-diagonal elements in the diffusion coefficient matrix $\kappa_{qq'}$, the diffusion current of a particular charge $q$ can receive contributions from gradients of the chemical potentials of all charge types. This is a natural consequence of the fact that each constituent of the fluid may carry multiple types of charges, leading to intricate couplings between different charges. For instance, in the context of a (1+1)-dimensional system, where the dissipative effects from shear and bulk viscous pressures are ignored, it was observed in Ref.~\cite{Fotakis:2019nbq} that a gradient in baryon number can not only give rise to a baryon current but also induce currents in strangeness and electric charge. The study demonstrated that strangeness separation can manifest in a system initially in a strangeness-neutral state, primarily driven by its coupling to the baryon diffusion current.

In scenarios where the charges are not coupled, assuming that each constituent of the fluid carries only a single type of charge, the equation simplifies to:
\begin{equation}
    \tau_{q} \dot n_{q}^{\langle\mu\rangle}+n_{q}^\mu= \kappa_{q}\nabla^\mu \left(\frac{\mu_{q}}{T}\right)+(\textrm{higher-order terms})\,,\label{eq:single_charge}
\end{equation}
where the relaxation time $\tau_{q}\equiv\tau_{qq}$ and diffusion coefficient $\kappa_{q}\equiv\kappa_{qq}$ represent the diagonal elements of the respective matrices. This corresponds to the conventional scenario that investigations have predominantly focused on. Similar to the case of bulk viscous pressure and shear stress tensor, Eq.~\eqref{eq:single_charge} of relaxation type suggests that when the expansion rates are sufficiently small, such that $\nabla^\mu \left(\mu_{q}/T\right)$ evolves slowly, the charge diffusion current $n_q^\mu$ will asymptotically approach the values predicted by the Navier-Stokes theory at times much greater than the relaxation time $\tau_q$, i.e.,
\begin{equation}
    n_{q,\,\mathrm{NS}}^\mu \equiv \kappa_{q}\nabla^\mu \left(\frac{\mu_{q}}{T}\right)\,.\label{eq:NS_nq}
\end{equation}
This is the equation of motion derived from relativistic Navier-Stokes theory. It also resembles relativistic Fick's law, illustrating how diffusion currents arise due to gradients in the thermal potential associated with the charge. Notably, in a flat Minkowski space, where $\nabla^\mu=(0, -\partial_x, -\partial_y, -\partial_z)$ in the local rest frame, the minus sign in the spatial components indicates that diffusion currents work to smooth out the existing inhomogeneities that initially generated the current. Additionally, the diffusion coefficient $\kappa_{q}$ decreases as the temperature of the expanding fireball decreases. Due to these two factors, the baryon diffusion current is found to be significant only during the early stages of the evolution \cite{Du:2021zqz,Du:2021uxo}.

\begin{figure}[t]
    \centering
    \includegraphics[width=0.4\linewidth]{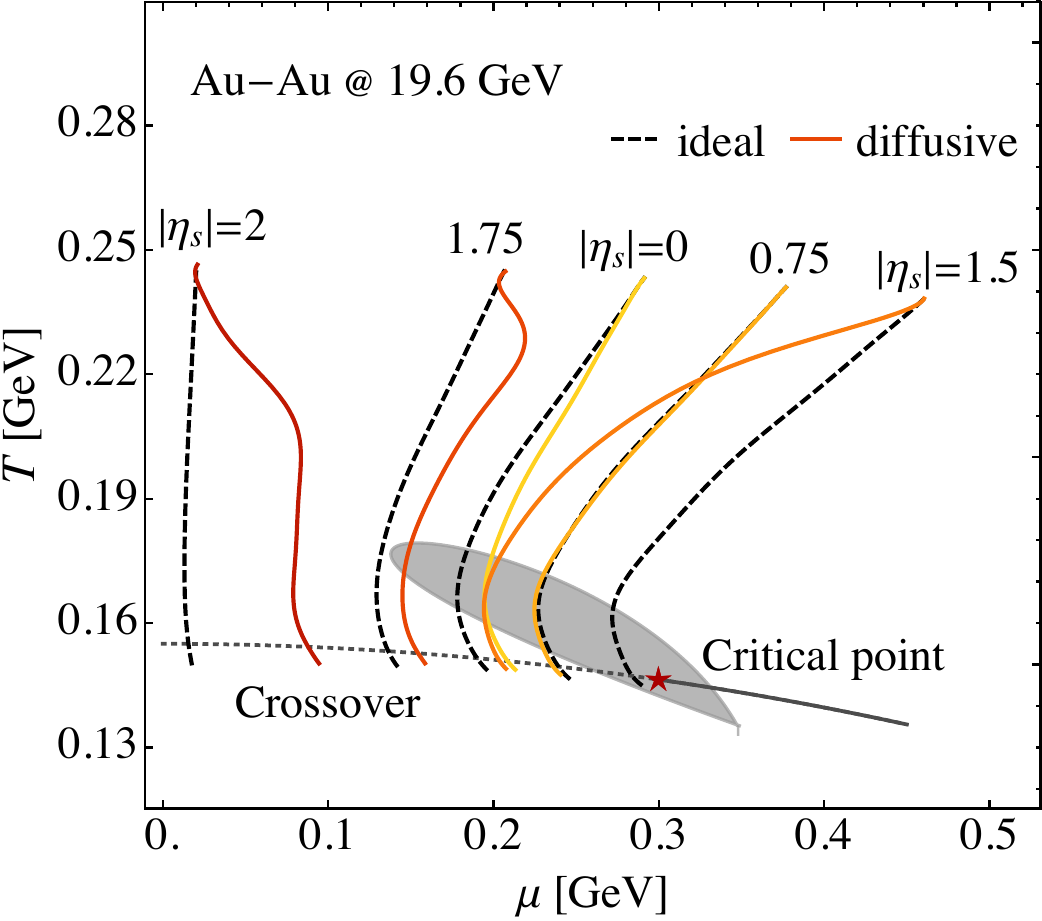}
	\hspace{4mm}
    \includegraphics[width=0.52\linewidth]{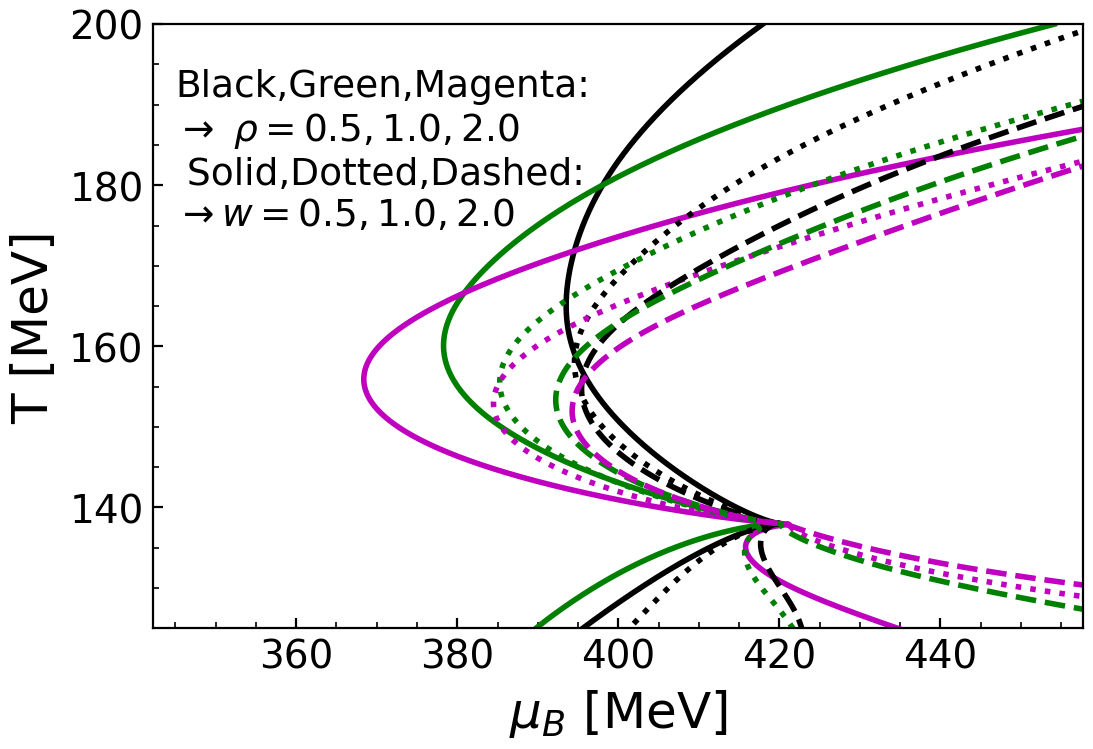}
    \caption{\textit{Left:} Phase diagram trajectories of the fireball within different rapidity windows, comparing ideal evolution (black dashed lines) with evolution including baryon diffusion (colored solid lines) \cite{Du:2021zqz}. \textit{Right:} Illustration of the critical lensing effect through isentropic trajectories crossing the critical point with different parameters controlling the critical region of the EOS \cite{Dore:2022qyz}.}
    \label{fig:phase_diagram_traj}
\end{figure}

So far, among the three types of charges, the baryon charge is the only one that has received significant attention in phenomenological studies, albeit to a much lesser extent compared to the investigations of viscous effects associated with shear stress tensor and bulk pressure. One of the primary reasons for the focus on the baryon charge is that both baryon density and temperature are among the most important controllable features that can be adjusted in heavy-ion collisions by varying the collision energy. Moreover, the QCD phase diagram is commonly depicted with a horizontal axis representing either baryon density or baryon chemical potential, and a vertical axis representing temperature. Consequently, the dynamics of the baryon current plays a pivotal role in determining the evolution of baryon-rich QCD matter within this diagram and in the search for the signatures of the QCD critical point (see the left panel of Fig.~\ref{fig:phase_diagram_traj}). Eqs.~\eqref{eq:single_charge}-\eqref{eq:NS_nq} encapsulate the fundamental information required to comprehend baryon diffusion current, which is essential for understanding the associated phenomenology.

We should also briefly mention a few studies as examples that have focused on developing hydrodynamic formalisms considering only baryon number (or a single-component charge). These approaches have utilized various methods, including the moment method \cite{Bouras:2010hm,Denicol:2012cn}, Chapman-Enskog method \cite{Jaiswal:2015mxa}, 14-moment method \cite{Jaiswal:2013npa}, and anisotropic hydrodynamics \cite{Florkowski:2017ovw,Almaalol:2018jmz}. Additionally, a newly developed Maximum Entropy method \cite{Everett:2021ulz,Chattopadhyay:2023hpd} and a novel Relaxation Time Approximation \cite{Rocha:2021zcw} show promise but has yet to be extended to systems with finite charges. Moreover, multiple hydrodynamic codes, such as \texttt{MUSIC} \cite{Denicol:2018wdp,De:2022yxq}, \texttt{BEShydro} \cite{Du:2019obx,Du:2021fyr}, \texttt{vHLLE} \cite{Karpenko:2013wva,Schafer:2021csj}, and \texttt{CLvisc} \cite{Pang:2018zzo,Wu:2021fjf}, have been developed and implemented to simulate the (3+1)-dimensional evolution of a baryon-charged fluid.


\subsubsection{Transport coefficients}\label{sec:coeff}

Transport coefficients are parameters characterizing the transport of various quantities—like energy-momentum and charges—within the QGP fluid. Some key first-order transport coefficients encompass shear ($\eta$) and bulk ($\zeta$) viscosity, alongside diffusion coefficients for conserved charges ($\kappa_{q}$) and the relaxation times governing these dissipative quantities (e.g., $\tau_{q}$). Second-order and even higher-order transport coefficients describe the interplay among these dissipative components in the higher-order terms of their equations of motion.\footnote{
Currently, second-order transport coefficients receive little attention, and in practice, they are often connected to thermodynamic quantities through relationships with first-order coefficients evaluated in kinetic theory or holography \cite{JETSCAPE:2020mzn,JETSCAPE:2020shq,Nijs:2020ors,Nijs:2020roc}.
}
These coefficients play a vital role in describing the macroscopic behavior of QCD matter by bridging microscopic properties, such as interactions between its constituents, with the system's overall transport behavior. Their dependence on temperature and chemical potentials encapsulates the distinctive properties of the QCD matter and the underlying microscopic interactions.

\begin{figure}[t]
    \centering
    \includegraphics[width=0.45\linewidth]{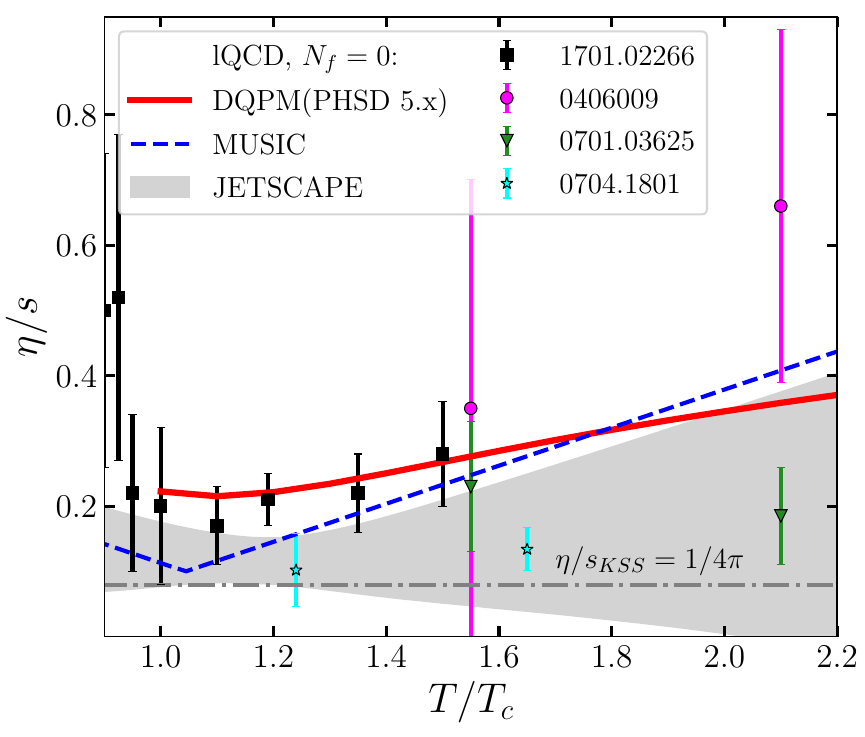}
    \includegraphics[width=0.53\linewidth]{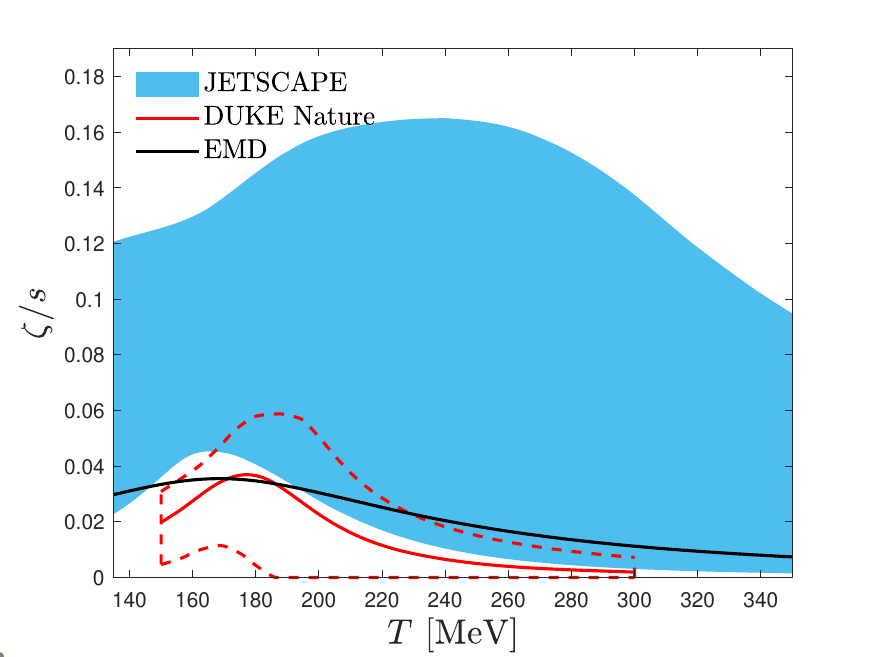}
    \caption{\textit{Left:} Different specific shear viscosities plotted as a function of the scaled temperature $T/T_c$ at zero chemical potential (figure from Ref.~\cite{Aarts:2023vsf}). \textit{Right:} The holographic EMD prediction for the specific bulk viscosity compared to the results favored by phenomenological multistage models at zero chemical potential (figure from Ref.~\cite{Rougemont:2023gfz}).}
    \label{fig:viscosities}
\end{figure}

Obtaining transport coefficients poses a great challenge \cite{Moore:2020pfu} and often demands sophisticated calculations and models. Within the field of heavy-ion physics, acquiring these coefficients for nuclear matter involves various methods (see Fig.~\ref{fig:viscosities}), with one of the most common being the application of kinetic theory to many-particle systems. This method involves comparing the macroscopic and microscopic definitions of thermodynamic quantities, integrating particle interactions as dynamic inputs into the expressions of transport coefficients. Usually, when deriving the hydrodynamic equations of motion, these expressions for transport coefficients in terms of particle interactions are naturally derived. For studies employing kinetic theory to derive transport coefficients, refer to, for example, Refs.~\cite{Gavin:1985ph,Jeon:1995zm,Denicol:2014vaa,Jaiswal:2015mxa,Albright:2015fpa,Mitra:2017sjo,Greif:2017byw,Fotakis:2021diq}. Although the coefficients estimated within kinetic theories can provide insights, they do not represent real QCD matter due to the oversimplified interactions often assumed.

Moreover, linear response theory serves as a framework to describe a system's behavior when subjected to small perturbations or external fields. Within this framework, the Kubo formula characterizes the system's response to perturbations and provides specific expressions for transport coefficients by relating the transport coefficients of a slightly non-equilibrium system to real-time correlation functions computed within an equilibrium thermal ensemble \cite{Kovtun:2018dvd}. This formula serves as a foundational tool for calculating transport coefficients from first principles using finite-temperature quantum field theory and has played a crucial role in determining the transport coefficients of nuclear matter \cite{Jeon:1994if,Carrington:1999bw,Moore:2008ws,Jeon:2015dfa,Rose:2020sjv,Hammelmann:2023fqw}. Additionally, transport coefficients can be extracted using lattice QCD calculations  \cite{Meyer:2011gj,Ding:2015ona,Ratti:2018ksb,Kaczmarek:2022ffn} through conserved current correlators. These Euclidean correlators are connected, via an integral transform, to spectral functions. The small-frequency form of these spectral functions determines the transport properties using the Kubo formula.

The development of the AdS/CFT correspondence, also known as the gauge/gravity duality more broadly, has introduced a new paradigm for exploring strongly coupled gauge theories through weakly coupled gravitational systems. This duality has proven invaluable for investigating both thermal and hydrodynamic properties of field theories at strong coupling. Utilizing holographic methods, the specific shear viscosity $\eta/s$ has been bounded by a minimum value of $1/4\pi$ \cite{Kovtun:2004de}. The remarkable order-of-magnitude agreement with RHIC data has spurred extensive efforts to employ holographic techniques in computing various transport coefficients of the QCD medium \cite{Policastro:2001yc,Buchel:2007mf,Buchel:2010wf,Cremonini:2012ny,Li:2014dsa,Finazzo:2014cna,Rougemont:2015ona,Rougemont:2017tlu} (for comprehensive reviews, see \cite{Casalderrey-Solana:2011dxg,Cremonini:2011iq,Rougemont:2023gfz}).

Despite the advancements in various methods for calculating transport coefficients, the arguably most reliable approach to determining their values remains model-to-data comparison phenomenologically \cite{Moore:2020pfu}. Early recognition highlighted the sensitivity of hadronic observables, like the distributions and correlations of hadrons in heavy-ion collisions, to the shear and bulk viscosities of QCD matter \cite{Romatschke:2007mq,Song:2007fn,Denicol:2009am,Gale:2012rq}. Qualitatively, a significant bulk viscosity tends to isotropically reduce the momenta of hadrons, whereas shear viscosity diminishes their azimuthal momentum asymmetry. Initial works aimed at constraining these coefficients via hydrodynamic simulations of heavy-ion collisions primarily focused on the specific shear viscosity $\eta/s$, often approximated as a constant (as reviewed in Refs.~\cite{Heinz:2013th,Gale:2013da}). Contemporary endeavors seek to characterize their dependencies on the equilibrium properties of the system, typically focusing on constraining the temperature dependence of the specific shear and bulk viscosities  \cite{Denicol:2015nhu,Niemi:2015qia,Ryu:2015vwa,Ryu:2017qzn}, denoted as $(\eta/s)(T)$ and $(\zeta/s)(T)$, respectively. The functional forms of the dependence are often guided by insights obtained from the theoretical calculations \cite{Csernai:2006zz,Kharzeev:2007wb,Karsch:2007jc,Noronha-Hostler:2008kkf,Arnold:2006fz}. 

As we move into an era of demanding precision, the challenge lies in quantitatively constraining viscosities with quantified uncertainties derived from measurements. The hydrodynamic expansion of the deconfined QGP stands as one among several phases within heavy-ion collisions, accompanied by various stages of many-body dynamics before and after this fluid phase. The evolution itself involves intricate variations in flow velocity and temperature profiles, varying from collision to collision. These factors pose significant challenges in disentangling the contributions of shear and bulk viscosities via hadronic measurements. Consequently, theoretical uncertainties at each stage of the collision can significantly influence estimations of the viscosities. Achieving a meaningful constraint on QGP viscosities from collider measurements demands extensive model-to-data comparisons encompassing multiple collision stages, meticulously modeling each phase, and contrasting resulting predictions against extensive and diverse experimental data. 

Utilizing Bayesian statistical analysis to systematically constrain QGP transport properties, with precisely quantified uncertainties, has emerged as a successful approach within various theoretical frameworks over the last decade \cite{Novak:2013bqa,Pratt:2015zsa,Bernhard:2015hxa,Bernhard:2016tnd,Moreland:2018gsh,Bernhard:2019bmu,JETSCAPE:2020mzn,JETSCAPE:2020shq,Nijs:2020ors,Nijs:2020roc,Parkkila:2021tqq,Parkkila:2021yha,Heffernan:2023gye,Heffernan:2023utr,Liyanage:2023nds} (for comprehensive overviews, see Refs.~\cite{Shen:2020mgh,Paquet:2023rfd}). These studies have significantly advanced model-data inference in heavy-ion physics, offering diverse perspectives and insights. For instance, pioneering works in Refs.~\cite{Bernhard:2016tnd,Bernhard:2019bmu} conducted the first Bayesian inference on temperature-dependent shear and bulk viscosities. Additionally, deeper insights into the temperature dependence of transport coefficients and the event-by-event nucleon substructure were gained using measurements from $p$-Pb and Pb-Pb collisions in Ref.~\cite{Moreland:2018gsh}. The JETSCAPE Collaboration presented Bayesian analyses utilizing $p_T$-integrated measurements at top RHIC and LHC energies, offering robust constraints on the temperature dependence of QGP shear and bulk viscosities around the crossover temperature \cite{JETSCAPE:2020mzn,JETSCAPE:2020shq}. Notably, this study also examined theoretical uncertainties stemming from various off-equilibrium corrections (see Sec.~\ref{sec:off-eq}) during the particlization stage for the first time. Further constraints using $p_T$-differential observables were explored in Refs.~\cite{Nijs:2020ors,Nijs:2020roc}. Moreover, Bayesian quantifications with color glass condensate initial conditions \cite{Heffernan:2023gye,Heffernan:2023utr} and anisotropic hydrodynamics \cite{Liyanage:2023nds} were separately undertaken, contributing to a comprehensive understanding of QGP dynamics.

At the BES energies, the consideration of $BQS$ charges becomes essential, and thus understanding the transport coefficients' dependence on equilibrium properties necessitates considering the finite chemical potentials associated with these charges. Current efforts have primarily focused on the dependence on the baryon chemical potential ($\mu_B$), although progress is significantly less compared to investigations at zero $\mu_B$. The presence of baryon chemical potential can significantly modify the system's transport properties.\footnote{
In the critical region, fluctuations significantly modify the physical transport coefficients, leading to their correlation length dependence. As a result, the transport coefficients could exhibit singularities at the critical point. A concise discussion on this topic can be found in Ref.~\cite{Du:2021fyr}.
}
For example, at finite $\mu_B$, the system's fluidity is assessed based on the ratio of shear viscosity to the enthalpy multiplied by temperature, $\eta T/(\epsilon+p)$, rather than relying on the specific shear viscosity $\eta/s$ \cite{Liao:2009gb}. Calculations using Chapman-Enskog theory within a simple hadronic model revealed a substantial reduction in $\eta T/(\epsilon+p)$ compared to $\eta/s$ at nonzero $\mu_B$ for a hadron resonance gas \cite{Denicol:2013nua}. This suggests that systems generated at lower collision energies could exhibit fluid-like behavior, displaying an effective fluidity similar to that observed near the phase transition at the topc RHIC energy. Studies like Refs.~\cite{Denicol:2015nhu,Shen:2020jwv} demonstrated that measurements of rapidity-differential anisotropic flows could constrain the shear viscosity's temperature dependence in regions where $\mu_B$ plays a significant role. Additionally, several investigations have explored constraints on $\eta/s$ at various beam energy using measurements at the BES \cite{Karpenko:2015xea,Auvinen:2017fjw,Shen:2023awv,Yang:2023apw}.

In addition to phenomenological constraints on $\eta/s$ at finite $\mu_B$, various approaches have been employed to evaluate it. For example, calculations have been conducted using the gauge/gravity duality \cite{Rougemont:2017tlu,Hoyos:2020hmq,Grefa:2022sav}, the dynamical quasiparticle model (DQPM) by explicitly computing parton interaction rates \cite{Soloveva:2019xph}, the extended $N_f{\,=\,}3$ Polyakov Nambu-Jona-Lasinio (PNJL) model \cite{Soloveva:2020hpr}, and through the HRG model \cite{Rose:2017bjz,McLaughlin:2021dph,Hammelmann:2023fqw}. Regarding the specific bulk viscosity $\zeta/s$, although there are some phenomenological constraints at finite $\mu_B$ \cite{Shen:2023awv,Yang:2023apw}, it has received less attention in many phenomenological studies at the BES. However, several of the aforementioned approaches that evaluate $\eta/s$ at finite $\mu_B$ also examine $\zeta/s$ under the same conditions \cite{Rougemont:2017tlu,Hoyos:2020hmq,Grefa:2022sav,Soloveva:2019xph}. The Bayesian inference conducted in Ref.~\cite{Shen:2023awv} suggests a non-monotonic relationship between the QGP's specific bulk viscosity $\zeta T/(\epsilon+p)$ and the collision energy, although further systematic analysis is required to confirm this observation.

At the BES, the diffusion coefficients $\kappa_q$ associated with $BQS$ charges, along with the cross diffusion coefficients $\kappa_{qq'}$, hold significant importance. Among these, the baryon diffusion coefficient $\kappa_B$ assumes a crucial role in the evolution of charged QGP fireball. Its significance lies in determining the trajectory of systems across the QCD phase diagram and is particularly relevant for interpreting potential signatures of the QCD critical point, especially concerning proton cumulant measurements. The baryon diffusion coefficient has been derived through various methodologies. Expressions for $\kappa_B$ for the net baryon diffusion current were obtained using Grad’s 14-moment and Chapman-Enskog methods \cite{Denicol:2012cn,Monnai:2012jc,Jaiswal:2015mxa,Albright:2015fpa,Mitra:2016zdw,Denicol:2018wdp}. Additionally, evaluations of $\kappa_B$ have been performed within holographic models \cite{Rougemont:2015ona,Rougemont:2017tlu,Hoyos:2020hmq,Grefa:2022sav} and the dynamical quasiparticle model \cite{Soloveva:2019xph}. Cross diffusion coefficients $\kappa_{qq'}$ were also derived employing kinetic approaches \cite{Greif:2017byw,Fotakis:2019nbq,Fotakis:2022usk} and various transport models \cite{Fotakis:2021diq,Rose:2020sjv,Hammelmann:2023fqw}. 

Currently, we face significant challenges in precisely constraining the baryon diffusion coefficient, particularly relying on net proton yields in rapidity \cite{Denicol:2018wdp,Li:2018fow,Du:2021zqz,Du:2023gnv}. This observable is intricately entangled with both the initial baryon distribution and the hydrodynamic transport. The former is particularly sensitive to baryon stopping during the prehydrodynamic stage, while the latter is particularly sensitive to the longitudinal gradient of baryon density and $\kappa_B$. Consequently, understanding and accurately constraining $\kappa_B$ necessitate a thorough disentanglement of these intertwined effects, particularly improving our comprehension of the initial baryon distribution, as discussed in Sec.~\ref{sec:prehydro}. Additional observables at the BES, such as baryonic directed flow $v_1(y)$ \cite{Du:2022yok} and the polarization of $\Lambda$ hyperons \cite{Wu:2022mkr}, might also offer insights into constraining $\kappa_B$. Additionally, when explored as functions of relative azimuthal angle, balance functions exhibit sensitivity to the diffusivity of light quarks, serving as independent observables for constraining the charge diffusion constants \cite{Kapusta:2017hfi,Pratt:2019pnd,De:2020yyx,Pratt:2021xvg}. However, achieving a robust phenomenological constraint on $\kappa_B$ and its temperature and baryon chemical potential dependence necessitates a comprehensive model-data inference approach incorporating diverse measurements, especially those in rapidity.

\subsubsection{Equations of state at finite chemical potentials}\label{sec:eos}

The EOS of QCD characterizes the thermodynamic properties of strongly interacting matter, providing a relationship among key thermodynamic variables such as pressure, temperature, and chemical potentials. In hydrodynamic simulations, the EOS plays a crucial role in reducing the number of unknown variables to close the system of equations and is sometimes essential when determining the values of transport coefficients. Much like transport coefficients, the EOS encapsulates the fundamental properties of QCD. The mapping of the QCD phase diagram, essentially dictated by probing the EOS, stands as one of the primary objectives of Beam Energy Scan programs \cite{Bzdak:2019pkr}. Presently, the high-energy nuclear physics and astrophysics communities are in a unique position to establish highly stringent constraints on the EOS of strongly interacting matter, drawing from both heavy-ion collisions and neutron star observations through a comprehensive, multi-disciplinary strategy \cite{Baym:2017whm,Dexheimer:2020zzs,Sorensen:2023zkk}.

For a comprehensive quantitative description of nuclear collisions, it is imperative to have an EOS as input for hydrodynamic simulations. Particularly, simulating the dynamic evolution of the collision fireball created at various beam energies, which transitions from a deconfined partonic fluid to a confined hadronic gas, necessitates an EOS that captures the thermodynamic properties across a wide range of temperatures and chemical potentials. Just like the transport coefficients, an EOS can be derived within the same kinetic approach used for deriving hydrodynamic equations of motion, but it is not truly representative of QCD. A modern approach to constructing such an EOS involves a smooth interpolation around the phase transition between a lattice QCD EOS (effective at high temperatures) and an HRG EOS (effective at low temperatures) \cite{Bluhm:2007nu,Huovinen:2009yb} (for a comprehensive overview, see Refs.~\cite{Dexheimer:2020zzs,Monnai:2021kgu}). Utilizing such an EOS is essential for maintaining the conservation of energy-momentum and charges during the particlization process at the freeze-out surface, marking the transition from a continuous macroscopic hydrodynamics framework to a discrete microscopic particle transport approach (see Sec.~\ref{sec:fzout}).

Deriving the EOS of QCD at finite baryon chemical potential through conventional Monte Carlo simulations in lattice QCD faces significant challenges. This is primarily due to the fermion sign problem, a fundamental technical obstacle characterized by complex exponential \cite{Philipsen:2012nu}. However, recent years have seen the proposal of alternative methods to extract the properties of QCD matter at low baryon chemical potential \cite{Bazavov:2020bjn,HotQCD:2012fhj,Borsanyi:2012cr,Borsanyi:2021sxv}. These methods include Taylor expansion around $\mu_B=0$ and analytic continuation from imaginary $\mu_B$ (for a comprehensive overview, see Ref.~\cite{Philipsen:2012nu,Soltz:2015ula,Ratti:2018ksb}). Furthermore, holographic approaches have been developed to match QCD thermodynamics at $\mu_B=0$, successfully reproducing all the Taylor expansion coefficients available from lattice QCD. These holographic methods are then extended to high baryon density, a regime currently beyond the reach of lattice simulations \cite{Critelli:2017oub}. For a comprehensive overview of holographic descriptions, particularly those based on the class of gauge-gravity Einstein–Maxwell–Dilaton (EMD) effective models, refer to Ref.~\cite{Rougemont:2023gfz}.
Note that, in these methods, the investigation of the EOS of QCD has mostly focused on finite $\mu_B$, while keeping $\mu_Q=\mu_S=0$.

In heavy-ion collisions at the BES, it is reasonable to assume that light quarks ($u$, $d$, and $s$) can achieve thermalization in the QGP, making the EOS dependent on non-zero chemical potentials of baryon number, electric charge, and strangeness \cite{Noronha-Hostler:2019ayj,Monnai:2019hkn}. The chemical potentials of $BQS$ are interconnected with those of the relevant quarks through the following relationships:
\begin{equation}\label{eq:quark_poten}
\mu_u = \frac{1}{3}\mu_B+\frac{2}{3}\mu_Q\,,\quad \mu_d = \frac{1}{3}\mu_B-\frac{1}{3}\mu_Q\,,\quad \mu_s = \frac{1}{3}\mu_B-\frac{1}{3}\mu_Q-\mu_S\,.
\end{equation}
The lattice QCD EOS at non-zero chemical potentials can be constructed using a Taylor expansion around vanishing chemical potentials:
\begin{equation}\label{eq:taylorexp_multi}
    \frac{p_\mathrm{lat}}{T^4} = \frac{p(T)}{T^4} + \sum_{l,m,n} \frac{\chi^{BQS}_{l,m,n}(T)}{l!m!n!} \left(\frac{\mu_B}{T}\right)^{l}\left(\frac{\mu_Q}{T}\right)^{m}\left(\frac{\mu_S}{T}\right)^{n}\,,
\end{equation}
where $p(T)$ represents the pressure and the expansion coefficients $\chi^{BQS}_{l,m,n}(T)$ are the susceptibilities at vanishing chemical potentials, calculated from lattice QCD simulations. The susceptibilities at finite chemical potentials are defined as
\begin{equation}\label{eq:suscept}
    \chi^{BQS}_{l,m,n} = \frac{\partial^l\partial^m\partial^n (p/T^4)}{\partial(\mu_B/T)^l\partial(\mu_Q/T)^m\partial(\mu_S/T)^n}\,.
\end{equation}
Here matter-antimatter symmetry requires $l+m+n$ to be even. 

In low-temperature regions, HRG models have been developed to incorporate hadronic interactions, aiming to enhance agreement with lattice QCD observables at temperatures below the chiral pseudocritical temperature $T_{pc}$. In the commonly used ideal HRG model, the system is represented as a non-interacting, multi-component gas of known hadrons and resonances. While this model describes the thermodynamic functions of lattice QCD at $\mu_B=0$ up to temperatures around $T_{pc}$, deviations arise in susceptibilities of conserved charges, introduced in Eq.~\eqref{eq:suscept}, which involve derivatives of the pressure function with respect to the chemical potentials and provide insights into the finer details of the EOS \cite{Borsanyi:2011sw,HotQCD:2012fhj}. In lattice QCD at the physical point, susceptibilities show rapid deviations from the ideal HRG in the vicinity, and even below, $T_{pc}$. These deviations may stem from hadronic interactions that are not dominated by resonance formation. Some attractive interactions, for instance, cannot be accurately described by simply adding resonances as free particles. Additionally, it has been suggested that deviations in lattice data on higher-order susceptibilities from the uncorrelated hadron gas baseline can be attributed to repulsive interactions \cite{Vovchenko:2016rkn}. Consequently, HRG models incorporating repulsive interactions have gained interest in the context of lattice QCD calculations of fluctuations and correlations of conserved charges. For a comprehensive overview of HRG with van der Waals interactions, see Ref.~\cite{Vovchenko:2020lju}.

Taking the ideal HRG as an example, its hydrostatic pressure can be expressed as \cite{Monnai:2019hkn}:
\begin{eqnarray}
p_\mathrm{hrg} &=& \pm T \sum_i \int \frac{d^3p}{(2\pi)^3} \ln [1 \pm e^{-(E_i-\mu_i)/T} ]\nonumber\\
&=& \sum_i \sum_k (\mp1)^{k+1} \frac{1}{k^2} \frac{1}{2\pi^2} m_i^2 T^2 e^{k\mu_i/T} K_2\bigg(\frac{k m_i}{T}\bigg), \label{eq:P_had}
\end{eqnarray}
Here, $i$ is the index for particle species, $m_i$ is the particle's mass, and $K_2(x)$ is the modified Bessel function of the second kind. The index $k$ describes the expansion of quantum distributions around the classical ones. The upper signs are for fermions, and the lower signs are for bosons. The hadronic chemical potential of species $i$ is given by Eq.~\eqref{eq:hadron_chem}. Note that to ensure energy and momentum conservation during the transition from hydrodynamics to a hadronic transport approach at particlization, it is essential that the particle species in the HRG EOS align with those included in the transport model \cite{Alba:2017hhe,JETSCAPE:2020mzn}.

Different approaches exist for interpolating the EOS between lattice QCD and the HRG. In Ref.~\cite{Monnai:2019hkn}, the EOS at non-zero $(\mu_B, \mu_Q, \mu_S)$ for the high-temperature region is constructed using lattice QCD calculations of $\chi^{BQS}_{l,m,n}(T)$. Subsequently, it is interpolated with the HRG EOS at low temperatures using a suitable interpolating function. In contrast, Ref.~\cite{Noronha-Hostler:2019ayj} first introduces interpolations for $\chi^{BQS}_{l,m,n}(T)$ between the HRG and lattice QCD calculations across the entire relevant temperature range. The full EOS at non-zero chemical potentials is then constructed using Eq.~\eqref{eq:taylorexp_multi} with the interpolated $\chi^{BQS}_{l,m,n}(T)$. Commonly, the interpolation between the two equations of state is centered around the crossover/phase transition line. 

To simulate nuclear collisions at the BES energies, it is essential to employ relativistic hydrodynamics that accounts for multiple conserved charges and incorporates such an EOS constructed at nonzero $BQS$ chemical potentials. Simultaneously, the evolution of charge transport must be taken into account, considering their interplay as outlined in Eq.~\eqref{eq:charges_evo}. However, doing so is computationally demanding, particularly when conducting spacetime evolution in (3+1) dimensions which is necessary for collisions at the BES. Up to now, most simulations for the BES have focused on nonzero baryon charge alone, assuming $\mu_Q=\mu_S=0$, and at times, even neglecting its diffusion current. Nevertheless, the consideration of nonzero electric charge and strangeness effects can be simplified in simulations that only account for nonzero baryon charge by deducing $QS$ charges from $B$ charge. This can be achieved by imposing constraints on the 4-dimensional EOS in $(T, \mu_B, \mu_Q, \mu_S)$ and subsequently projecting it onto 2-dimensional planes in $(T, \mu_B)$. The values of $(\mu_Q, \mu_S)$ can then be obtained from $(T, \mu_B)$ utilizing their relationships associated with the imposed constraints \cite{Monnai:2019hkn, Monnai:2021kgu}.

In nuclear collisions, the absence of valence quarks of strangeness in the colliding nuclei implies that the average strangeness density in the produced systems should be zero, denoted as $\langle n_S\rangle=0$, a condition known as strangeness neutrality. Moreover, the electric density $n_Q$ is linked to the net baryon density $n_B$ taking into account the proton-to-nucleon ratio $Z/A$ of the colliding nuclei. For instance, in the case of Au and Pb nuclei with $Z/A$ around 0.4, one can assume $\langle n_Q\rangle=0.4\,\langle n_B\rangle$ for the systems produced in the collisions of these nuclei \cite{Noronha-Hostler:2019ayj,Monnai:2019hkn}. Based on these considerations, Ref.~\cite{Monnai:2019hkn} imposes various constraining conditions on the EOS and constructs three types: NEOS-B (assuming vanishing strangeness and electric charge chemical potentials, $\mu_Q=\mu_S=0$), NEOS-BS (assuming strangeness neutrality $\langle n_S\rangle=0$ and vanishing electric charge chemical potential $\mu_Q=0$), and NEOS-BQS (assuming strangeness neutrality $\langle n_S\rangle=0$ and a fixed electric charge-to-baryon ratio $\langle n_Q\rangle=0.4\,\langle n_B\rangle$). Comparing results obtained with different equations of state can elucidate the effects arising from imposing these distinct constraints.

\begin{figure}[t]
\centering
    \hspace{0.5mm}\includegraphics[width=.325\linewidth]{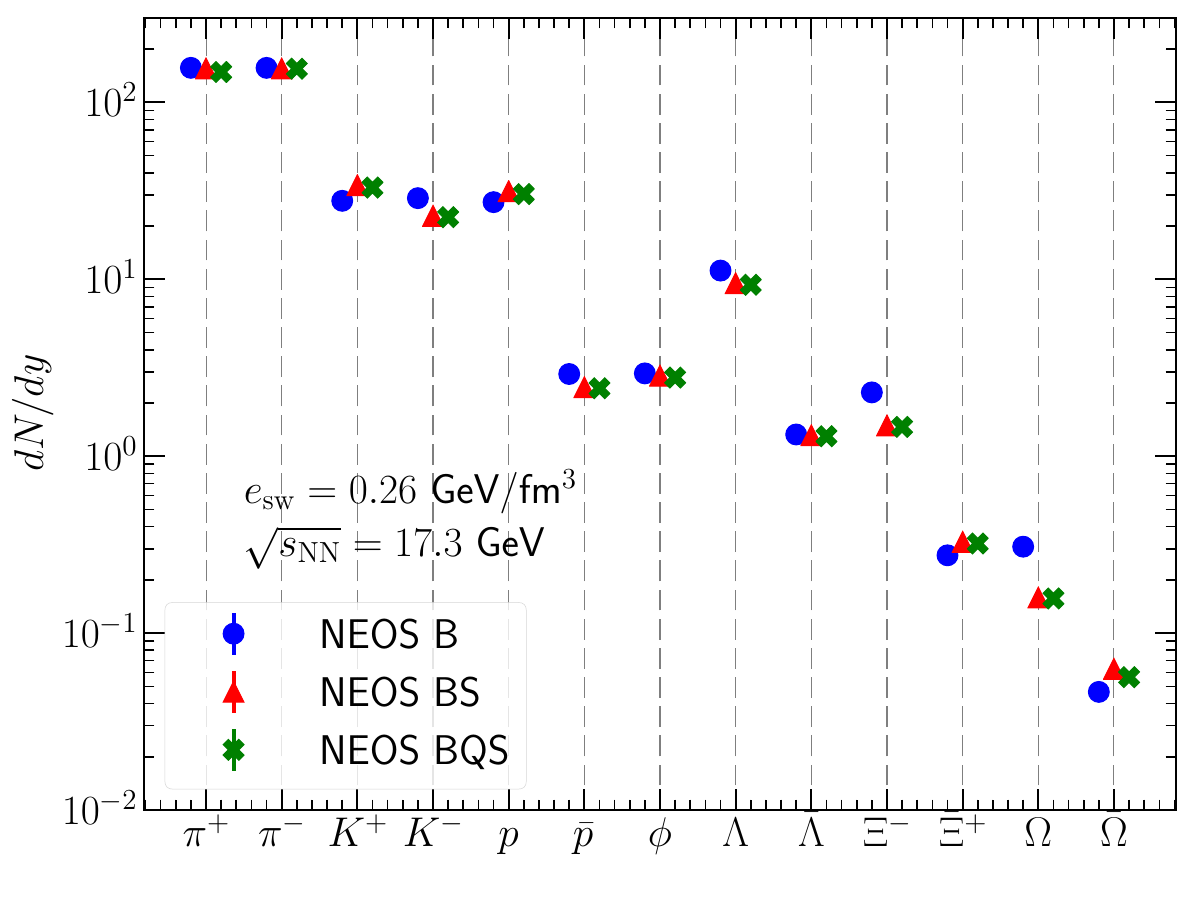}
    \includegraphics[width=.325\linewidth]{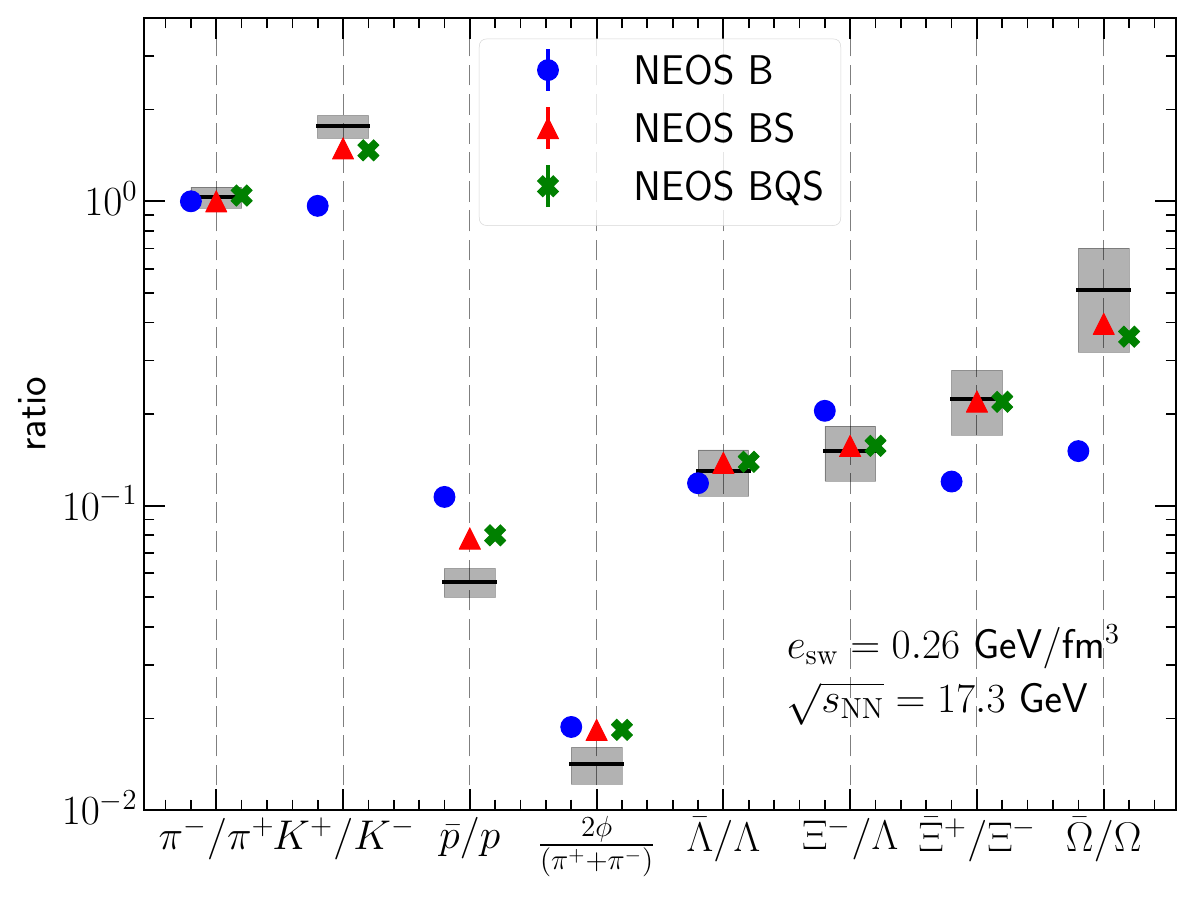}
    \includegraphics[width=.325\linewidth]{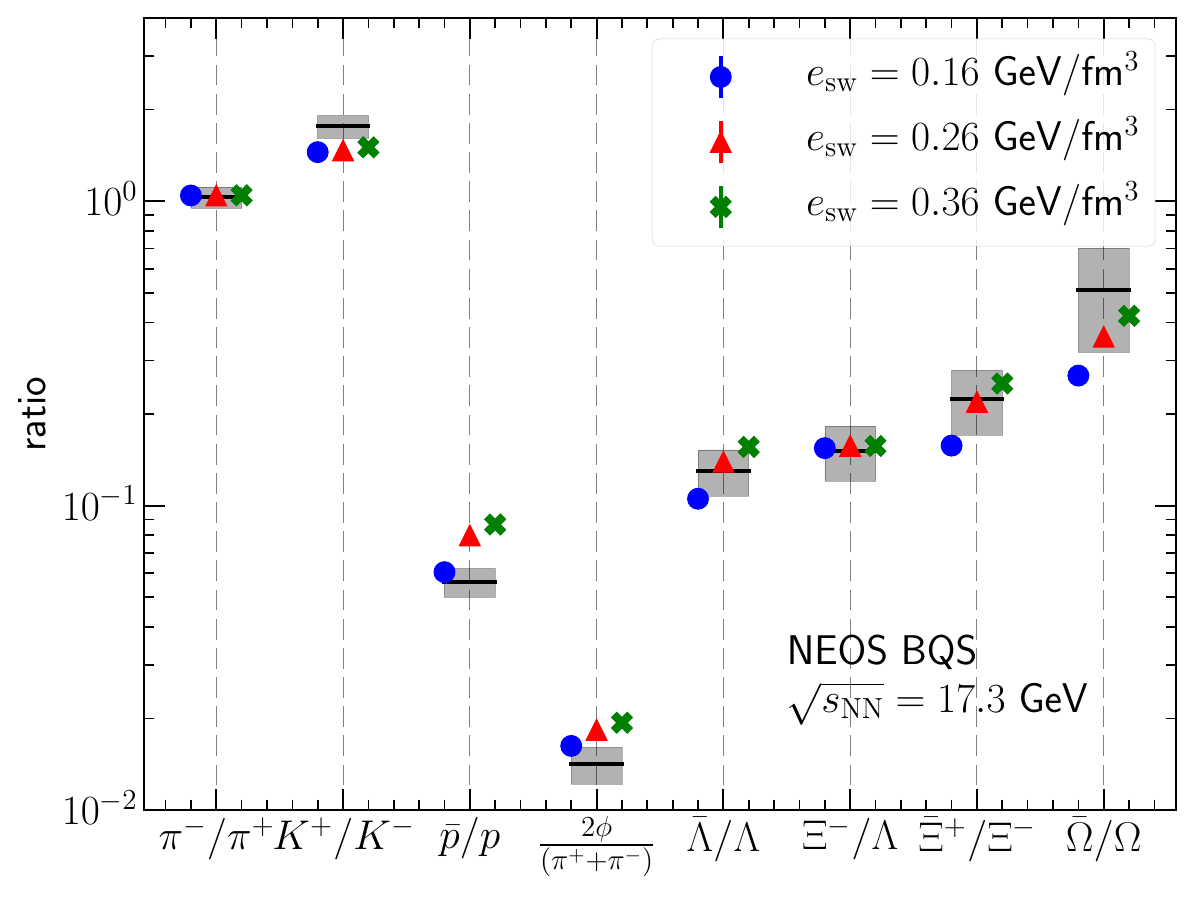}\\
    \vspace{1.5mm}
    \includegraphics[width=\linewidth]{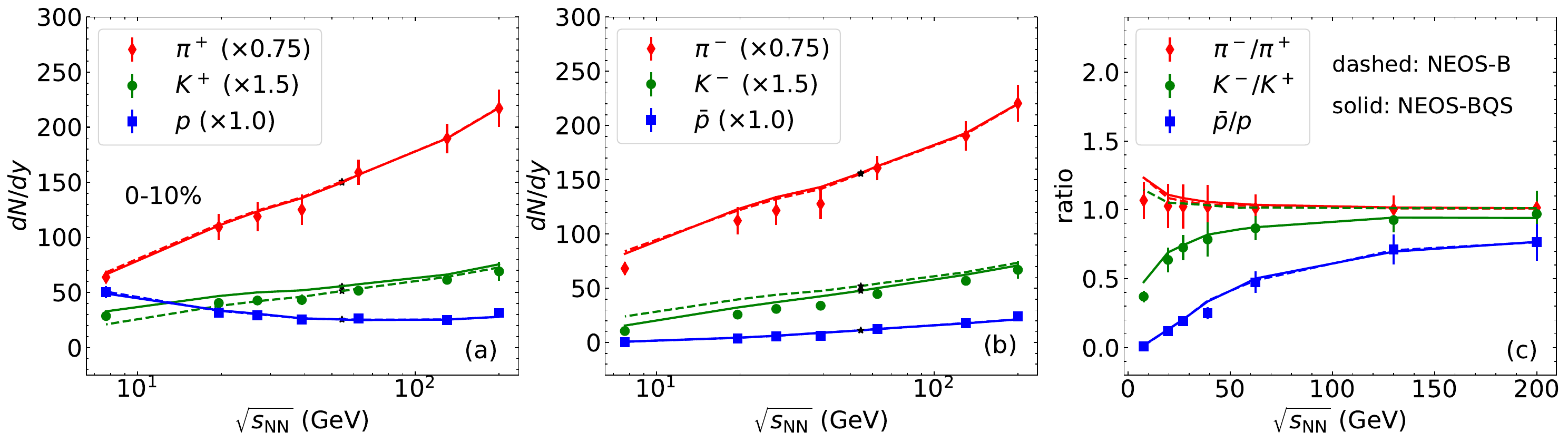}
    \caption{\textit{Top:} Particle yields and ratios for central Pb+Pb collisions at $\snn=17.3$ GeV with various freeze-out energy densities \cite{Monnai:2019hkn}. \textit{Bottom:} Particle yields and ratios for central Au+Au collisions at various beam energies $\snn$ with a fixed freeze-out energy density \cite{Du:2023efk}. The calculations incorporate equations of state with different constraints.}
    \label{fig:eos_yields}
\end{figure}

These constraining conditions interrelate the $BQS$ chemical potentials in various ways and thereby can impact the ratios between final yields of various hadron species carrying different hadronic chemical potentials. For instance, studies such as Refs.~\cite{Monnai:2019hkn, Du:2023efk} have indeed observed a significant improvement in the agreement between theoretical calculations and experimental measurements (see Fig.~\ref{fig:eos_yields}), particularly in the particle yield ratio such as $K^-/K^+$ as a function of beam energy when utilizing NEOS-BQS. However, it is crucial to acknowledge that the conditions $\langle n_S\rangle=0$ and $\langle n_Q\rangle=0.4\,\langle n_B\rangle$ should be applied to averaged quantities across the colliding nuclei. Using these EOS enforces these constraints on each fluid cell, potentially leading to overly stringent limitations. For instance, studies like Refs.~\cite{Du:2022yok, Du:2023efk} demonstrate that the implementation of NEOS-BQS could disrupt the agreement between theoretical predictions and experimental measurements for some rapidity-dependent observables obtained using NEOS-B, particularly in the directed flows, $v_1(y)$, for identified hadrons with strangeness (e.g., $\Lambda$ and $K^+$ carrying opposite $\mu_S$). This illustrates that the choice of an appropriate EOS, whether a 2-dimensional EOS with specific constraints to save computational time or a 4-dimensional EOS with multicharge evolution for interpreting particular observables, may highly depend on the specific experimental measurements and relevant physics of interest. Regardless, it is crucial to bear in mind that heavy-ion collisions at the BES explore the $T$-$\mu_B$-$\mu_Q$-$\mu_S$ space instead of the $T$-$\mu_B$ plane with $\mu_Q=\mu_S=0$. Caution should be exercised when drawing conclusions solely in terms of $T$-$\mu_B$.

In phenomenological studies employing multistage hydrodynamic simulations, the EOS is often fixed as an input, without accounting for potential theoretical uncertainties. However, uncertainties associated with the EOS can arise from various factors, including uncertainties in lattice QCD results and different methods of interpolating the EOS between lattice QCD and the HRG. These uncertainties can propagate into the results of model calculations, impacting extracted QCD properties such as transport coefficients \cite{Bluhm:2007nu,Alba:2017hhe,Auvinen:2020mpc,Steinheimer:2022gqb}. To address these uncertainties, Bayesian constraints \cite{Pratt:2015zsa,Moreland:2015dvc,Oliinychenko:2022uvy,OmanaKuttan:2022aml} and machine learning techniques \cite{Pang:2016vdc,Wang:2020tgb} have been employed in the determination of EOS with consideration for various measurements (for a comprehensive review, see Ref.~\cite{Zhou:2023pti}). For example, in Ref.~\cite{Pratt:2015zsa}, Bayesian techniques were applied to a combined analysis of a large number of observables, allowing for variations in model parameters, which encompassed those related to the EOS. The resulting posterior distribution over possible equations of state was found to be consistent with results from lattice QCD EOS, providing valuable insights into constraining the QCD EOS.

Near the critical point, the thermodynamic properties of QCD matter are expected to exhibit singularities. To accurately describe dynamics in the vicinity of the critical point, the EOS should incorporate correct singular behaviors. The QCD critical point belongs to the same universality class as the 3D Ising model \cite{Guida:1996ep,Berges:1998rc, Halasz:1998qr}, from which the universal static critical behavior of the QCD critical point can be inferred. Starting with the pressure of the 3D Ising model, denoted as $p^\mathrm{Ising}(r, h)$, where $r$ is the reduced temperature and $h$ is the magnetic field, one must map it to the pressure of QCD as a function of $(T, \mu_B)$. However, the mapping is not universal, and different mappings can result in different shapes for the critical region, where the critical pressure has a significant contribution \cite{Nonaka:2004pg, Parotto:2018pwx, Karthein:2021nxe, Pradeep:2019ccv,Kapusta:2022pny}. Moreover, the global scale of the critical pressure is also unknown, with larger values corresponding to larger critical regions. To account for these non-universalities, one can parametrize the critical contribution to the pressure in the critical region, once a mapping between $(r, h)$ and $(T, \mu_B)$ is chosen:
\begin{equation}
p_\mathrm{critical}^\mathrm{QCD}(T, \mu_B) \sim p^\mathrm{Ising}\bigl(r(T, \mu_B), h(T, \mu_B)\bigr)\,,
\end{equation}
where a function of $(T, \mu_B)$ can be added to set the overall scale. The full pressure is then expressed as the sum of the Ising contribution (i.e., $p_\mathrm{critical}^\mathrm{QCD}$) and the non-Ising one, the latter of which is not known {\it a priori}. Refs.~\cite{Parotto:2018pwx, Karthein:2021nxe} construct the non-Ising pressure using the same Taylor expansion method described in Eq.~\eqref{eq:taylorexp_multi}, and the expansion coefficients at $\mu_B=0$ are calculated by subtracting the Taylor coefficients of the Ising model (with a factor relevant to the overall function mentioned above) from the ones calculated by lattice QCD. 

The construction of the EOS with a critical point involves large uncertainties and many parameters. Constraining these parameters from a model-to-data comparison, especially within a complex multistage hydrodynamic framework, is certainly not straightforward. Adding the critical point in an EOS can deform hydrodynamics trajectories, leading them to converge toward the critical point, a phenomenon known as the critical lensing (see the right panel of Fig.~\ref{fig:phase_diagram_traj}). This occurrence has been observed in both equilibrium and out-of-equilibrium evolutions, resulting in a higher number of trajectories passing through the vicinity of the critical point \cite{Dore:2022qyz}. This potentially enhances the likelihood of detecting the critical point in experimental settings. Finally, we also note that initial off-equilibrium effects associated with the shear stress tensor and bulk viscous pressure can significantly influence the trajectories on the phase diagram in a non-trivial manner \cite{Dore:2020jye,Chattopadhyay:2022sxk}.

\subsection{Freeze-out and particlization}\label{sec:fzout}

As the QGP expands and cools below the pseudocritical temperature, \textit{hadronization} takes place, transforming the deconfined partons into confined hadrons. During this transition, color neutralization takes place, leading to an increase in the mean-free path and rendering the hydrodynamic description ineffective. In multistage descriptions, we transition from a continuous macroscopic hydrodynamics framework to a discrete microscopic particle transport approach known as \textit{particlization}. This transition occurs on a freeze-out hypersurface denoted as $\Sigma^\mu(x)$, which is defined based on various criteria. One approach is to use a constant switching temperature $T_\mathrm{sw}$ when the matter is baryon-neutral \cite{JETSCAPE:2020mzn}, another involves a constant energy density $e_\mathrm{sw}$ when the matter is baryon-charged \cite{Denicol:2018wdp}, and yet another method selects a surface of constant Knudsen number \cite{Ahmad:2016ods}. These choices are motivated by our understanding of the phase transition line or the applicability of hydrodynamics. Determining the freeze-out hypersurface for particlization from a discretized hydrodynamic evolution can be a complex task, often accomplished using routines like the well-known Cornelius routine \cite{Huovinen:2012is}. Particles sampled on the hypersurface further propagate through the hadronic cascade stage, where the chemical and kinetic freeze-out processes are subsequently automatically performed.\footnote{%
Hadronization, chemical freeze-out, and kinetic freeze-out processes are physical phenomena, while particlization and the freeze-out hypersurface are not inherently physical but rather practical or computational concepts.
} 
Modeling the emission of hadrons on this hypersurface is a crucial component of the multistage descriptions of heavy-ion collisions.

\subsubsection{Cooper-Frye prescription}\label{sec:CF}

The Cooper-Frye prescription \cite{Cooper:1974mv,PhysRevD.11.192} establishes a connection between the hydrodynamic fields and momentum distributions of identified particles, yielding the Lorentz-invariant momentum distribution for species $i$ on a freeze-out surface denoted as $\Sigma_\mu (x)$, expressed as:
\begin{equation}\label{eq:distr}
    p^0\frac{d^3N_i}{d^3p}=\frac{1}{(2\pi)^3}\int_\Sigma d^3\sigma_\mu(x) p^\mu f_i(x, p)\,.
\end{equation}
Here, $p^\mu$ represents the four-momentum of the particle, $d^3\sigma_\mu(x)$ corresponds to the normal vector for a freeze-out surface element at the point $x$, and $f_i(x, p)$ is the one-particle distribution function for species $i$. 
If we refrain from integrating over the hypersurface in Eq.~\eqref{eq:distr} and perform momentum integration over the formula, it becomes evident that the Cooper-Frye prescription conserves various types of charges across the surface element, as indicated by:
\begin{equation}\label{eq:CF_Nq}
    N^\mu_q(x)=\sum_i q_i \int \frac{d^3p}{(2\pi)^3p^0} p^\mu f_i(x, p)\,,\quad q\in\{B,\,Q,\,S\}
\end{equation}
where $q_i$ represent $BQS$ charges carried by species $i$. This equation describes the local conservation of net charge $q$ at surface element $d^3\sigma_\mu(x)$ with spacetime coordinates $x$. 
Moreover, the distribution function $f_i(x, p)$ should be such that it conserves the energy-momentum tensor for an element at $x$ on the hypersurface as well: 
\begin{equation}\label{eq:CF_Tmunu}
    T^{\mu\nu}(x)=\sum_i \int \frac{d^3p}{(2\pi)^3p^0} p^\mu p^\nu f_i(x, p)\,.
\end{equation}
The distributions given by Eq. \eqref{eq:distr} are Monte Carlo sampled to generate an ensemble of particles with well-defined positions and momenta. These ensembles serve as the initial state for the hadronic afterburner, as described by transport models that allow for the decay of unstable resonances and their regeneration via hadronic rescattering (see Sec.~\ref{sec:hadronic_afterburner}).

\subsubsection{Off-equilibrium corrections}\label{sec:off-eq}

In a scenario where the QGP fluid is in a state of perfect local kinetic and chemical equilibrium, the distribution function $f_i(x, p)$ in Eq.~\eqref{eq:distr} adopts the form outlined in Eq.~\eqref{eq:fk_eq}. Within the context of particlization, the local temperature $T(x)$ and chemical potential $\mu_i(x)$ \eqref{eq:hadron_chem} should be considered as values on the freeze-out hypersurface. However, given the dissipative nature of the QGP fluid, which is indicated by the presence of terms such as $\pi^{\mu\nu}$, $\Pi$, and $n^\mu_q$ within the hydrodynamic description, the distribution function $f_i(x, p)$ includes off-equilibrium corrections. Without any hydrodynamic information guiding the decomposition of the fluid energy-momentum tensor $T^{\mu\nu}$ into contributions from different hadron species $i$, as written in Eq.~\eqref{eq:CF_Tmunu}, and lacking hydrodynamic information for all the infinitely many other momentum moments of the distribution functions (e.g., $\int _p p^\mu p^\nu p^\lambda f_i$), there exist infinitely many possibilities for the choice of distribution functions $f_i(x, p)$ \cite{JETSCAPE:2020mzn}.

Given that the dissipative terms, such as $\pi^{\mu\nu}$, $\Pi$, and $n^\mu_q$, are intended to capture deviations in the hadrons' momentum distributions and yields from local thermodynamic equilibrium, $f_{\mathrm{eq},i} \equiv f^0_{i, \mathbf{p}}$, due to dissipative corrections, the distribution functions $f_i(x, p)$ can be expressed as: 
\begin{equation}\label{eq:off_equi_dist}
    f_i(x, p)\equiv f_{\mathrm{eq},i}+\delta f_{i}= f_{\mathrm{eq},i}+\delta f_{\pi,i}+\delta f_{\Pi,i}+\delta f_{n_q,i}\,.
\end{equation}
Here the terms $\delta f_{\pi,i}$, $\delta f_{\Pi,i}$, and $\delta f_{n_q,i}$ correspond to the phase space distributions that encapsulate off-equilibrium effects originating from the dissipative terms $\pi^{\mu\nu}$, $\Pi$, and $n^\mu_q$, respectively. In a formal context, to deduce the off-equilibrium corrections from the dissipative terms, we require the inverse mapping provided by Eqs.~\eqref{eq:fk_pimunu}-\eqref{eq:fk_nqmu}. The constraints from these dissipative terms are insufficient to fully determine all the off-equilibrium corrections associated with every hadron species, given the count of known and unknown variables. In principle, the same methods used to derive the hydrodynamic equations of motion (see Eq.~\eqref{eq:df_expansion}), such as the Grad’s method or Chapman-Enskog method, can also be employed to obtain these off-equilibrium terms in the Cooper-Frye prescription. However, similar to the transport coefficients and EOS derived directly from underlying kinetic theory, which fundamentally differ from those of QCD as previously discussed, there is often a lack of clear theoretical guidance regarding which formalism of $\delta f_{i}$ to utilize for implementing particlization. 

Assuming the simultaneous applicability of dissipative fluid dynamics and relativistic kinetic theory on the freeze-out hypersurface, various models of viscous corrections to the local equilibrium distribution function have been developed. These include Grad’s method (also known as the 14-moments method in the relativistic context) \cite{Grad:1949zza,Teaney:2003kp}, the first-order Chapman-Enskog (CE) expansion in the relaxation time approximation \cite{Anderson:1974nyl,Jaiswal:2014isa}, the Pratt-Torrieri-Bernhard (PTB) modified equilibrium distribution \cite{Pratt:2010jt}, the Pratt-Torrieri-McNelis (PTM) modified equilibrium distribution \cite{McNelis:2019auj}, the Pratt-Torrieri-McNelis modified anisotropic distribution (PTMA) \cite{McNelis:2021acu}, and the newly proposed Maximum Entropy distribution \cite{Everett:2021ulz,Chattopadhyay:2023hpd}. Brief summaries of the features of these various $\delta f_{i}$ models can be found in Refs.~\cite{McNelis:2019auj, JETSCAPE:2020mzn, McNelis:2021acu}. Once $\delta f_{i}$ and, consequently, $f_i$ are determined, generating an ensemble of particles with positions and momenta at particlization involves interpreting the Cooper-Frye integrand in Eq.~\eqref{eq:distr} as a probability density in phase-space and sampling it stochastically, recognizing that the integrand may not always be positive-definite. Two potential sources of negative contributions arise from instances where either $f_i(x, p)$ becomes negative or $p^\mu d^3\sigma_\mu(x)$ turns negative.

For instance, the distribution functions $f_i(x, p)$ can go negative when a linearized form of $\delta f_{i}$ is used, such as in the Grad and Chapman-Enskog methods, which assume $|\delta f|\ll f_\mathrm{eq}$. These two models provide corrections that are linear in the dissipative currents $\pi^{\mu\nu}$ and $\Pi$, and the corrections scale either linearly or quadratically with the hadron momentum $p$. Consequently, there are values of $\pi^{\mu\nu}$ and $\Pi$ for which $|\delta f| > f_\mathrm{eq}$, even for moderate momenta. Moreover, even for small values of $\pi^{\mu\nu}$ and $\Pi$, $|\delta f| > f_\mathrm{eq}$ at sufficiently large momenta. In hydrodynamic simulations of heavy-ion collisions, it is not uncommon to encounter situations where $|\delta f| > f_\mathrm{eq}$ or $f_i(x, p)$ becomes negative in certain phase-space regions. This is particularly noticeable when particlization is performed near the phase transition, where the bulk viscosity peaks, leading to significant $\delta f_{i}$ corrections. These regions, where the assumption $|\delta f|\ll f_\mathrm{eq}$ of these models is often pushed to the limit or even beyond, are generally small enough to not significantly contribute to experimental observables. Commonly, this issue is addressed by regulating the Grad or Chapman-Enskog viscous corrections to prevent $|\delta f| > f_\mathrm{eq}$ and by ignoring the fluid cells where $f_i(x, p)$ becomes negative. Although this approach may violate conservation laws (\ref{eq:CF_Nq},\ref{eq:CF_Tmunu}) at hadronization, the deviations of the regulated $f_i(x, p)$ from the original spectra are usually small, except for very soft or high-momentum particles.

The necessity for regulating linearized viscous corrections has motivated the development of models aiming to resum the viscous corrections to all orders and provide positive-definite ``modified equilibrium distributions.'' The PTB, PTM, and PTMA distributions \cite{Pratt:2010jt,McNelis:2019auj,McNelis:2021acu} serve this purpose by incorporating the effects of viscous pressures into the arguments of the exponential function characterizing the equilibrium distribution outlined in Eq.~\eqref{eq:fk_eq}. This is achieved by introducing effective temperature and chemical potential, effectively transforming Eq.~\eqref{eq:fk_eq} into a quasiequilibrium distribution. In practice, the PTMA distribution is more suitable for use on freeze-out hypersurfaces constructed from anisotropic hydrodynamic simulations. It is worth noting that, in contrast to the unregulated linearized Grad and Chapman-Enskog distributions, the PTB, PTM, and PTMA distributions do not exactly satisfy the matching constraint Eq.~\eqref{eq:CF_Tmunu} when the viscous stresses are large.  A comparison of particle spectra and $p_T$-differential elliptic flow coefficients from the Cooper-Frye formula, computed with the modified equilibrium distribution and with linearized viscous corrections, is presented in Ref.~\cite{McNelis:2021acu}. The Maximum Entropy distribution, whose features are yet to be fully explored, constitutes a novel type of positive-definite particle distribution. It introduces dissipative corrections to the local thermal equilibrium distribution by maximizing entropy and does not rely on any uncontrolled assumptions about the microscopic dynamics of the distribution function.

Another source of negative contributions to Eq.~\eqref{eq:distr} arises from regions on the hypersurface with a spacelike normal vector (i.e., $d^3\sigma\cdot d^3\sigma<0$), where certain momenta satisfy $p\cdot d^3\sigma<0$. This indicates that the momenta are directed toward the surface elements, representing particles flowing into the hydrodynamic region and subsequently being reabsorbed by the fireball, i.e., the backflow. Various methods exist to handle these types of negative contributions \cite{Pratt:2014vja}. The first approach is simply to neglect the negative contribution \cite{Huovinen:2012is}. However, since contributions from all momenta are required to construct the conserved currents at a position $x$, discarding contributions from momenta with $p\cdot d^3\sigma<0$ would violate current conservation. A second method for handling the backflow has been to remove particles from the hadronic cascade that cross back into the hydrodynamic region \cite{Pratt:2008sz}. However, the numerical cost associated with tracking the backflow would be high, as it is challenging to find a robust algorithm that avoids constant checks for backflows. A third method is, instead of tracking particles relative to the interface, to simulate the evolution of particles emitted with negative weights \cite{Pratt:2014vja}. Studies on negative contributions stemming from backflow can be found in Refs.~\cite{Huovinen:2012is,Oliinychenko:2014tqa,Oliinychenko:2020cmr}.

\begin{figure}[t]
    \centering
    \includegraphics[width= 0.99\linewidth]{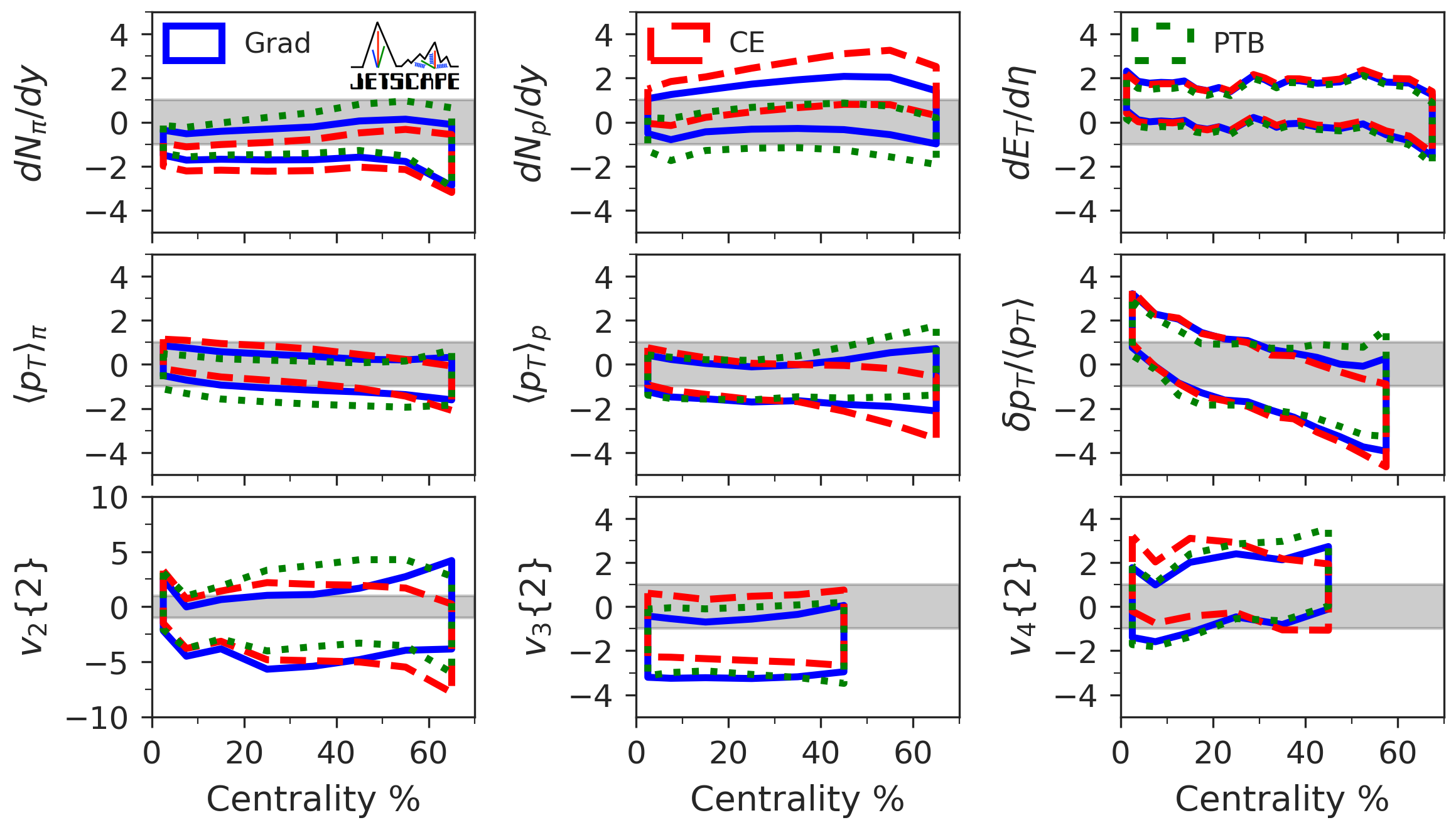}
    \caption{The 90\% credible intervals of the posterior distribution of observables for Pb+Pb collisions at the LHC as functions of centrality, considering various particlization models. The vertical axis represents the model discrepancy in units of the experimental standard deviation. Figure from Ref.~\cite{JETSCAPE:2020shq}.}
    \label{fig:off-eqn}
\end{figure}

It is worth noting that, given an off-equilibrium correction model, one can calculate $f_i(x, p)$ in Eq.~\eqref{eq:off_equi_dist}, and thus determine the current $N^\mu(x)$ and energy-momentum tensor $T^{\mu\nu}(x)$ using the {\it smooth} Cooper-Frye distribution. In this scenario, there is no need to sample particles, eliminating the requirement to interpret Eq.~\eqref{eq:distr} as a probability distribution. Consequently, there is no concern about it going negative, and thus, no need to regulate $\delta f_i$ or discard any fluid cells. One can assess the performance of models for $\delta f_i(x, p)$ by comparing the constructed $N^\mu(x)$ and $T^{\mu\nu}(x)$ from $f_i(x, p)$ with those obtained from the hydrodynamic evolution on the hypersurface, as discussed in \cite{Pratt:2010jt,McNelis:2021acu}. A well-designed model of $\delta f_i(x, p)$ should yield a good match between them. Finally, it is crucial to acknowledge that the selection of a dissipative correction model involves irreducible theoretical ambiguity. Therefore, it is logical to inquire about the constraints that experimental data may provide on models of dissipative corrections to the local equilibrium distribution function. For example, a recent model-to-data analysis quantifying the theoretical uncertainties associated with the particlization process in heavy-ion collisions was reported in Ref.~\cite{JETSCAPE:2020mzn,JETSCAPE:2020shq} (see Fig.~\ref{fig:off-eqn}).

\subsubsection{Statistical ensembles and sampling algorithms}\label{sec:sampling}

To calculate the off-equilibrium corrections $\delta f_i$ and subsequently apply the Cooper-Frye prescription \eqref{eq:distr} for obtaining the momentum distribution of particle species, several key elements are required. These include the hydrodynamic fields, such as $(T, \mu_B, \mu_Q, \mu_S, u^\mu)$, as well as dissipative variables like $\pi^{\mu\nu}$, $\Pi$, $n_q^\mu$, at each spacetime point $x$, on the freeze-out hypersurface. Furthermore, the spacetime-dependent normal vector of the hypersurface itself, $d^3\sigma_\mu(x)$, is also essential for this process. The aforementioned Cornelius routine can construct such a 4-dimensional hypersurface $\Sigma^\mu(x)$ with the above required elements in the context of (3+1)-dimensional hydrodynamic simulations. In practice, on each fluid cell on the hypersurface, the distributions given by Eq.~\eqref{eq:distr} are Monte Carlo sampled to generate particles with well-defined positions and momenta. The primary task of a hadronic sampler is to faithfully sample particles whose probabilities follow these distributions. Several samplers exist in the field, including \texttt{Therminator} \cite{Kisiel:2005hn}, \texttt{iSS} \cite{Shen:2014vra}, \texttt{iS3D} \cite{McNelis:2019auj}, \texttt{Microcanonical Sampler} \cite{Oliinychenko:2019zfk}, and \texttt{FIST-Sampler} \cite{Vovchenko:2022syc}.

In the most common sampling procedure, the yields of hadrons and resonances are usually sampled independently in each fluid cell from a Poisson distribution, with the particle number obtained from the Cooper-Frye formula interpreted as its mean \cite{Schwarz:2017bdg}. Since the Poisson distribution is additive, this implies that the yields of all hadron species in the entire space on the hypersurface follow a Poisson distribution as well. Consequently, the samples obtained belong to a grand canonical ensemble (GCE), wherein, when comparing the hydrodynamic quantities and the sampled particles, only the temperature, chemical potentials, and volume are fixed, while the energy, momentum, and charges are conserved only on average. To enhance statistics and reduce statistical uncertainties, particles are often sampled multiple times for each hydrodynamic hypersurface, a method known as the over-sampling method. With over-sampling, grand canonical sampling is computationally efficient and can provide good approximations for bulk observables, such as transverse momentum spectra and rapidity distributions. Various sampling routines, considering the yields, momenta, viscous corrections, and quantum statistics of different hadron species with various masses, have been developed \cite{Petersen:2008dd,Pratt:2014vja,Shen:2014vra,Schwarz:2017bdg,Denicol:2018wdp,Oliinychenko:2019zfk,McNelis:2019auj}. A robust sampling method should yield particle distributions obtained from a large number of oversamples that agree with those from smooth Cooper-Frye distributions. This serves as a common validation method to assess the performance of a hadronic sampler \cite{Shen:2014vra,Denicol:2018wdp,McNelis:2019auj,Oliinychenko:2020cmr,Vovchenko:2022syc,Du:2023efk}.

While grand canonical sampling is arguably consistent with the hydrodynamic description, as seen in the usage of lattice QCD EOS assuming the GCE, recent studies emphasize caution and highlight its potential effects on observables sensitive to event-by-event fluctuations \cite{Oliinychenko:2016vkg,Steinheimer:2017dpb,Schwarz:2017bdg,Oliinychenko:2019zfk,Oliinychenko:2020cmr}.\footnote{%
The fluctuations due to sampling on a hydrodynamic hypersurface differ from the fluctuations arising from an ensemble of hydrodynamic events, which result from factors such as initial condition fluctuations and thermal fluctuations.
}
For example, on an event-by-event level, grand canonical sampling has the potential to violate conservation laws at particlization, leading to fluctuations in final particle multiplicity. These fluctuations may subsequently impact the selection of centrality classes \cite{Huovinen:2012is}. A comparison with cases that account for global conservation in each event, as demonstrated in Ref.~\cite{Schwarz:2017bdg}, reveals that the differences in bulk observables, such as transverse momentum spectra and rapidity distributions, between the two approaches are expected to be more pronounced for lower collision energies, smaller collision systems, and rare hadron species. The potential effects on observables involving fluctuations and correlations are of particular interest, especially in the context of searching for the critical point. Therefore, we require a particlization routine that can avoid introducing unphysical fluctuations due to the sampling method, while preserving the physical fluctuations, whether critical or thermal, from the hydrodynamic evolution properly.

\begin{figure}[t]
    \centering
    \includegraphics[width= 0.24\linewidth]{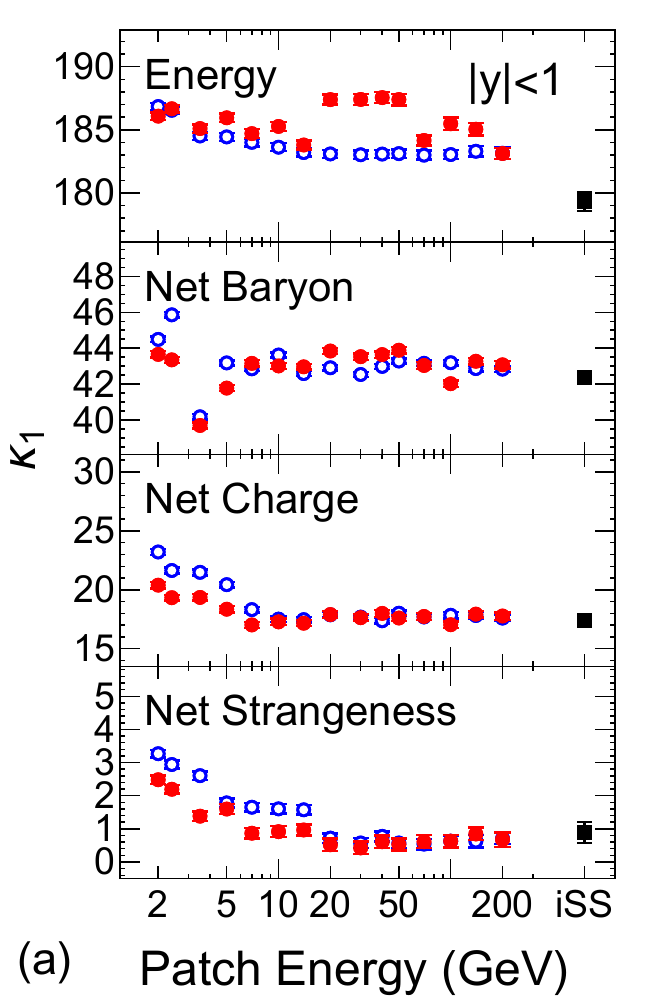}
    \includegraphics[width= 0.24\linewidth]{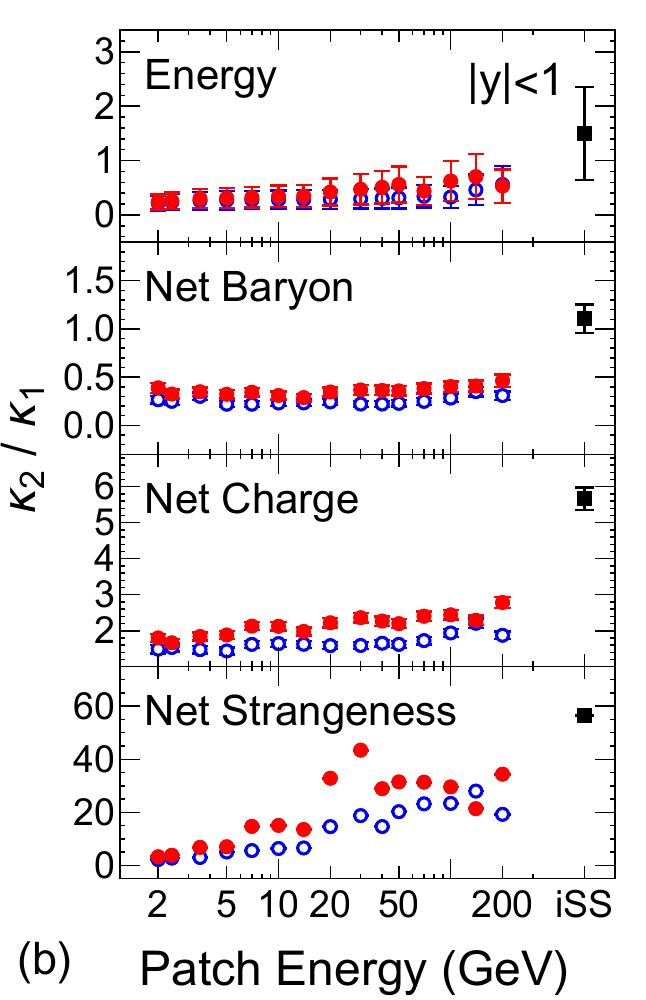}
    \includegraphics[width= 0.24\linewidth]{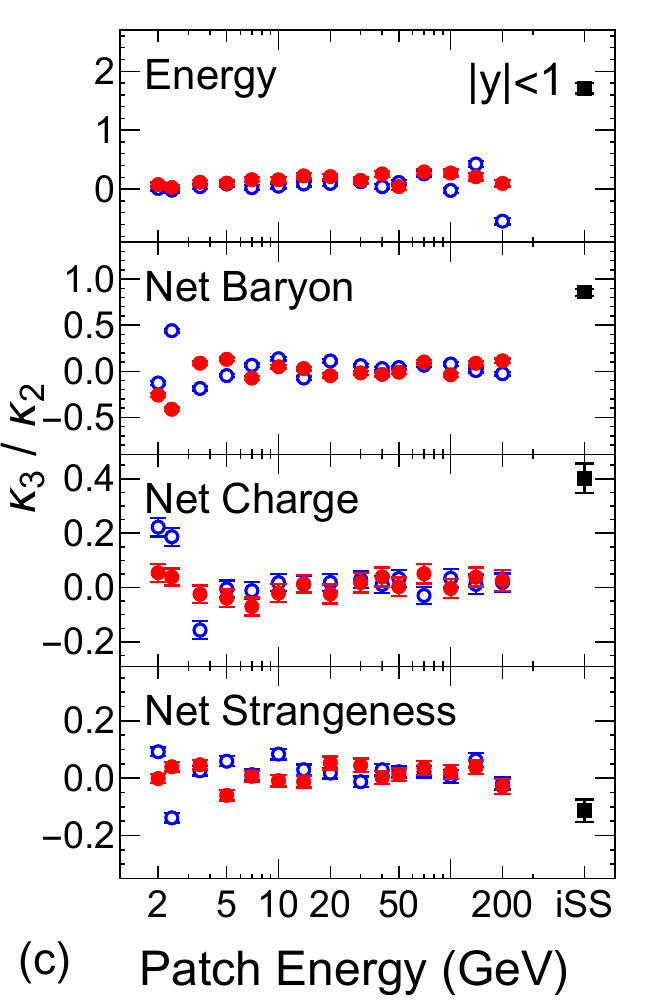}
    \includegraphics[width= 0.24\linewidth]{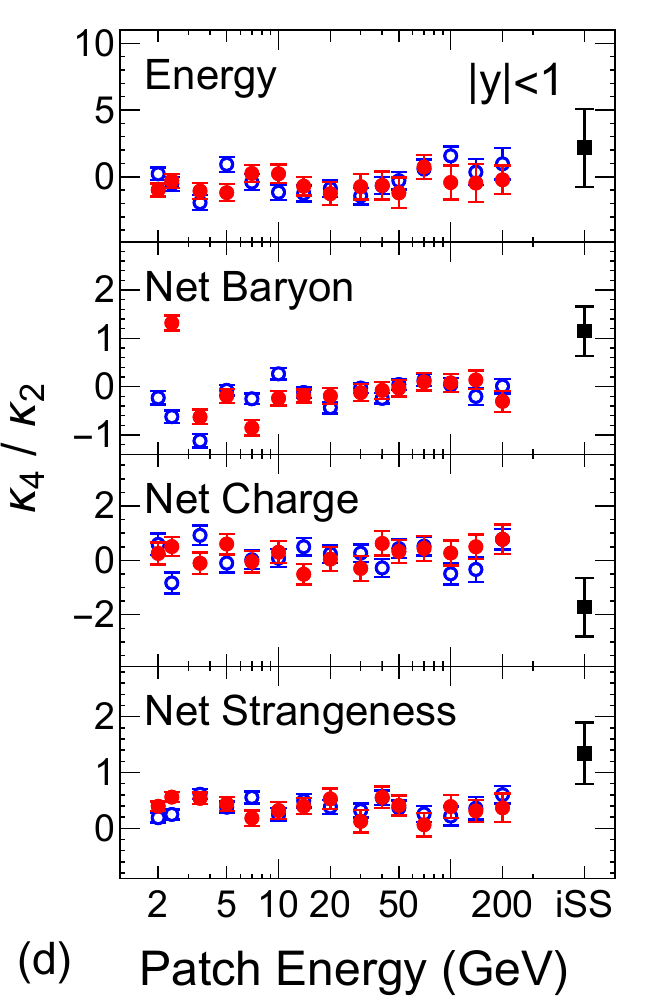}
    \caption{Cumulants or cumulant ratios of conserved quantities for particles within a rapidity range $|y| < 1$ as a function of patch energies. The results obtained by the \texttt{iSS} sampler assume the GCE sampling method. From top to bottom: energy, net baryon number, net electric charge, and net strangeness, respectively. Figure from Ref.~\cite{Oliinychenko:2020cmr}.}
    \label{fig:sampler}
\end{figure}

The effects of assuming different statistical ensembles have been studied by comparing grand canonical sampling to other cases, such as microcanonical sampling, where conservation laws are fulfilled for the entire freeze-out hypersurface globally in every event \cite{Petersen:2008dd,Huovinen:2012is,Schwarz:2017bdg}. If this holds not only for the entire hypersurface but also for smaller spacetime regions, it is referred to as local microcanonical sampling. Refs.~\cite{Oliinychenko:2019zfk,Oliinychenko:2020cmr} implemented local microcanonical sampling, where conservation laws are satisfied on ``patches'' which are space-time regions smaller than the entire surface but still containing many particles. Due to the conservation laws, the sampled particles of different species are found to be correlated, deviating from the conventional grand canonical sampling method with no such constraints (see Fig.~\ref{fig:sampler}). Various effects, including suppression in $p_T$ spectra and enhancement in $v_2(p_T)$ for protons at high $p_T$, stronger suppression of fluctuations, and enhancement of correlations, were observed. They found that many effects, particularly fluctuations and correlations, do not vanish even in the thermodynamic limit. The effects of over-sampling on the net baryon number cumulants were also studied \cite{Steinheimer:2017dpb}. It was found that the commonly used finite particle number sampling procedure introduces an additional Poissonian contribution when global baryon number conservation is absent or multinomial if global baryon number conservation is enforced. This introduces an increase in the extracted moments of the baryon number distribution. 

The above sampling methods of the Cooper-Frye formula, even when considering off-equilibrium corrections, generate an ensemble of particles corresponding to an ideal hadron resonance gas with no interactions. Certainly, the particles could exhibit correlations due to the collective motion inherited from the hydrodynamic expansion. Therefore, these existing methods are not suitable for analyzing fluctuation signals beyond the physics of an ideal hadron gas. Extensions of the Cooper-Frye procedure are necessary to incorporate additional physics, such as extending the ideal HRG to include short-range repulsive interactions \cite{Vovchenko:2020lju}. A particlization routine appropriate for describing event-by-event fluctuations encoded in the EOS of such interacting HRG models is implemented in \texttt{FIST-Sampler} \cite{Vovchenko:2020kwg,Vovchenko:2022syc}. The short-range repulsion between particles is implemented through a rejection sampling step, which prohibits any pair of particles from overlapping in coordinate space. This step effectively models the effect of hard-core repulsion.

Moreover, it is crucial to maintain the critical correlations in coordinate space, which are propagated during the hydrodynamic evolution, and translate them into correlations in momentum space among the particles. This preservation is essential for the search for the critical point through model-data comparison. Attempts have been made to calculate (without sampling) correlations of multiplicity fluctuations at different rapidities from off-equilibrium critical fluctuations propagated in the hydrodynamic stage (see Sec.~\ref{sec:fluctuations}).

\subsection{Hadronic afterburner}
\label{sec:hadronic_afterburner}

The last component of a multistage approach to describing heavy-ion collision dynamics, known as the hadronic afterburner, evolves hadrons from particlization to the last scattering (kinetic freeze-out). 
Physically, this stage describes a gas, consisting of hundreds of hadronic species, which is expanding and cooling. 
Throughout this process, inelastic collisions become infrequent and, consequently, chemical equilibration times become much longer than the lifetime of the hadronic stage, which means that chemical equilibrium ceases to be preserved~\cite{Prakash:1993bt,Rose:2020lfc}. 
Similarly, local kinetic equilibrium is also difficult to maintain because with the expanding volume~$V$, heavy particles cool faster ($T \propto 1/V^{2/3}$) than nearly massless particles ($T \propto 1/V^{1/3}$). 
Because of these factors, the role of the hadronic afterburner is best fulfilled by dynamic microscopic transport models, which handle the non-equilibrium evolution taking place after the system is particlized.

\begin{figure}[t]
\centering
    \includegraphics[width=0.99\linewidth]{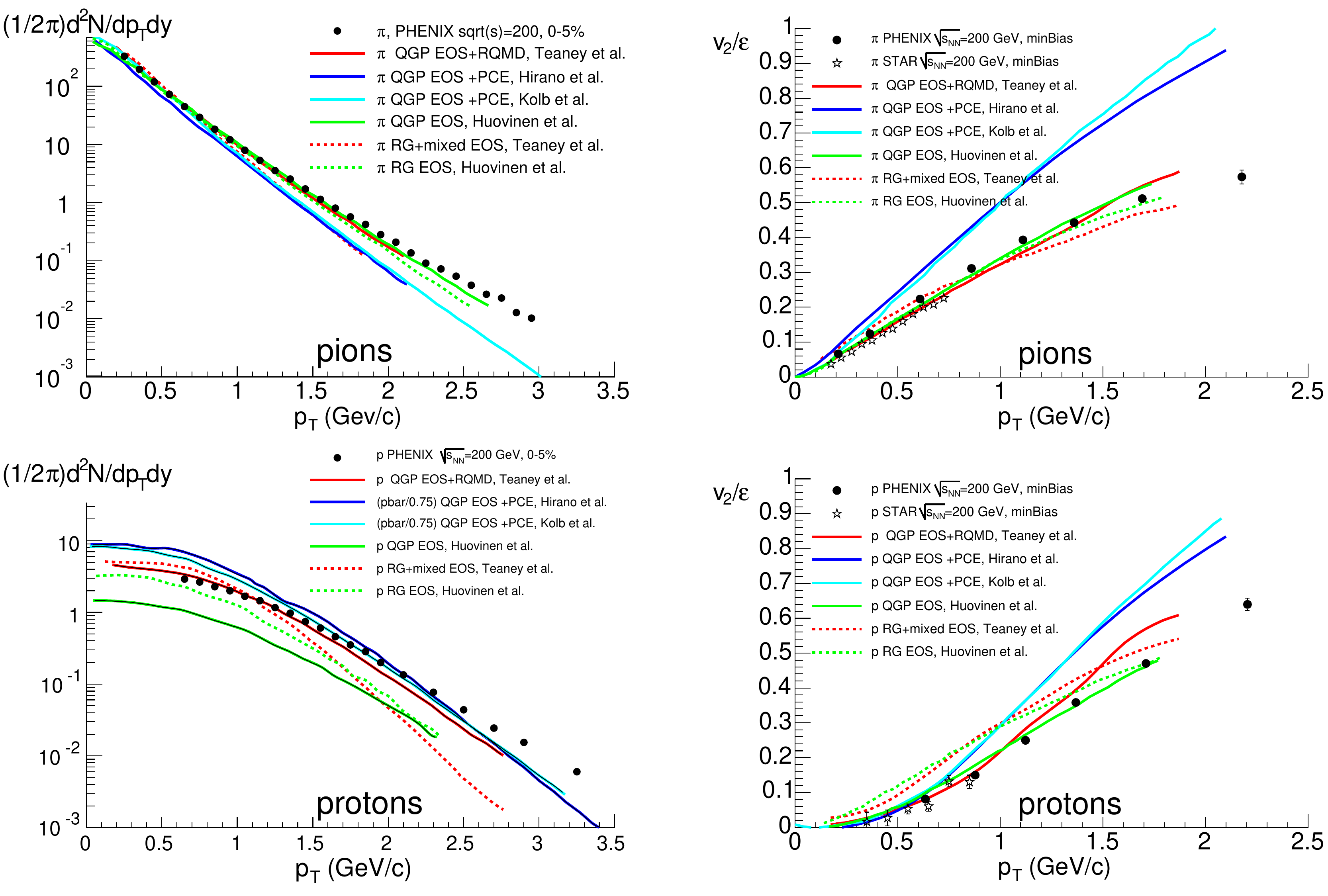}
    \caption{The $p_T$-spectra (\textit{left}) and $p_T$-differential elliptic flow $v_2$ scaled by the initial eccentricity $\varepsilon$ (\textit{right}) for pions (\textit{top}) and protons (\textit{bottom}). The PHENIX data~\cite{PHENIX:2004vcz} is compared with results from simulations within several hydrodynamic models~\cite{Teaney:2000cw,Huovinen:2001cy,Kolb:2002ve,Hirano:2002ds,Hirano:2004en}. Figure from~\cite{Sorensen:2009cz}.
  	}
\label{fig:hydro_compilation}
\end{figure}
The crucial influence of including a hadronic afterburner in simulations was discovered when dynamic models attempted to simultaneously describe multiple observables measured at the top RHIC energy of $\snn = 200~\rm{GeV}$. 
A compilation~\cite{Sorensen:2009cz} of such attempts within a selection of
models~\cite{Teaney:2000cw,Huovinen:2001cy,Kolb:2002ve,Hirano:2002ds,Hirano:2004en} is shown in Fig.~\ref{fig:hydro_compilation} against measurements from the STAR~\cite{STAR:2005gfr} and PHENIX~\cite{PHENIX:2004vcz} experiments.
The only model describing all shown data simultaneously (labeled as ``$\pi$ QGP EOS+RQMD, Teaney et al.'') used a hybrid approach, novel at the time, consisting of a hydrodynamic simulation followed by a hadronic afterburner~\cite{Teaney:2000cw,Teaney:2001av}. 
This allowed the model to use a soft EOS in the hydrodynamic stage, producing results for the differential elliptic flow~$v_2(p_T)$, where $p_T$ is the transverse momentum, agreeing with the experimental data, while also obtaining relatively hard $p_T$-spectra, likewise consistent with experimental measurements, thanks to effects due to rescattering in the afterburner stage.

\begin{figure}[t]
    \centering
    \includegraphics[width=0.99\linewidth]{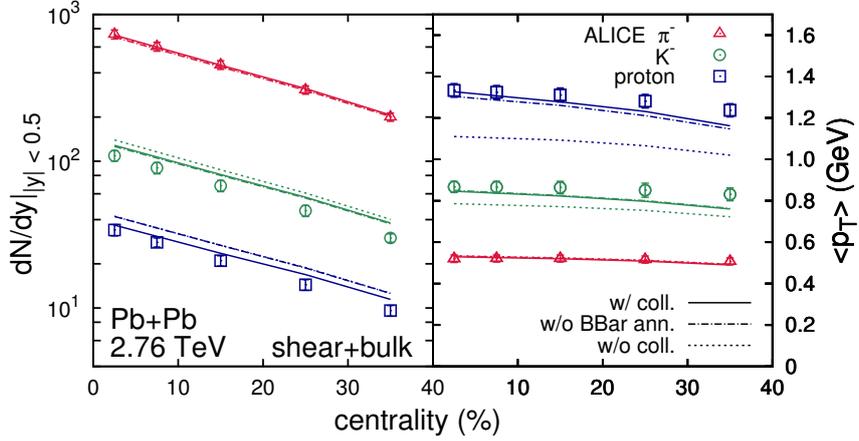}
    \caption{The centrality dependence of midrapidity yields (\textit{left}) and mean $p_T$ (\textit{right}) of pions, kaons, and protons in Pb+Pb collisions at $\snn = 2.76~\rm{TeV}$. Results from simulations including full \texttt{UrQMD} (labeled ``w/~coll.''), \texttt{UrQMD} without $B\bar{B}$ annihilations (labeled ``w/o~BBar ann.''), and \texttt{UrQMD} without both collisions and annihilations (labeled ``w/o~coll.'') are shown by the solid, dash-dotted, and dotted curves, respectively, with the ALICE data~\cite{ALICE:2013mez} also shown for comparison. Figure from~\cite{Ryu:2017qzn}.
	}
    \label{fig:hadronic_afterburner_change_in_yields}
\end{figure}

Elements of hadronic transport approaches as well as recent advancements in modeling are discussed in Sec.~\ref{sec:microscopic_transport}.
Below, we highlight the role of the afterburner stage in arriving at the final results of multistage dynamical simulations.
Notably, this role becomes ever more prominent as collision energies decrease and the hadronic stage describes an ever larger fraction of the total evolution time.

\subsubsection{Particle yields}

Hadronic afterburners change the final particle yields, with the extent of this effect depending on the particle species. 
At top collider energies, pion and kaon yields are only changed by a few percent, while proton yields can be changed, mainly due to annihilation processes (whose importance is expected to be proportional to the collision energy), by 10\% to 30\%~\cite{Ryu:2017qzn}, see the left panel in Fig.~\ref{fig:hadronic_afterburner_change_in_yields}.

\begin{figure}[t]
    \centering
    \includegraphics[width=0.99\linewidth]{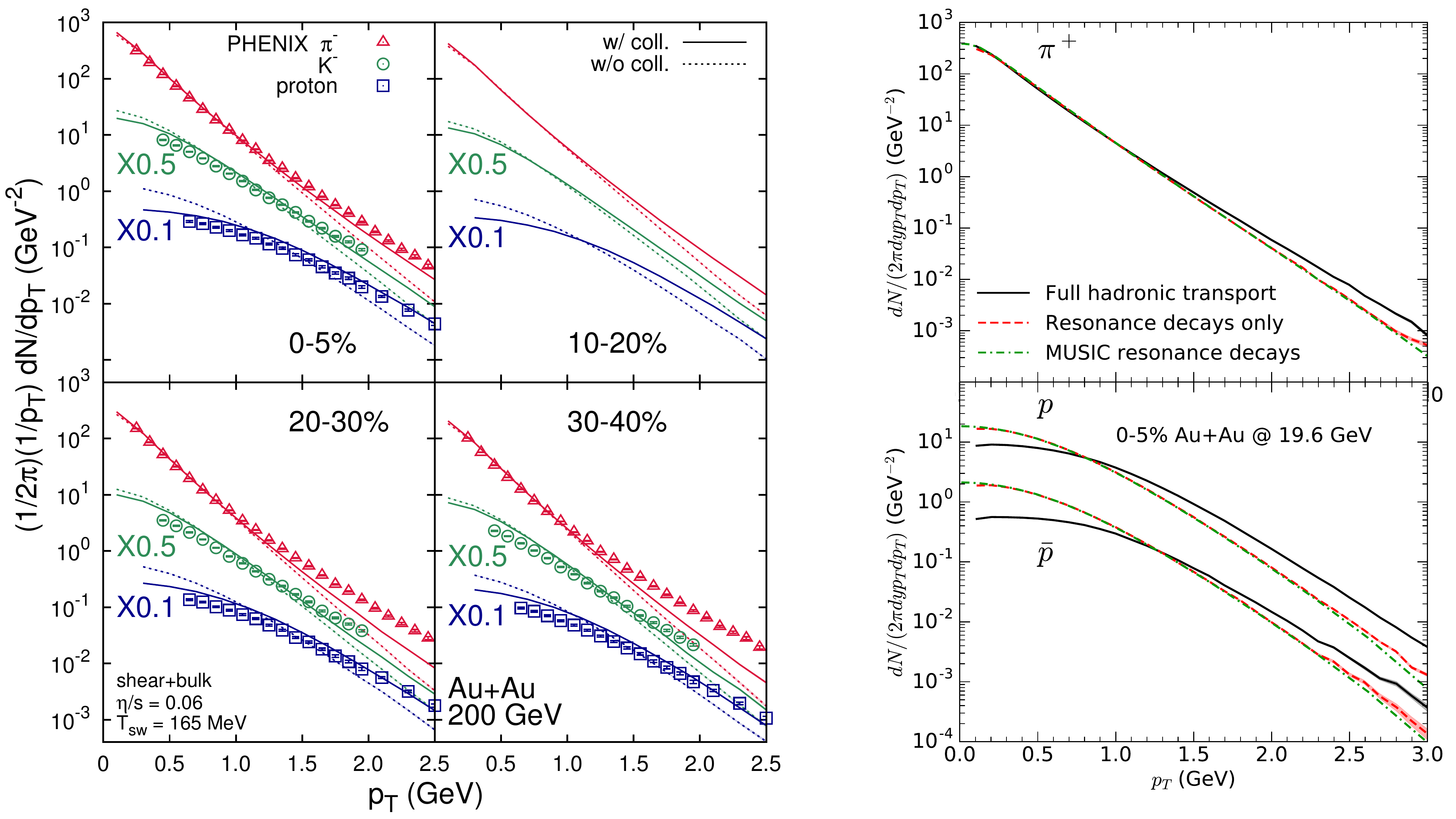}
    \caption{\textit{Left:} The $p_T$-differential spectra of pions, kaons, and protons for centrality classes 0--5\%, 10--20\%, 20--30\%, and 30--40\% in Au+Au collisions at $\snn = 200~\rm{GeV}$. Results from simulations including full \texttt{UrQMD} (labeled ``w/~coll.'') and \texttt{UrQMD} without collisions (labeled ``w/o~coll.'') are shown by the solid and the dashed curves, respectively, with the PHENIX data~\cite{PHENIX:2003iij} also shown for comparison. Figure from~\cite{Ryu:2017qzn}. \textit{Right:} The $p_T$-differential spectra of pions, protons, and antiprotons for centrality class 0--5\% in Au+Au collisions at $\snn = 19.6~\rm{GeV}$. Results from simulations in \texttt{MUSIC} followed by resonance decays, \texttt{MUSIC} followed by hadronic transport with resonance decays only, and \texttt{MUSIC} followed by full hadronic transport with both decays and scatterings are shown as dash-dotted green, dashed red, and solid black lines, respectively. Figure from~\cite{Denicol:2018wdp}.
	}
    \label{fig:hadronic_afterburner_change_in_spectra}
\end{figure}

\subsubsection{Particle spectra}

The hadronic stage also shifts spectra, where in particular proton spectra are significantly depleted in the low-$p_T$ region owing to baryon-antibaryon~($B\bar{B}$) annihilations and, at the same time, they are shifted toward higher transverse momenta due to the large cross sections between nucleons and highly energetic pions~\cite{Elfner:2022iae} (this effect is sometimes referred to as the ``pion wind''), see Fig.~\ref{fig:hadronic_afterburner_change_in_spectra}. 
Comparisons at the top RHIC energy show that using an afterburner leads to a better description of the experimental data. Similarly, effects due to scatterings in the hadronic stage are found to be crucial at BES energies~\cite{Denicol:2018wdp}, with $B\bar{B}$ annihilations also contributing at a significant level~\cite{Monnai:2019hkn}.

\begin{figure}[t]
    \centering
    \includegraphics[width=0.99\linewidth]{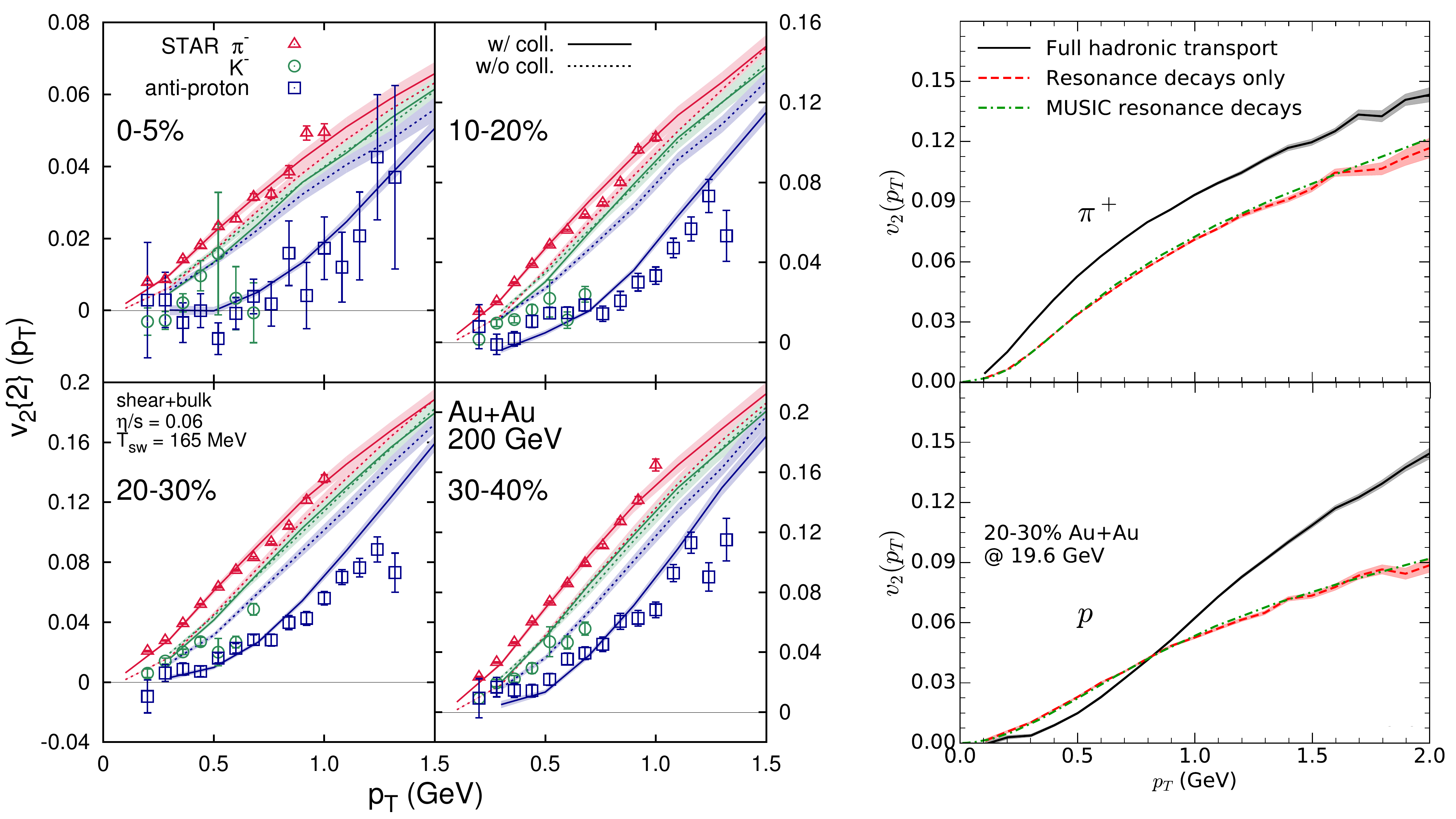}
    \caption{\textit{Left:} The $p_T$-differential elliptic flow $v_2\{2\}$ of identified hadrons for centrality classes 0--5\%, 10--20\%, 20--30\%, and 30--40\% in Au+Au collisions at $\snn = 200~\rm{GeV}$. Results from simulations including full \texttt{UrQMD} (labeled ``w/~coll.'') and \texttt{UrQMD} without collisions (labeled ``w/o~coll.'') are shown by solid and dashed curves, respectively, with the STAR data~\cite{STAR:2004jwm} also shown for comparison. Figure from~\cite{Ryu:2017qzn}. \textit{Right:} The \mbox{$p_T$-differential} elliptic flow~$v_2$ of pions (\textit{top}) and protons (\textit{bottom}) for centrality class 20--30\% in Au+Au collisions at $\snn = 19.6~\rm{GeV}$. Results from simulations in \texttt{MUSIC} followed by resonance decays, \texttt{MUSIC} followed by hadronic transport with resonance decays only, and \texttt{MUSIC} followed by full hadronic transport with both decays and scatterings are shown as dash-dotted green, dashed red, and solid black lines, respectively. Figure from~\cite{Denicol:2018wdp}.
	}
    \label{fig:hadronic_afterburner_change_in_flow}
\end{figure}

\subsubsection{Collective flow}

At the top RHIC energy, the afterburner stage slightly increases the \mbox{$p_T$-differential} elliptic flow~$v_2(p_T)$ of light mesons while it significantly decreases $v_2(p_T)$ of protons~\cite{Ryu:2017qzn}, which leads to an overall better description of experimental data, see the left panel in Fig.~\ref{fig:hadronic_afterburner_change_in_flow} (note that at ultrarelativistic energies, proton and antiproton flow are nearly equal). 
In contrast, at $\snn = 19.6~\rm{GeV}$, the afterburner stage increases the $p_T$-differential elliptic flow $v_2(p_T)$ of pions significantly; moreover, in the case of $v_2(p_T)$ of protons and antiprotons, the afterburner significantly enhances the medium-to-high-$p_T$ region, while it reduces the flow in the low-$p_T$ region~\cite{Denicol:2018wdp}, see the right panel in Fig.~\ref{fig:hadronic_afterburner_change_in_flow}. 
These significant increases in flow may be explained by the additional time that the afterburner stage gives the system to convert the remaining spatial eccentricity (which at the BES energies, in contrast to the top RHIC energy, remains substantial at particlization) into the particle momentum anisotropy. The depletion of the proton and antiproton $v_2(p_T)$ at low values of $p_T$ can be ascribed to the corresponding modification of the proton spectra~\cite{Denicol:2018wdp} (see the right panel in Fig.~\ref{fig:hadronic_afterburner_change_in_spectra}).

\subsubsection{Future directions}

It is important to note that in its use as an afterburner, microscopic transport is usually employed without any mean-field effects on particles' trajectories (in this mode, it is often referred to as the ``hadronic cascade''). 
This is typically motivated by the corresponding significant reduction in the computational cost as well as by the fact that mean-field effects tend to be negligible at high collision energies (note, however, that this may be a feature of the design: since mean fields used in microscopic transport are most often parametrized in terms of baryon density $n_B$, it is not surprising that they do not exert any appreciable effects in systems characterized by $n_B \approx 0$). 
As the collision energy is reduced and both the length of the afterburner stage 
as well as densities probed during that stage 
increase, however, mean-field effects are bound to become more important. 
Consequently, studies of the influence of the hadronic stage on final state observables which include effects due to mean-field particle interactions are vital for BES physics.
Such studies should also include investigations of the mean-field influence on correlation and fluctuation observables, which carry a considerable significance in the BES program (see Sec.~\ref{sec:fluctuations}) and which have been previously shown to be strongly affected by processes typical to the afterburner stage such as resonance decays and baryon annihilation~\cite{Werner:2010aa,Steinheimer:2012rd,Hammelmann:2022yso,Hammelmann:2023aza}. 
Initial studies along this direction have been recently presented~\cite{Savchuk:2022msa}.

%% file: microscopic_transport.tex

Using heavy-ion collisions to constrain the Quantum Chromodynamics~(QCD) equation of state~(EOS) at baryon densities~$n_B$ larger than the saturation density of nuclear matter~$n_0$ is a long-standing objective in nuclear physics. 
Initially motivated by the goal of describing nuclear matter at densities reached in neutron stars~\cite{Baym:2001in,Bear_Mountain_workshop_report}, studies were further fueled by the discovery of asymptotic freedom~\cite{Gross:1973id,Politzer:1973fx} which led to the postulation of a hypothesized state of matter in which hadrons are packed so tightly that quarks and gluons are practically free~\cite{Collins:1974ky,Chapline:1976gq}. 
Following works suggesting~\cite{Chapline:1973kkq} that highly energetic heavy-ion collisions may probe regions allowing for the study of the Hagedorn ``limiting temperature''~\cite{Hagedorn:1965st} (which was also interpreted as a sign of a second-order phase transition~\cite{Cabibbo:1975ig}) and calculations considering the production of ``quark matter''~\cite{Chapline:1978kg,Chin:1978gj}, ``hadronic plasma''~\cite{Shuryak:1977ut}, or ``quark-gluon plasma''~\cite{Shuryak:1978ij} in such collisions, experiments colliding heavy-ions at relativistic velocities became a means to probe the phase diagram of QCD.

At this time, studies of the dense nuclear matter EOS are pursued within several approaches. 
Microscopic many-body calculations within the chiral effective field theory~\cite{Epelbaum:2008ga,Machleidt:2011zz} allow one to obtain state-of-the-art predictions on the nuclear matter EOS with quantified uncertainties~\cite{Drischler:2021kxf,Sorensen:2023zkk,MUSES:2023hyz}. 
However, these calculations are only applicable at relatively low temperatures and moderate densities, and in particular are characterized by large uncertainties for densities $n_B \gtrsim 1.5 n_0$. 
Astrophysical observations of neutron stars can provide constraints on the EOS of neutron-rich matter at zero temperature up to densities several times that of normal nuclear matter~\cite{Ozel:2016oaf}, with a particular sensitivity for densities above $2n_0$~\cite{Legred:2021hdx,Miller:2019nzo}, while additionally neutron star mergers are expected to probe temperatures on the order of tens of MeV~\cite{Radice:2020ddv}.

In comparison, heavy-ion collisions at the nucleon-nucleon center-of-mass beam energies from about $\snn \approx 1.9~\rm{GeV}$ (equivalent to incident kinetic energies, excluding the rest mass, of $\ekin \approx 50~\rm{MeV/nucleon}$) up to $\snn \approx 7.7~\rm{GeV}$ (equivalent to $\ekin \approx 30~\rm{GeV/nucleon}$) probe nearly isospin-symmetric nuclear matter at densities from a few tenths to about 5 times $n_0$~\cite{Sorensen:2023zkk}. 
While, at these energies, a large fraction of the collision evolution is out-of-equilibrium, preventing one from using thermodynamic quantities in a well-defined way, one can connect the explored energy densities to temperatures ranging from a few to a few hundreds of MeV~\cite{Sorensen:2023zkk}. 
As a result, collisions of heavy nuclei at these energies probe nuclear matter over an impressive range of densities and temperatures, both overlapping with and complementary to other approaches, and, moreover, present the only means of probing the dense nuclear matter EOS in controlled laboratory experiments. 
The average density and temperature (the latter to the extent that it can be well-defined) of the collision systems are varied by changing the beam energy and collision geometry\footnote{Note, however, that they cannot be varied independently, as compression is always accompanied by heating~\cite{Arsene:2006vf,Oliinychenko:2022uvy}.}. 
Moreover, certain control of the isospin asymmetry is also possible by varying the isotopic composition of the target and projectile~\cite{FRIB400,Sorensen:2023zkk}.

A list of selected recent, current, and upcoming efforts using heavy-ion collisions to probe high-density nuclear matter is given in Table~\ref{tab:experiments}. Among currently run or analyzed experiments, the Solenoidal Tracker at RHIC (STAR) experiment's Fixed-Target~(FXT) campaign of the Beam Energy Scan (BES) II program at the Relativistic Heavy Ion Collider (RHIC) performs gold-gold (Au+Au) collisions in the energy range $\snn \in [3.0$--$13.1]~\rm{GeV}$ ($\ekin \in [2.92$--$89.6]~\rm{GeV/nucleon}$). 
The High Acceptance Di-Electron Spectrometer (HADES) experiment at GSI, Germany, collides gold nuclei at $\snn = 2.42~\rm{GeV}$ ($\ekin = 1.24~\rm{GeV/nucleon}$) and silver nuclei at $\snn = 2.55~\rm{GeV}$ ($\ekin = 1.58~\rm{GeV/nucleon}$), while an upcoming campaign at HADES will collide gold nuclei at $\snn \in \{1.97, 2.07, 2.16, 2.24\}~\rm{GeV}$ ($\ekin \in \{0.2, 0.4, 0.6, 0.8\}~\rm{GeV/nucleon}$) and carbon nuclei at $\snn \in \{2.16, 2.24\}$ GeV ($\ekin \in \{0.6, 0.8\}~\rm{GeV/nucleon}$).
The future Compressed Baryonic Mater (CBM) experiment at the Facility for Ion and Antiproton Research (FAIR), Germany, will explore collisions of a variety of stable nuclei at $\snn \in [2.7$--$4.9]~\rm{GeV}$ ($\ekin \in [ 2.0, 10.9]~\rm{GeV/nucleon}$). 

\begin{table}[t]
    \centering
    \begin{tabular}{c|c|c|c}
    experiment & $E_{\rm{kin}}$ [GeV/$u$] & $\sqrt{s_{\rm{NN}}}$ [GeV] &  ion species \\
    \hline
    \hline
    STAR FXT & 2.92 -- 98.1   & 3.0 -- 13.7   &  Au   \\
    \hline
    HADES & 1.23, 1.58   & 2.42, 2.55   &  Au, Ag    \\
	\hline
    HADES & 0.2 --  0.8   & 1.97 -- 2.24   & Au   \\
    (2024) &  0.6, 0.8  & 2.16, 2.24  & C   \\	
    \hline
    CBM    & 2.0 -- 10.9   & 2.7 -- 4.9   &  Au  \\
	(2028)    &   &   &  other stable  \\	
    \hline
    S$\pi$RIT    & 0.27  & 2.01  &  $^{108,132}$Sn+$^{112,124}$Sn  \\
	\hline
    EOS@FRIB    & 0.175  & 1.96  &  $^{56,70}$Ni+$^{58,64}$Ni   \\
    (2025)   & 0.2  & 1.97  &  $^{108,132}$Sn+$^{112,124}$Sn   \\
    \hline
    EOS@FRIB400    & 0.4  & 2.07  &  $^{104,134}$Sn+$^{112,124}$Sn  \\
	(2030?)& & & $^{36,52}$Ca+$^{40,48}$Ca  \\
    & & & other unstable+stable
    \end{tabular}
    \caption{Comparison of different experimental conditions in selected recent, current, and upcoming heavy-ion collision experiments; dates in the brackets indicate the expected timeframe for the upcoming efforts.}
    \label{tab:experiments}
\end{table}

Colliding nuclei of varied isotopic composition is a particular strength of facilities where neutron- or proton-rich radioactive beams can be collided on different stable targets, leading to a range of probed isospin densities. Within the recent efforts, the SAMURAI Pion Reconstruction and Ion-Tracker (S$\pi$RIT) experiment at the Radioactive Isotope Beam Factory (RIBF) at RIKEN, Japan, performed collisions of tin isotopes at $\snn = 2.01~\rm{GeV}$ ($\ekin = 0.27~\rm{GeV/nucleon}$). In the near future, the recently commissioned Facility for Rare Isotope Beams (FRIB) will provide unique capabilities, colliding the widest range of isotopes available worldwide at up to $\snn = 2.01~\rm{GeV}$ ($\ekin = 0.27~\rm{GeV/nucleon}$), which is expected to yield tight constraints on the isospin-dependence of the EOS up to $n_B \approx 1.5n_0$. 
The proposed FRIB400 upgrade to $\snn = 2.07~\rm{GeV}$ ($\ekin = 0.4~\rm{GeV/nucleon}$) will further expand this range up to $n_B \approx 2n_0$~\cite{FRIB400}.

The EOS can be studied in heavy-ion collisions by measuring the final state products of the experiments and constructing observables sensitive to bulk properties of nuclear matter. 
Any quantitative, and often also qualitative, interpretation of these observables can only be obtained by comparing experimental measurements with results of dynamic simulations. 
Conversely, simulations are also used to identify promising observables. 
Since collisions in the above-listed energy ranges not only proceed out-of-equilibrium for considerable fractions of the collision time, but are also characterized by prolonged initial compression and final rescattering stages, the appropriate framework for describing the evolution can be provided by microscopic transport simulations. 
These approaches, which in contrast to hydrodynamics do not require an assumption of local equilibrium, can be used to describe the entire span of a heavy-ion collision: from the initial state of the colliding nuclei, through initial compression and growth of the compression zone, to the development of collective behaviors, decompression
of the fireball, and particle scatterings and decays in the final dilute stage of the process.

At the same time, the inherent complexity of heavy-ion collisions in this energy range makes one-to-one comparisons between theory and experiment challenging, especially when attempting to extract an equilibrium property of dense nuclear matter such as the EOS at different densities, temperatures, and isospin asymmetries. 
To reliably constrain the dense nuclear matter EOS, multiple complex aspects of the simulations have to be well controlled and understood. 
The effort required in addressing this challenge is well motivated by the fact that heavy-ion collisions have the potential to provide information on the EOS in a region of the QCD phase diagram which cannot be accessed by any other means.

The structure of this section is as follows: 
In Sec.~\ref{sec:selected_observables}, we give an overview of selected observables identified as promising for constraining the EOS of symmetric nuclear matter. 
In Sec.~\ref{sec:microscopic_transport_models}, we briefly review the theoretical foundations of microscopic transport approaches. 
In Sec.~\ref{sec:recent_developments_relevant_to_EOS}, we highlight some of the recent developments in modeling low- and intermediate-energy heavy-ion collisions with microscopic transport.
Finally, in Sec.~\ref{sec:constraints_on_the_EOS} we discuss selected recent efforts to constrain the dense nuclear matter EOS using comparisons between transport models and experimental data.


\subsection{Selected prominent observables for extracting the equation of state}
\label{sec:selected_observables}

Relativistic heavy-ion collisions proceed through several stages which include the initial impact, development of the compressed region and heating, decompression and cooling, and late-stage rescattering. 
The influence of the EOS is prominent in nearly all elements of the evolution. 
The initial state of the collisions is affected by the EOS through its influence on the structure of the colliding nuclei~\cite{Cao:2010bc,Bally:2022vgo}
. 
Furthermore, whether the EOS is stiff or soft (describing matter that is relatively more or less incompressible, respectively) impacts not only the degree of the compression and the corresponding heating~\cite{Arsene:2006vf,Oliinychenko:2022uvy}, but also the duration of the compression and decompression stages. 
Likewise, the EOS affects the yields of particles produced throughout the evolution, in particular influencing scatterings and decays in the late dilute stages of the collision
.

Through theoretical considerations and simulations, a number of observables have been identified as sensitive to the EOS, including collective flow, subthreshold meson production, dilepton production, femtoscopic correlations, and light cluster production. 
Among these, certain observables are more promising for collisions at specific beam energies; for example, subthreshold kaon production is an effective probe of the EOS in collisions at energies where subthreshold production dominates.
Below, we review some of these prominent observables used to constrain the EOS.

\subsubsection{Remarks about sensitivity}
\label{sec:remarks_about_sensitivity}

While the sensitivity of a given measurement to the EOS can often be demonstrated relatively easily, the fact that heavy-ion observables are constructed out of the final-state particle distributions means that they are often \textit{also} sensitive to multiple additional elements influencing the dynamics of the collisions. 
Notable examples of such elements include elastic and inelastic scattering cross sections or the initial structure of the colliding nuclei. 
As a result, quantitative constraints on the EOS depend not only on the particular way in which the EOS is modeled, but also, often substantially, on other features of the collision dynamics.

This has profound consequences for efforts to extract the EOS from comparisons of experimental data to simulation results. 
Simulations of heavy-ion collisions, especially at lower energies where the system is out of equilibrium for the majority (if not the entirety) of the evolution, must include many complex aspects of the collisions which interplay with each throughout the evolution. 
For many of these aspects, there is either no clear theoretical guidance as to how they should be modeled, or the theoretical guidance cannot be straightforwardly implemented in simulations, creating the need to rely substantially on phenomenology. 
Since there are many choices available when choosing how to model a given aspect of the evolution, consequently there are differences in both phenomenological approaches and numerical solutions employed, which can lead to substantial differences in obtained results. 

Because of the above, extracting the EOS from heavy-ion collisions ultimately  requires assigning systematic errors connected with both the way in which the EOS is modeled and modeling of other aspects of the collision. 
In this endeavor, Bayesian analyses of simulation results and experimental data (allowing one to understand the relative importance of different aspects of modeling for final results) as well as controlled comparisons between different transport codes will play a pivotal role. 
Both of these research directions have been gaining prominence in recent years. 
Bayesian analyses have already led to a deeper understanding of the interplay between different aspects of the collision dynamics and constraints (with meaningful uncertainties) on, among others, the EOS~\cite{Pratt:2015zsa,Sangaline:2015isa,Oliinychenko:2022uvy,OmanaKuttan:2022aml}. 
Controlled comparisons between different simulation frameworks are being done within the Transport Model Evaluation Project (TMEP) collaboration, which has extensively studied the influence of both phenomenological assumptions and numerical realizations on numerous components of the simulated collision dynamics and, consequently, observables~\cite{TMEP:2022xjg}. 
Notably, the TMEP effort has been primarily addressing aspects of simulations which are of importance for studies of the isospin dependence of the EOS, performed at relatively low collisions energies ($\snn \lesssim 3~\rm{GeV}$, or $\ekin \lesssim 2.9~\rm{GeV/nucleon}$). 
A similar effort exploring the Beam Energy Scan regime would be invaluable to explorations of the QCD EOS and phase diagram by providing key insights into the interpretation of currently available and upcoming measurements.

\subsubsection{Collective flow}

Through its importance for the development and evolution of the compression zone, the EOS influences the transverse expansion of the system. 
The latter, in turn, is reflected in the final state particle distributions, where in particular angular particle distributions in the transverse plane $dN_i/d\phi$, where the subscript $i$ denotes particle species, are often considered. 
Given the invariant particle distribution $\infrac{d^3 N_i}{ (p_T dp_T~ dy~d\phi)}$, where $p_T$ is the transverse momentum, $y$ is the rapidity, and $\phi$ is the azimuthal angle, it is particularly convenient to decompose $dN_i/d\phi$ in terms of its Fourier coefficients $v_n$,
\begin{align}
\frac{d^3 N_i}{ p_T dp_T~ dy~d\phi} = \frac{d^2 N_i}{ p_T dp_T~ dy} ~ \frac{1}{2\pi} \left[ 1 + \sum_{n=1}^{\infty} 2v^{(i)}_n(p_T,y) ~ \cos(n\phi)  \right] ~,
\end{align}
from which it follows that
\begin{align}
v_n^{(i)} (p_T, y) &\equiv \frac{  \int_0^{2\pi} d\phi ~ \cos(n\phi) ~   \frac{d^3 N_i}{ p_T dp_T~ dy~d\phi}    }{\int_0^{2\pi} d\phi ~  \frac{d^3 N_i}{ p_T dp_T~ dy~d\phi}  } \\
& = \big\langle \cos(n \phi) \big \rangle = \frac{1}{N_i} \sum_{k=1}^{N_i} \cos(n\phi_k) \Big|_{p_T, y}~,
\end{align}
where the sum in the last expression is performed over all particles of species~$i$ which have transverse momentum $p_T$ and rapidity $y$. 
Since the Fourier coefficients $v_n^{(i)}(p_T, y) $ reflect the direction of particles' momenta averaged over the entire system, they are referred to as ``collective flow''. 
To tease out broad characteristics of this collective motion, flow observables $v_n^{(i)}(p_T, y) $ are often further integrated over $p_T$, $y$, or both.

The influence of the EOS on the collective flow is especially prominent at collision energies for which two time scales associated with the collisions are comparable. 
The first of those is the time necessary for the compression and decompression of matter originating from participant nucleons (that is nucleons trapped in the collision region), while the second is the time of propagation of spectator nucleons (that is nucleons whose trajectories do not directly cross the initial transverse overlap of the colliding nuclei) past the collision region. 
The two time scales are similar for collision energies ranging from around $\snn \approx 2~\rm{GeV}$ ($\ekin \approx 0.2~\rm{GeV/nucleon} $) to $\snn \approx 10~\rm{GeV}$ ($\ekin \approx  50~\rm{GeV/nucleon}$), and consequently in this regime the development of flows $v_n^{(i)}$ is heavily affected by the interplay between the initial geometry of the colliding system, the gradual creation of the compression zone from incoming participants and its subsequent expansion, and the movement of spectator nucleons. 
It has been shown in numerous hydrodynamic ~\cite{Stoecker:1980vf,Ollitrault:1992bk,Rischke:1995pe,Stoecker:2004qu,Brachmann:1999xt,Csernai:1999nf,Ivanov:2014ioa} and microscopic transport~\cite{Hartnack:1994ce,Li:1998ze,Danielewicz:2002pu,LeFevre:2015paj,Wang:2018hsw,Nara:2021fuu,Oliinychenko:2022uvy,Steinheimer:2022gqb} models that the resulting collective motion of the system, which can be measured to a very high precision, is very sensitive to the EOS, placing flow among the central observables used for constraining the EOS.

\begin{figure}[t]
    \centering
	\includegraphics[width=0.99\linewidth]{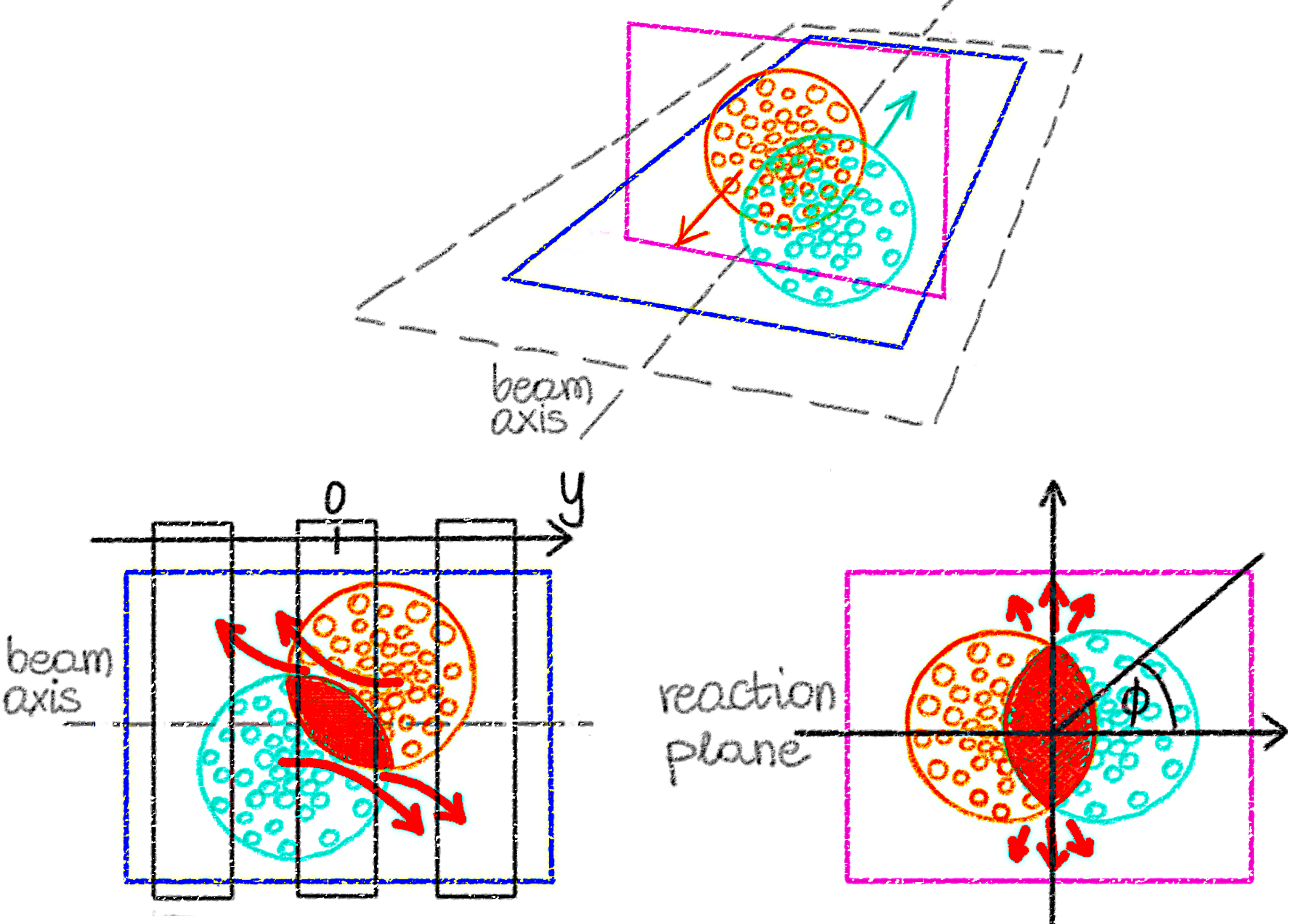}
    \caption{A sketch picturing the geometry of a heavy-ion collision system at the moment of impact (\textit{top}) as well as the development of the directed (\textit{bottom left}) and elliptic (\textit{ bottom right}) flow. The solid red regions indicate the compressed matter created in the aftermath of the impact, while the red arrows indicate the motion of mater in response to interaction with the dense collision zone (\textit{middle}) and during its subsequent decompression (\textit{right}). The presented picture is valid in collisions at energies for which the time scale associated with the compression and decompression of the collision zone and the time scale of propagation of spectator nucleons past the collision region are comparable. Note that two nuclei moving at relativistic speeds should be Lorentz contracted, however, the sketch omits the contraction for clarity.
	}
    \label{fig:flow_sketch}
\end{figure}

An asymmetry in the compression and expansion of the collision zone is necessary for the development of finite $v_n$. 
A substantial asymmetry is obtained in off-central collisions where the initial overlap of the colliding nuclei has an approximately elliptical shape (see Fig.~\ref{fig:flow_sketch}), which leads to pressure gradients within the collision zone (that are larger along the short axis of the ellipsoid) 
and supports the development of a non-zero $v_1(p_T, y)$, referred to as the directed flow, and $v_2(p_T, y)$, known as the elliptic flow. 
The particular values of such generated flow coefficients depend on the aforementioned intrinsic time scales of the collision and the EOS. 
In the case in which the spectator nucleons leave the collision region before the collision zone starts to expand, nothing obstructs the expansion of the created fireball to follow the directions of the largest pressure gradients, which results in a preferential emission in directions parallel to the reaction plane (``in-plane'' emission), directly corresponding to the shape of the collision region which can be quantified, e.g., by its eccentricity~\cite{Sorensen:2009cz}.
This, in turn, leads to positive values of $v_2 (p_T, y)$, while the values of $v_1 (p_T, y)$ in this case are zero (up to event-by-event contributions from random fluctuations) due to the symmetry. 
On the other hand, if the spectators remain in the close vicinity of the collision region, they may block the in-plane motion of the expanding fireball and thus lead to negative values of $v_2(p_T,y)$ (``out-of-plane'' emission). 
The degree of this blocking depends strongly on the EOS, with more blocking (and a more negative $v_2$) occurring for a stiffer (more repulsive) EOS.
Moreover, while the directed flow at midrapidity, $v_1(p_T, y=0)$, remains equal to zero (up to random event-by-event fluctuations) also in this case due to symmetry, it acquires non-zero values at finite values of rapidity due to the interaction between the spectators and the collision zone~\cite{Voloshin:2008dg}. 
Here, if the spectators whose trajectories coincide with the positive beam direction are deflected in such a way that the values of $v_1(p_T, y >0)$ are positive, then conversely the spectators going in the negative beam direction will be deflected in the opposite direction, thus leading to negative values of $v_1(p_T, y <0)$. 
(Note that because of the symmetry, one should have $v_1(p_T, y = y') = - v_1(p_T, y = -y')$, again up to random fluctuations, as well as $\int dy ~v_1(p_T, y)$ = 0.)
The value of the directed flow at a given value of $y$ as well as its slope at midrapidity, $\infrac{dv_1}{dy}\big|_{y=0}$, depend on the EOS, with stiffer (softer) EOSs leading to larger (smaller) values of $\infrac{dv_1}{dy}\big|_{y=0}$. 
Notably, several studies~\cite{Stoecker:2004qu,Zhang:2018wlk} even suggest $\infrac{dv_1}{dy}\big|_{y=0} < 0$ for EOSs exhibiting a first-order phase transition, for which the interactions between the collision zone and the spectators may become attractive instead of repulsive.

While the sensitivity of the directed and elliptic flow to the EOS has been extensively explored and confirmed in multiple studies considering various possible behaviors of the EOS with baryon density~\cite{Oliinychenko:2022uvy,Steinheimer:2022gqb}, it is also well-known that the extraction of the EOS from flow observables is heavily affected by the momentum dependence of interactions~\cite{Aichelin:1987ti,Pan:1992ef,Danielewicz:1999zn,Danielewicz:2002pu,Nara:2021fuu}. 
Therefore, studies using flow observables to extract the EOS should also address the influence of the momentum dependence, which can be independently constrained using data from peripheral collisions~\cite{Pan:1992ef,Danielewicz:2002pu} where the effects due to the unknown density-dependence of the EOS diminish due to lower probed densities. 
Similarly, flow results are also affected by the modification of elementary cross-sections in the medium, which can be calculated in models~\cite{Persram:2001dg,Muther:2000qx,Pandharipande:1992zz,Li:2005jy,Chen:2013bwa,Sammarruca:2005tk} and the necessity of which can be inferred from measured stopping observables that tend to be overpredicted in models using free-space cross sections~\cite{Westfall:1993zz,Persram:2001dg,Zhang:2007gd,Li:2022wvu}.
Still, the in-medium dependence of cross-sections is not well constrained, especially at energies relevant to the BES program. 
Flow results have also been shown in the past to be influenced by the degrees of freedom considered in simulations, in particular by inclusion of high-mass resonant states. 
These states are produced in significant amounts at collision energies $\snn \gtrsim 2.3~\rm{GeV}$ ($\ekin \gtrsim 1~\rm{GeV/nucleon}$) and, by providing additional production channels which can use some part of the collision energy, affect the degree of initial compression and the spectrum of emitted and absorbed mesons, with non-negligible implications for the collective flow~\cite{Hombach:1998wr}.
Consequently, simulations aimed at inferring the EOS from flow observables should include all relevant resonant species as well as provide a reliable description of meson production.
Finally, flow results can be also affected by production algorithms for deuteron, triton, and other light clusters included in simulations~\cite{Mohs:2020awg}. 
While several modeling choices are available~\cite{Danielewicz:1991dh,Li:2015pta,Ikeno:2016xpr,Oliinychenko:2018ugs,Ono:2019jxm,Staudenmaier:2021lrg}, they still display certain shortcomings: models based on coalescence typically do not take into account the influence of light clusters on the evolution, while approaches based on dynamical production of clusters through the collision term in transport equations are impeded by increasing complexity when attempting to include heavier clusters which are characterized by more and more production channels.
Moreover, cluster production is ultimately related to intermediate-range correlations and quantum effects, and as such touches on fundamental properties of nuclear matter reaching beyond the classical one-body description typical for microscopic transport.
Nevertheless, identifying successful phenomenological approaches to cluster production can help shed light on the underlying fundamental mechanisms.

\subsubsection{Light cluster production}

Recently, light nuclei production itself has been suggested as a probe of the EOS and, in particular, the existence and location of the QCD first-order phase transition region with the corresponding critical point~\cite{Sun:2017xrx,Sun:2018jhg}.
The key observation made was that fluctuations in the phase-space density, which are expected to be significantly enhanced both inside the spinodal region of a first-order phase transition and in the vicinity of a critical point and (see Sec.~\ref{sec:fluctuations}), may lead to an enhanced production of light nuclear clusters. 
Specifically, based on the coalescence model, it was observed\cite{Sun:2017xrx} that the yield ratio $N_t N_p / N_d^2$, where $N_t$, $N_p$, and $N_d$ are triton, proton, and deuteron yields, respectively, is sensitive to relative fluctuations in neutron density $\Delta n_n \equiv \infrac{\big\langle (\delta n_n)^2 \big\rangle }{ \langle n_n \rangle ^2}$, 
\begin{align}
\frac{N_t N_p}{N_d^2} &= \frac{1}{2\sqrt{3}} \frac{1 +  2 C_{np}  + \Delta n_n }{ (1 + C_{np} )^2}~,
\end{align}
where $C_{np} \equiv \infrac{\langle \delta n_n \delta n_p \rangle}{ \big(\langle n_n \rangle \langle n_p \rangle}\big)$ denotes proton-neutron density correlation.
Moreover, combined heavy-ion collision data spanning the energy range $\snn \approx 3$--$ 200~\rm{GeV}$ ($\ekin \approx 3$--$21,300~\rm{GeV/nucleon}$)~\cite{NA49:2010lhg,Blume:2007kw,NA49:2016qvu,STAR:2022hbp,STAR:2023uxk} shows a nonmonotonic behavior in the dependence of the $\infrac{N_t N_p}{N_d^2} $ ratio on collision energy, which suggests enhanced spatial density fluctuations in the region of the phase diagram probed by the corresponding beam energies. 
However, firm conclusions are prevented by the fact that the available models (both dynamical and thermal) disagree on the $\infrac{N_t N_p}{N_d^2} $ ratio even when using the same EOS \textit{without} a critical point, thus disagreeing about the trivial baseline for the measurement. 
Thus, using this observable to constrain the EOS requires further studies on cluster production, with a particular emphasis on dynamical studies at intermediate and low collision energies where the intriguing behavior of the light nuclei yield ratio occurs.
Some work in this direction has already been done~\cite{Sun:2020uoj,Hillmann:2021zgj,Sun:2022cxp}.

\subsubsection{Femtoscopic correlations}

Two-particle correlations at small relative momentum, also known as femtoscopic or Hanbury-Brown--Twiss (HBT) correlations, have been proposed as a sensitive probe of the EOS. 
Given that correlations due to final-state interactions are stronger between particles that are emitted close to each other in space or time, HBT correlations provide a tool to investigate the spatial and temporal properties of particle emission as well as the dynamics of the emitting source. 
In particular, information about not only the spatial, but also temporal extent of the source can be extracted from measurements of femtoscopic radii $R_{\rm{long}}$, $R_{\rm{out}}$, $R_{\rm{side}}$~\cite{Heinz:1999rw}.
Here, $R_{\rm{long}}$ is a measure of the source size along the longitudinal (beam) axis, $R_{\rm{out}}$ measures the source size in the direction of the average transverse momentum of the correlated pair, while $R_{\rm{side}}$ measures the source size in the direction perpendicular to both $R_{\rm{long}}$ and $R_{\rm{out}}$.
Since the femtoscopic radii provide information about the space-time extent of the studied source, they are often also referred to as interferometry.

It was pointed out~\cite{Pratt:1986cc,Bertsch:1988db,Rischke:1996em} that if a heavy-ion collision evolves through a first-order phase transition, which both allows the system to compress more and prolongs the expansion of the fireball due to the softening of the EOS in the transition region, then this will increase the size of the emission region in the radial direction.
In that case, $R_{\rm{out}}$ is comparatively larger, and in particular the femtoscopic radii will satisfy $R_{\rm{out}} \gg R_{\rm{side}}$. 
As a consequence, both the difference $R_{\rm{out}}^2 - R_{\rm{side}}^2$ and the ratio $R_{\rm{out}} / R_{\rm{side}}$ can be used to constrain the EOS. 
(We note here that while the difference of the femtoscopic radii can be directly linked to the emission duration $\Delta \tau$ through $R_{\rm{out}}^2 - R_{\rm{side}}^2 = \beta_T^2 \Delta \tau$~\cite{Bertsch:1989vn,Pratt:1990zq,Heinz:1996qu}, where $\beta_T$ is the transverse velocity of the emitted particles, this relationship is only reliable for systems without transverse collective flow; if flow is present, extracting $\Delta \tau$ is impeded by significant model dependence~\cite{Lisa:2005dd,STAR:2014shf,Retiere:2003kf}.)

The sensitivity of femtoscopic correlations to the EOS has been demonstrated in a Bayesian analysis of high-energy heavy-ion collisions, constraining the EOS at near-zero baryon density~\cite{Sangaline:2015isa,Pratt:2015zsa}. 
Recent work~\cite{Li:2022iil} has addressed the possibility of using HBT correlations to constrain the EOS of dense nuclear matter in collision energy range $\snn = 2.4$--$7.7~\rm{GeV}$, explored in the HADES experiment, the FXT campaign of the RHIC BES II program, and the future CBM experiment.

On the experimental side, femtoscopic measurements have been dominated by pion interferometry~\cite{E895:2000bxp,ALICE:2011dyt,STAR:2014shf,HADES:2019lek,STAR:2020dav}. 
However, recently more attention has been given to femtoscopic correlations between nucleons~\cite{Siejka:2019ptk} as well as light nuclei~\cite{Stefaniak:2024eux}.
Preliminary results from the HADES experiment~\cite{Stefaniak:2024eux} show significant differences between proton--proton, proton--deuteron, and proton--\mbox{Helium-3} as well as proton--\mbox{Helium-3} and proton--triton correlations, suggesting changes in strong interaction effects between these systems.
Moreover, the results are shown to be affected by decays of exotic \mbox{Lithium-4} and excited \mbox{Helium-4} states, stressing the need for modeling light nuclei production in this energy region.


\subsection{Microscopic transport models}
\label{sec:microscopic_transport_models}

Microscopic transport models can be divided into models of the Boltzmann-Uehling-Uhlenbeck (BUU) type (also referred to as the Boltzmann-Vlasov type) and models of the Quantum Molecular Dynamics (QMD) type. 
In both cases, with the exception of very limited simple considerations
, the complexity of the underlying non-linear equations means that the models cannot solved analytically.
Instead, insights into the evolution of described systems is gained through dynamical simulations. 
Reviews discussing many of the currently used models, including particular choices of phenomenological approaches and details of implementation as well as their consequences for obtained results, are available~\cite{TMEP:2022xjg,Xu:2019hqg,Bertsch:1988ik,Bass:1998ca}.
Below, we give a brief description of each type of models, largely based on similar summaries~\cite{TMEP:2022xjg,Yao:2023yda}, to facilitate the discussion that follows. 

\subsubsection{Boltzmann-Uehling-Uhlenbeck models}

In BUU approaches, the evolution of a system of particles, involving $N$ particle species indexed by $n$, is described by a set of coupled differential equations for the evolution of the phase space distribution function of each species $f_n(t, \bm{x}, \bm{p})$,
\begin{align}
\left[ \parr{}{t} + \derr{\bm{x}}{t}\parr{}{\bm{x}} + \derr{\bm{p}}{t}\parr{}{\bm{p}}  \right] f_n (t, \bm{x}, \bm{p}) = I_{\rm{coll}} \big[ \{f_n\}\big] ~, 
\label{eq:BUU_eq}
\end{align}
where the terms on the left-hand side are equivalent to a total time derivative of $f_n(t, \bm{x}, \bm{p})$ and describe the drift of particles in the phase space, while $I_{\rm{coll}}\big[ \{f_n\}\big]$ is known as the collision term or the collision integral. 
In the absence of particle-particle collisions as well as particle production (i.e., resonance excitation) and decays, the right-hand side of the above equation is identically equal to zero. This, in fact, follows from the Liouville theorem and can be intuitively understood by realizing that the total time derivative on the left-hand side, $\inderr{f_n(t, \bm{x}, \bm{p})}{t} $, describes changes that $f_n(t, \bm{x}, \bm{p})$ undergoes in time from the perspective of an observer traveling along the system's trajectory, $\big( \bm{x}(t), \bm{p}(t)\big)$. 
When collisions, particle production, and decays, which either add or remove particles from a given infinitesimal volume of the phase space, are absent, the distribution function remains constant in the subvolume moving along the trajectories of particles. 
(We additionally note here that these trajectories are realized both by changes in particles' positions due to propagation as well as by changes in particles' momenta due to forces acting on the particles, and as such may be far from trivial.) 
On the other hand, if collisions, particle production, or decays are allowed to occur, the corresponding changes in the distribution function are reflected in~$I_{\rm{coll}}\big[ \{f_n\}\big]$ which sums over all possible cross sections and production or decay channels between considered particle species. 
Typically, within the derivation of Eq.~\eqref{eq:BUU_eq} from the Bogoliubov-Born-Green-Kirkwood-Yvon (BBGKY) hierarchy~\cite{Kardar_Statistical_physics_of_particles}, the contribution to the collision integral due to two-body scattering with a particle of species~$k$, $p_n + p_k \to p_n' + p_k'$, formally takes the form
\begin{align}
I_{\rm{coll}} \big[ \{f_n\}\big] &= \sum_k g_k \int \frac{d^3p_k}{(2\pi)^3} ~ d\Omega' ~ \nonumber \\
& \hspace{-15mm}\times ~ v_{nk} \frac{d\sigma_{nk}}{d\Omega'} ~ \Big[f_n' f_k' \big(1 - af_n\big) \big(1 - af_k\big) - f_n f_k \big(1 - af_n'\big)\big(1 - af_k'\big)  \Big] ~,
\end{align}
where the sum runs over all considered particle species, $g_k$ is the degeneracy (which takes into account spin and may also include isospin degeneracy), the distribution functions are all taken at the same spacetime point $\big(t, \bm{x})$, the outgoing momenta $p_n'$ and $p_k'$ are determined by the scattering angle $\Omega'$ and energy-momentum conservation, $\infrac{d\sigma_{nk}}{d\Omega'}$ is the differential cross section for scattering, and $v_{nk}$ is the relative velocity,
\begin{align}
v_{nk} = \frac{\sqrt{ (p_n \cdot p_k)^2 - m_n^2 m_k^2 }}{p_n^0 p_k^0} ~,
\label{eq:relative_velocity}
\end{align}
where $m_n$ and $m_k$ are bare masses of particle species $n$ and $k$, respectively. 
Furthermore, the first term in the square bracket is known as the gain term, given that it provides the phase space factor related to particles scattering \textit{into} the state with momentum $p_n$, while the second term is known as the loss term since it describes particles scattering \textit{out of} the $p_n$ state; here, effects due to Pauli blocking or Bose enhancement are taken into account by factors of the form $(1 - af)$, where $a = 1$ for fermions and $a = -1$ for bosons.

Numerically, Eq.~\eqref{eq:BUU_eq} is solved based on the method of test particles, which can be formally derived~\cite{Leupold:1999ui} as well as introduced in a more pedestrian way~\cite{Oliinychenko:2022uvy}. 
Within this approach, the continuous distribution function $f_n(t, \bm{x}, \bm{p})$ describing $A$ particles of species $n$ is approximated by a discrete distribution of a large number $N_T A$ of test particles occupying phase space coordinates $(t, \bm{x}_i, \bm{p}_i)$,
\begin{align}
f_n(t, \bm{x}, \bm{p}) \approx \frac{1}{N_T} \sum_{i = 1}^{N_T A} \delta^3 \big( \bm{x} - \bm{x}_i(t) \big) \delta^3 \big( \bm{p} - \bm{p}_i(t) \big) ~,
\label{eq:fn_test_particle_approximation}
\end{align}
where $N_T \gg 1$ is the number of test particles per particle.
Note that the introduction of the normalization factor of $1/N_T$ preserves the interpretation of $f_n$ as the number of ``real'' particles in a given infinitesimal volume of the phase space. 
The phase space evolution of test particles proceeds according to single-particle equations of motion, which are obtained by inserting Eq.~\eqref{eq:fn_test_particle_approximation} into Eq.~\eqref{eq:BUU_eq},
\begin{align}
& \derr{\bm{x}_i}{t} = \frac{\bm{p}_i}{p^0_i} \equiv \derr{\varepsilon_i}{\bm{p}_i} ~,
\label{eq:dxdt} \\
& \derr{\bm{p}_i}{t}  = - \bm{\nabla}U_i \equiv -\derr{\varepsilon_i}{\bm{x}_i} ~,
\label{eq:dpdt}
\end{align}
where $\varepsilon_i$ is the single-particle energy (including a possible mean-field potential term), and by enabling the test particles to undergo allowed scatterings, resonance excitations, and decays using a chosen algorithm for simulating the collision term of Eq.~\eqref{eq:BUU_eq}. 
Such algorithms can be based on, e.g., the geometric criterion for scattering~\cite{Bass:1998ca,Kodama:1983yk} or stochastic scattering rates~\cite{Danielewicz:1991dh,Cassing:2001ds,Xu:2004mz} informed either by experimental cross section measurements or calculated from, e.g., the principle of detailed balance.
To preserve the physical average number of scatterings per particle in a heavily ``oversampled'' system of $N_T A$ test particles, one further scales the cross sections as $\sigma \to \sigma/N_T$, where $\sigma$ is a given physical cross section.

Altogether, this yields the time evolution of a system of test particles which follow the trajectories of ``real'' particles in the phase space.
In other words, using Eq.~\eqref{eq:fn_test_particle_approximation} to approximate the phase space distribution at the initial time $t_0$, $f_n(t_0, \bm{x}, \bm{p})$, and evolving thus obtained system of test particles effectively solves Eq.~\eqref{eq:BUU_eq} (i.e., yields the time evolution of the continuous phase space distribution $f_n(t, \bm{x}, \bm{p})$), since at any time \mbox{$t >t_0$} the evolved test particles can be always used to again approximate $f_n(t, \bm{x}, \bm{p})$ provided that the evolution of each test particle appropriately mimics the evolution of a ``real'' particle under the same conditions.

Unless a very large number of test particles per particle, on the order of $N_T = \mathcal{O}(1000)$, is used, the approximation of the phase space distribution function given in Eq.~\eqref{eq:fn_test_particle_approximation} fluctuates strongly due to effects related to the finite number of samples (often also referred to as the numerical noise). 
This is of great importance to the success of the method, as the accuracy of calculating local values of $f_n$ is crucial for several key elements of the simulation, including the evaluation of single-particle mean-field potentials $U_i$ entering the equations of motion, Eq.~\eqref{eq:dpdt}, and the corresponding mean-field dynamics~\cite{TMEP:2021ljz} or calculation of Pauli blocking factors~\cite{TMEP:2017mex} which inform the probability of a particle to scatter or decay into a given final state.
On the other hand, using very large numbers of test particles per particle $N_T$ comes at a significant numerical cost which is especially prominent for algorithms performing scatterings, which scale with the square of the total number of particles in an event, $\propto N_T^2$. 
For this reason, generalizations of Eq.~\eqref{eq:fn_test_particle_approximation} are often used where the Dirac delta functions are substituted by profile functions (also known as smearing functions) in coordinate and momentum space,
\begin{align}
f_n(t, \bm{x}, \bm{p}) \approx \frac{1}{ N_T} \sum_{i = 1}^N G_{\bm{x}}\Big( \bm{x} - \bm{x}_i(t) \Big) G_{\bm{p}}\Big( \bm{p} - \bm{p}_i(t) \Big) ~,
\label{eq:fn_test_particle_approximation_smearing}
\end{align}
which ``distribute'' the contribution of a test particle to the phase space density over a finite volume and as a result reduce local fluctuations due to a finite number of samples, thus allowing for successfully employing smaller values of $N_T$.
Most often, non-Dirac-delta profile functions are used only for distributing the particles over the coordinate space, usually over volumes on the order of $\Delta V = \mathcal{O}(1~\rm{fm}^3)$.
Popular choices for the shapes of the smearing profiles include triangular functions and Gaussians, where the latter have been also used to achieve a fully covariant description~\cite{Oliinychenko:2015lva}.
Note that because Eq.~\eqref{eq:fn_test_particle_approximation} was used to derive the single-particle equations of motion, Eq.~\eqref{eq:fn_test_particle_approximation}, from the BUU equation, Eq.~\eqref{eq:BUU_eq}, using a more general form as shown in Eq.~\eqref{eq:fn_test_particle_approximation_smearing} will, in general, lead to different equations of motion for the test particles. 
A systematic way of accounting for the form of the smearing function both in calculating the phase space density and in single-particle equations of motion has been developed and is known as the lattice-Hamiltonian method~\cite{Lenk:1989zz}.
Besides formal consistency, this approach yields equations of motion that strictly preserve conservation of the total energy~\cite{Wang:2019ghr}, which otherwise can be more or less broken~\cite{Sorensen:2020ygf,Sorensen:2021zxd}.
Nevertheless, while multiple codes adopt the lattice-Hamiltonian method, its use is far from widespread~\cite{TMEP:2022xjg}.

We note here that while the BUU equation presented in Eq.~\eqref{eq:BUU_eq} is not written in a manifestly covariant way, it can be shown that it is relativistically covariant. In the collision term, $p_k^0$ in the denominator of $v_{nk}$ can be combined with the momentum integration to obtain an invariant integration measure, while the spin-averaged phase space distribution is a Lorentz scalar. 
Furthermore, both sides of Eq.~\eqref{eq:BUU_eq} can be multiplied by $p^0$, yielding a covariant derivative on the left-hand side. 
Consequently, any equations following from the BUU equation, including the single-particle equations of motion, are also relativistically covariant~\cite{Blaettel:1993uz,Sorensen:2021zxd,Oliinychenko:2022uvy}. Nevertheless, in practice, covariance is always broken by other elements of the simulation. 
A trivial example here is using non-relativistic single-particle potentials, however, covariance is also disturbed simply by solving the Boltzmann equation with a \textit{finite} number of test particles, which leads to, e.g., collisions between particles occurring at a finite distance~\cite{Cheng:2001dz} or calculations of the mean field at a given point based on particles present within a finite volume around that point, both of which imply instantaneous interaction over a finite distance and explicitly break causality. Formally, these effects disappear in the limit of an infinite number of test particles, and a full Lorentz covariance is restored in that limit.

\subsubsection{Quantum Molecular Dynamics models}

In QMD models, the many-body state of the system is described by a product wave function of single-particle states (which may~\cite{Ono:1998yd} or may not~\cite{Aichelin:1991xy} be antisymmetrized), the latter of which are usually taken to assume a Gaussian form. 
Through their use of localized wave packets, QMD models include the description of classical many-body correlations. 
The time evolution is obtained by evolving the centroids of the wave packets, representing the particles in the system, according to classical molecular dynamics, as well as by introducing collisions and decays in a way similar to that employed in BUU approaches.
While QMD models can, in principle, employ two-, three-, and many-body interactions, instead particles are usually evolved with a mean-field potential (similarly to BUU). 
As such, QMD may be interpreted as derived from the time-dependent Hartree method, where the product trial wave function
\begin{align}
\Psi (\bm{x}_1, \dots, \bm{x}_A; t) = \prod\limits_{i=1}^A \phi_i (\bm{x}_i; t) 
\label{eq:product_wave_function}
\end{align}
employs Gaussian single-particle states,
\begin{align}
\hspace{-2mm}\phi_i (\bm{x}_i; t) &= \frac{1}{\big[ 2\pi (\Delta x)^2 \big]^{3/4}} \nonumber\\
& \hspace{5mm} \times~ \exp \Bigg[ - \left( \frac{ \bm{x}_i - \bm{X}_i(t)}{2 \Delta x} \right)^2\Bigg]~ \exp\Big[ i \bm{P}_i(t) \cdot \big(\bm{x}_i - \bm{X}_i(t) \big) \Big] ~,
\label{eq:Gaussian_packet}
\end{align}
and the centroid positions $\bm{X}_i(t)$ and momenta $\bm{P}_i(t)$ are identified as variational parameters. 
With Eq.~\eqref{eq:product_wave_function}, one can then calculate the 1-body Wigner function, 
\begin{align}
f(\bm{x}, \bm{p}) = \sum_{i=1}^A \frac{1}{\big(\Delta x \Delta p\big)^3}  ~ \exp  \Bigg[ - \frac{\big( \bm{x} - \bm{X}_i (t) \big)^2}{ 2 (\Delta x)^2}   - \frac{\big( \bm{p} - \bm{P}_i (t) \big)^2}{ 2 (\Delta p)^2} \Bigg]~,
\end{align}
where the sum is performed over all single-particle contributions.
The QMD ansatz, Eq.~\eqref{eq:product_wave_function}, and the time-dependent variational principle then lead to the following single-particle equations of motion,
\begin{align}
& \derr{\bm{X}_i}{t} = \parr{ \langle H \rangle}{\bm{P}_i} ~ ,
\label{eq:dxdt_QMD} \\
& \derr{\bm{P}_i}{t} = -\parr{ \langle H \rangle}{\bm{X}_i} ~ ,
\label{eq:dpdt_QMD}
\end{align}
where $\langle H \rangle$ is the expectation value of the many-body Hamiltonian $H$ containing both kinetic and potential energy contributions.
In the classical limit, the expectation value of the many-body Hamiltonian can be taken as the sum of single-particle Hamiltonians, $H = \sum_i H_i = \sum_i \big(K_i + V_i \big)$, where $K_i$ and $V_i$ denote the kinetic and potential energy of the $i$th particle. 
Given $V_i$ dependent on local baryon density $V_i\big( n_B(\bm{X}_j) \big)$ (which can be also thought of as the \textit{average} potential of a particle at position~$\bm{X}_i$), the change in particle's momentum can be then written out explicitly as
\begin{align}
\derr{\bm{P}_i}{t} 
&= - \parr{ V_i\big( n_B \big) }{n_B(\bm{X}_i)} \parr{n_B(\bm{X}_i)}{\bm{X}_i} - \sum_{j\neq i} \parr{ V_j\big( n_B \big) }{n_B(\bm{X}_j)} \parr{n_B(\bm{X}_j)}{\bm{X}_i} ~,
\end{align}
where the first term accounts for the change in the potential at the location of the $i$th particle due to a local gradient of density~$\inparr{n_B}{\bm{X}_i}$, while the second term accounts for the change in the potential at the locations of all \textit{other} particles due to changes in~$\bm{X}_i$. 
This underscores the fact that QMD is an $n$-body theory describing interactions between $n$ nucleons.
The local computational frame density at the position of the $i$th particle~$\bm{X}_i$, also sometimes referred to as local interaction density, is calculated by summing contributions from all particles except the $i$th particle, which given the Gaussian form of the wave packet, Eq.~\eqref{eq:Gaussian_packet}, yields
\begin{align}
n_B (\bm{X}_i) = \bfrac{2  (\Delta x )^2}{\pi}^{3/2} \sum_{j \neq i}  B_j  \exp \Bigg[ -  \frac{ \big(\bm{X}_i - \bm{X}_j\big)^2}{2  (\Delta x )^2}\Bigg]~,
\end{align}
where $B_j$ is the baryon number of the $j$th particle. 
Note that as a result, $2  (\Delta x )^2$ is the effective range parameter of the interaction.
Finally, while the propagation of particles in QMD, Eq.~\eqref{eq:dxdt_QMD}, can and often is described respecting the laws of special relativity, the change in the momenta of the particles due to a density-dependent potential, Eq.~\eqref{eq:dpdt_QMD}, is calculated in a non-relativistic way.

\subsubsection{Degrees of freedom}

Microscopic transport models are mostly used to describe systems composed of baryons and mesons, in which case the corresponding frameworks are usually referred to as ``hadronic transport''.
Within this group, simulation codes often differ between each other with respect to the number of degrees of freedom considered. 
Codes used for the description of heavy-ion collisions at the lowest energies (up to the order of 50 MeV/nucleon in the fixed-target frame) may consider only nucleons and nuclear clusters, including both light clusters (up to alpha particles) and heavy clusters (nuclei with mass numbers on the order of tens of nucleons
).
Above around 200 MeV/nucleon, pion production starts to play a role, necessitating their inclusion as degrees of freedom.
Similarly, around 1 GeV/nucleon there is a need to include kaons in the description of collisions; at the same time, heavy clusters start becoming less abundant in that region.
In general, as the energy is increased, more and more hadronic species are considered, including resonant states, while the importance of even light clusters diminishes.
However, which particular degrees of freedom play a dominant role depends on the set of observables that one wants to describe. 
For example, deuterons may remain important for the description of some proton observables as long as their yield binds a substantial fraction of protons available in the system, which is the case up to rather high energies of $\snn \approx 10~\rm{GeV}$ ($\ekin \approx 50~\rm{GeV/nucleon}$)~\cite{NA49:2016qvu}.

Microscopic transport also lends itself to descriptions of parton dynamics. 
Such approaches include simulations of jet-medium interactions~\cite{Cao:2016gvr} and of heavy-quark dynamics coupled to a dense strongly-interacting medium~\cite{Ke:2018tsh,Yao:2020xzw}, where the underlying bulk evolution is provided by relativistic hydrodynamics, and can yield predictions for, e.g., production or elliptic flow of quarkonia.
There are also microscopic transport frameworks that can be used for the description of the entire evolution of a heavy-ion collision at high energies~\cite{Lin:2004en,Xu:2013sta,Linnyk:2015rco,Aichelin:2019tnk}. In these simulations, the dense initial state of the collisions (which, depending on the framework, can be obtained either from an initial state model or from hadronic transport evolution) transitions to a system of strongly-interacting partons which subsequently transition back to hadronic degrees of freedom as the system expands and cools. 
Such approaches are an alternative to using multi-stage simulations (see Sec.~\ref{sec:hydro}) for describing heavy-ion observables even at high and very high energies.

While both of the above-mentioned uses provide for exciting and developing research directions, we will not address them in this review.


\subsection{Recent developments relevant to constraining the dense nuclear matter equation of state}
\label{sec:recent_developments_relevant_to_EOS}

As already mentioned in Sec.~\ref{sec:remarks_about_sensitivity}, and as we will again discuss in Sec.~\ref{sec:constraints_and_discussion}, putting reliable constraints on the EOS does not involve a single observable or an isolated process affecting the evolution. 
Conversely, there are numerous elements of simulations which may influence the extraction of the EOS from experimental data.
Given the above, virtually \textit{any} development of the chosen simulation framework, also beyond the explicit treatment of the EOS, can affect constraints on the EOS. 
Covering all such recent developments and their influence on the overall performance of the relevant models is beyond the scope of this review.
Below, we concentrate on developments related to two most prominent decisions that one needs to make before using heavy-ion collision data to extract the EOS: the choice of the simulation framework and the choice of the paradigm for modeling the EOS.

\subsubsection{Transport codes for the Beam Energy Scan physics}

While there are many microscopic transport codes available~\cite{TMEP:2022xjg}, relatively few of them are suited for use either as an afterburner (see Sec.~\ref{sec:hadronic_afterburner}) or as an independent simulation framework for the range of energies explored in the RHIC BES program.

Since the early days of RHIC, the Ultrarelativistic Quantum Molecular Dynamics (\texttt{UrQMD}) transport model~\cite{Bleicher:1999xi,Bass:1998ca} has been widely used both as an independent simulation framework and, later on, as a hadronic afterburner. 
Suitably for attempts to describe collisions from the intermediate to ultrarelativistic energies, the model incorporates mean-field interactions, includes scattering, production, and decay processes for over 80 hadronic species, and describes string excitation and fragmentation for collisions at energies exceeding $\snn = 5~\rm{GeV}$. 
The code has been developed over the years, with improvements such as using \texttt{PYTHIA}~\cite{Sjostrand:2006za}, which employs the Lund string fragmentation model~\cite{Andersson:1983jt,Andersson:1983ia} for multiparticle production, for initial hard scatterings~\cite{Petersen:2008kb} or incorporating an intermediate hydrodynamic stage, based on the ideal 3+1D relativistic hydrodynamics simulation code \texttt{SHASTA} (SHarp And Smooth Transport Algorithm)~\cite{Rischke:1995ir,Rischke:1995mt}, for an intermediate fluid dynamic description of the hot and dense phase~\cite{Petersen:2008dd}, thus making \texttt{UrQMD} a self-contained hybrid approach.

Another framework suited for the RHIC era is the Parton-Hadron-String-Dynamics (\texttt{PHSD}) code~\cite{Linnyk:2015rco,Cassing:2009vt,Bratkovskaya:2011wp,Moreau:2019vhw}, which is a covariant approach, based on generalized off-shell transport equations for test particles~\cite{Cassing:1999mh,Cassing:1999wx}, that can describe both hadronic and partonic matter.
Given that \texttt{PHSD} propagates Green's functions which, in particular, include information about the spectral functions of the described degrees of freedom, it is an extension beyond the usual semi-classical BUU-type models.
The hadronic degrees of freedom include the baryon octet and decuplet, the $0^-$ and $1^-$ meson nonets, and certain higher resonances.
At low energies, hadronic reactions are performed using the corresponding cross sections, while at high energies multi-particle production is achieved by using event generators \texttt{FRITIOF}~\cite{Nilsson-Almqvist:1986ast,Andersson:1992iq} and \texttt{PYTHIA}~\cite{Sjostrand:2006za}. 
At high energies, partonic dynamics is realized by using the Dynamical Quasiparticle Model (DQPM)~\cite{Linnyk:2015rco}, fit to describe lattice QCD data~\cite{Moreau:2019vhw}.
As a result, \texttt{PHSD} can be used to describe the entire evolution of a relativistic heavy-ion collision, including the initial hard scatterings and string formation, transition to the dynamically evolving quark-gluon plasma~(QGP), hadronization, and subsequent hadronic phase.
Recently, an offshoot of the code called Parton-Hadron-Quantum-Molecular-Dynamics (\texttt{PHQMD})~\cite{Aichelin:2019tnk} has been developed which uses a QMD approach to propagating hadrons and thus includes $n$-body dynamical evolution.

The Jet AA Microscopic (\texttt{JAM}) transport model~\cite{Nara:1999dz} has likewise been designed to be used across a large range of energies. 
At low energies, the degrees of freedom of the model include hadrons and their excited states, and the nuclear mean field with momentum dependence can be included~\cite{Isse:2005nk}. 
String excitation is implemented according to the Heavy Ion Jet Interaction Generator (\texttt{HIJING}) model~\cite{Wang:1991hta,Wang:1996yf,Gyulassy:1994ew}, which also serves as a template for implementation of multiple mini-jet production, for which the jet cross sections and the number of jets are calculated using an eikonal formalism for perturbative QCD.
Hard parton-parton scatterings are simulated using \texttt{PYTHIA}~\cite{Sjostrand:1993yb}.

Another framework including both partonic and hadronic degrees of freedom is A Multi-Phase Transport (\texttt{AMPT}) model~\cite{Lin:2004en}.
The initial state of the evolution is obtained using \texttt{HIJING} and includes spatial and momentum distributions of minijet partons and soft string excitations. 
The evolution of partons in achieved by using Zhang’s Parton Cascade (\texttt{ZPC})~\cite{Zhang:1997ej}, while conversion of partons into hadrons is modeled either by the Lund string fragmentation model~\cite{Andersson:1983jt,Andersson:1983ia,Sjostrand:1993yb} or string melting with quark coalescence~\cite{Lin:2001zk,Lin:2002gc,Lin:2003iq}.
Further evolution of hadrons is described by A Relativistic Transport (\texttt{ART}) model~\cite{Li:1995pra,Li:2001xh} extended to include additional reaction channels of relevance for modeling high energy collisions, such as formation of antibaryon resonances, baryon-antibaryon production from mesons, and baryon-antibaryon annihilation. 
In subsequent developments, the model has been extended by both partonic~\cite{Ko:2012lhi} and hadronic~\cite{Xu:2012gf} mean fields.

A relatively recent addition is the \texttt{SMASH} (Simulating Many Accelerated Strongly-Interacting Hadrons) hadronic transport code~\cite{SMASH:2016zqf}. Building on the experience of other transport approaches to date, including \texttt{UrQMD}, \texttt{GiBUU}~\cite{Buss:2011mx}, and \texttt{pBUU}~\cite{Danielewicz:1991dh,Danielewicz:1999zn}, \texttt{SMASH} is a modern (utilizing \texttt{C++} and \texttt{ROOT}, among others), modular, open source and version-controlled code developed both to be used as an afterburner and as a stand-alone simulation framework for collisions at low energies.
The framework includes all hadrons and confirmed hadronic resonances with masses up to \mbox{$\approx 2~\rm{GeV}$} (over 120 hadronic species, where isospin states such as $\pi^{\pm}$ and $\pi^0$ or baryons and antibaryons are counted as one species). It also takes advantage of new experimental data for cross sections and resonance properties (with all hadronic reactions fulfilling the principle of detailed balance), includes a perturbative treatment of photon and dilepton emission~\cite{Schafer:2019edr,Staudenmaier:2017vtq}, and deuteron production via explicit reactions~\cite{Danielewicz:1991dh,Oliinychenko:2018ugs}.
Notably, \texttt{SMASH} often implements multiple phenomenological and algorithmic solutions (such as different collision criteria or various forms of mean-field potentials) that can be freely chosen by the user and which allow for extensive tests of given simulation inputs within the same framework. Similarly to the hybrid $\texttt{UrQMD}$ code, there also exists a modular hybrid approach \texttt{SMASH-vHLLE-hybrid}~\cite{Schafer:2021csj}, coupling 3+1D viscous hydrodynamics \texttt{vHLLE} (viscous Harten-Lax-van Leer-Einfeldt algorithm)~\cite{Karpenko:2013wva} to \texttt{SMASH}. A further improvement in the form of dynamical (gradual) initialization of the hydrodynamic phase in the \texttt{SMASH-vHLLE-hybrid} is currently under development~\cite{Hirayama_poster,HirayamaPaulinyovaInPreparation}.

\subsubsection{Modeling the equation of state}
\label{sec:modeling_the_EOS}

For both BUU- and QMD-type simulations, the EOS enters the evolution of the system primarily through single-particle equations of motion, Eqs.~(\ref{eq:dxdt}--\ref{eq:dpdt}) or Eqs.~(\ref{eq:dxdt_QMD}--\ref{eq:dpdt_QMD}), where it contributes to changes in particles' momenta through gradients of the single-particle energy. 
In the case of models including relativistic scalar-type interactions, the EOS also explicitly affects the propagation of particles through its effects on the particle effective mass. 
Additionally, the EOS can play a role in the scattering term, leading to, e.g., subthreshold production or reduced in-medium cross sections (note, however, that most in-medium cross section parametrizations currently in use are purely phenomenological with no explicit connection to the local potential).

Historically, development of transport simulations was driven by early experiments on relativistic heavy-ion collisions at incident kinetic energies per nucleon (excluding the nucleon mass) from a couple hundred of MeV/nucleon up to about $\ekin \approx 1~\rm{GeV/nucleon}$ ($\snn \approx 2.3~\rm{GeV}$).
These experiments probed densities from around that of normal nuclear matter $n_0$ up to at most 3 times $n_0$~\cite{Sorensen:2023zkk}, and consequently it was natural that simulations utilized EOSs known to correctly describe the established properties of nuclear matter in the ground state, such as saturation at $n_0 \approx 0.17\pm0.03~{\rm{fm}}^{-3}$ and binding energy of nuclear matter $B_0 \approx -16~\rm{MeV}$~\cite{Dutra:2012mb}. 
The single-particle potentials reproducing these properties can be cast in a particularly simple form often referred to as the Skyrme potential, in which they are parametrized as a function of local baryon density $n_B$
,
\begin{align}
U(n_B) = A \left( \frac{n_B}{n_0} \right) + B \left( \frac{n_B}{n_0} \right)^{\tau}~.
\label{eq:basic_Skyrme}
\end{align}
Here, $A$, $B$, and $\tau$ are free parameters which are fit so that the EOS reproduces the values  of $n_0$, $B_0$, and the incompressibility of nuclear matter at saturation $K_0$. 
With values of $n_0$ and $B_0$ relatively well known, the Skyrme EOSs differ between each other mostly by the value of $K_0$, the constraint on which is less stringent, $K_0 = 230\pm30~\rm{MeV}$~\cite{Dutra:2012mb}. 
In turn, variation in $K_0$ allows for variation between the EOSs at densities away from $n_0$.

Importantly, to maximize the range of possible behaviors of the EOS at high densities, $K_0$ is often varied well beyond the experimentally constrained region above, $K_0 \in [170, 380]~\rm{MeV}$.
In such cases, one should understand the used values of $K_0$ not as an assertion on the actual value of incompressibility at $n_0$, but rather as a parameter which specifies the behavior of the EOS for the range of densities relevant to a given study. 
As an example, in the case of collisions at energies which primarily probe the EOS at densities above $2n_0$, any conclusions on the value of $K_0$ which describes the data best only concerns the behavior of the EOS above $2n_0$, and is by no means a statement on the behavior of the EOS around $n_0$.

Note that the parametrization in terms of baryon density implies interactions of vector type, however, the potential shown in Eq.~\eqref{eq:basic_Skyrme} is non-relativistic. 
While one can always assume that $U(n_B)$ is simply the form of the potential in the local rest frame and boost the resulting equations of motion to account for relativistic effects (such as, e.g., contributions to the equations of motion from a non-zero baryon current), this needs to be done with care so that all relevant relativistic terms are recovered~\cite{Sorensen:2021zxd}.
Regardless of whether the modification of the equations of motion due to the potential is treated relativistically or not, the form of the potential shown in Eq.~\eqref{eq:basic_Skyrme} leads to a superluminal behavior at high densities for any $\tau \geq 1$~\cite{Zeldovich:1961sbr}, which is ubiquitous among Skyrme-type models~\cite{Dutra:2012mb}. 
To prevent such superluminal behavior, some codes~\cite{Danielewicz:2002pu} employ a more stable form 
\begin{align}
U(n_B) = \frac{A \left( \frac{n_B}{n_0} \right) + B \left( \frac{n_B}{n_0} \right)^{\tau}}{1 + \alpha\left( \frac{n_B}{n_0}\right)^{\tau - 1}}~,
\label{eq:generalized_Skyrme}
\end{align}
where $\alpha$ is a dimensionless parameter.

In addition to a density-dependent Skyrme parametrization, many codes also employ momentum-dependent terms in the potential~\cite{Gale:1987zz,Aichelin:1987ti,Welke:1988zz}. 
This is motivated both by fits to experimental data~\cite{Cooper:1987uy,Hama:1990vr} as well as model calculations~\cite{Botermans:1990qi}, showing that while a nucleon interacting with nuclear matter around $n_0$ feels attraction if it moves slowly, it feels repulsion once its kinetic energy exceeds $\approx 200~\rm{MeV}$.
Implementations of the momentum-dependence of the potential, in general, are of the following form,
\begin{eqnarray}
U_p (n_B, \bm{p}) = C \int \frac{d^3 p'}{(2\pi)^3} ~ \frac{f_{\bm{r}, \bm{p}'}}{1 + \left[\frac{\bm{p} - \bm{p}'}{\Lambda} \right]^2}~,
\label{eq:basic_momentum_dependence}
\end{eqnarray}
with the free parameters $C$ and $\Lambda$ fit to reproduce, e.g., the expected value of the effective mass at the Fermi surface~\cite{Gale:1987zz} or the measured values of $U_p$ at chosen nucleon momenta~\cite{Welke:1988zz}. 
Notably, to ease the significant numerical cost associated with the calculation of momentum-dependent potentials in simulations (which for each particle requires summing the contributions from all other particles), many approaches use approximated expressions for $U_p$, with sometimes non-trivial consequences for the behavior of the potential~\cite{Welke:1988zz}.
Alternatively to the form shown in Eq.~\eqref{eq:basic_momentum_dependence}, momentum dependence can be also included by employing scalar interactions, which lead to equations of motion including terms dependent on gradients of the local values of the effective mass; this can be achieved, among others, by using relativistic mean-field potentials, such as those based on the Walecka model~\cite{Blaettel:1993uz}.
A fully relativistic generalization of Eq.~\eqref{eq:basic_momentum_dependence} has been developed in Ref.~\cite{Weber:1993et}, however, due to formidable complexity its use in realistic simulations has not been pursued.

We note that while the Skyrme potential, Eq.~\eqref{eq:basic_Skyrme}, depends only on density, models using this potential can still exhibit non-trivial effects due to temperature. This is because the behavior of a given system depends not only on the interactions between the particles, but also on the momenta of the particles. In equilibrium, these momenta can often be described by an appropriate distribution with an explicit dependence on temperature: the Fermi or Bose distribution for temperatures where quantum statistics plays a role, or the Boltzmann distribution for higher temperatures. Consequently, even though the interaction terms governing the system do not depend on temperature, effects due to temperature are reflected in the kinetic part of the single-particle energies which in turn affect the evolution through single-particle equations of motion, Eqs.~\eqref{eq:dxdt} and \eqref{eq:dpdt} or \eqref{eq:dxdt_QMD} and \eqref{eq:dpdt_QMD}.

To illustrate this fact, one can consider the Skyrme potential as used to describe ordinary nuclear matter, in which case the nucleon pressure is the sum of the Fermi pressure and interaction pressure arising from the interaction term, Eq.~\eqref{eq:basic_Skyrme}. As already mentioned, for the model to describe nuclear matter one adjusts the interaction parameters to reproduce the well-established properties at $T=0$: saturation density~$n_0$, binding energy~$B_0$, and incompressibility~$K_0$. Note that satisfying these properties means that at $T=0$  and $n_B = n_0$, $P=0$ (because matter is in equilibrium) and $dP/dn_B> 0$ (because incompressibility is positive). We also naturally have $P (T=0, n_B = 0) = 0$ and, since the Fermi pressure is always positive and for very small densities the interaction terms are negligibly small, $P>0$ at small densities. Together, this means that at $T=0$ there is a region between $n_B = 0$ and $n_B = n_0$ where $P < 0$ and, consequently, there is also a region where $dP/dn_B <0$, that is a region where the matter is mechanically unstable. In other words, we are dealing with a system in which at $T=0$ there is a first-order phase transition between two phases characterized by different densities. Note that the occurrence of regions with $P<0 $ and $dP/dn_B < 0$ is an effect driven by the interaction pressure, as the Fermi pressure is always positive. In this model, changing the temperature affects only the kinetic (Fermi) contribution to the pressure; nevertheless, this is enough to yield further non-trivial features. As the temperature is increased, particles occupy higher momentum states and so the kinetic contribution to the pressure becomes larger. Gradually, the kinetic contribution starts to dominate the interaction pressure so that, for high enough temperatures, there isn't any region where $dP/dn_B < 0$, i.e., there is no density at which the system is unstable and, consequently, there is only one possible phase of the system. The temperature at which this happens is the critical temperature; as the critical temperature is approached from below, the density interval for which $dP/dn_B < 0$ collapses to a point, at the critical density, where \mbox{$dP/dn_B = 0$}. Thus, even though the interaction terms used in the model depend only on density and are fixed to reproduce the properties of nuclear matter at $T=0$, the system exhibits non-trivial behavior with changing temperature, including the existence of a critical point.

We stress that while the above simple example is considered using thermodynamic variables such as temperature or pressure, the arguments used can be easily generalized to systems out-of-equilibrium, and in particular to heavy-ion collision evolution as described by microscopic transport.
Since the evolution involves both particle velocities and gradients of mean-field interactions (where the former, in equilibrium, can be directly connected to temperature), transport simulations can exactly reproduce the thermodynamic behavior of the underlying EOS even though
there are no explicit temperature-dependent terms in the single-particle energies.
Here, the influence of the particle momenta leads to non-trivial effects and different outcomes in situations characterized by different momentum distributions, e.g., at varying collision energies or at different stages of collisions at a given energy which, if equilibrium could be achieved, would correspond to different temperatures. 
This effective temperature dependence is even more pronounced in case of interactions with momentum dependence such as, e.g., the interaction in Eq.~\eqref{eq:basic_momentum_dependence} or scalar interactions; in particular, scalar interactions can also lead to more complex structures in the phase diagram. At the same time, even though such an effective temperature dependence leads to non-trivial features, one can always wonder whether it is sufficient to capture all relevant effects; this remains to be seen.

The primary drawback of the Skyrme and momentum dependent potentials described above is that they only allow for modest variations in the interactions. 
This is naturally appropriate for describing experiments probing densities not too far from $n_0$, where the properties of nuclear matter are reasonably constrained. 
However, at higher densities and very high momenta there is no reason to expect the single-particle potentials to exhibit the same functional form which is found suitable to describe nuclear matter close to its ground state. 
Additionally, since higher densities are reached in collisions at higher energies, it becomes necessary to use potentials which are relativistically covariant. 
Consequently, describing experiments at $\ekin \gtrsim 1~\rm{GeV}$ ($\snn \gtrsim 2.3~\rm{GeV}$) calls for employing relativistic single-particle potentials which not only describe the known properties of ordinary nuclear matter, but also allow for the possibility of substantial changes to the behavior of the potential at high densities, momenta, and/or temperatures. 

Addressing this need has been the focus of a number of recent efforts. 
Relativistically covariant single-particle potentials based on an arbitrary number of vector-type interaction terms, referred to as the vector density functional (VDF) model, have been developed~\cite{Sorensen:2020ygf} with the goal of describing both the known properties of ordinary nuclear matter and its possible non-trivial behavior at high baryon densities. 
Starting from the relativistic Landau Fermi-liquid theory~\cite{Baym:1975va}, that work develops easily parametrizable, relativistically covariant, and thermodynamically consistent EOS which leads to single-particle energies of the form
\begin{align}
\varepsilon_{\bm{p}} = \sqrt{ m^2 +  \bm{\Pi}^2 } + \sum_{k=1}^K A_k^0~.
\end{align}
Here, $m$ is the particle mass, $\bm{\Pi}\equiv= \bm{p} - \sum_{k=1}^K \bm{A}_k$ is the kinetic momentum, and the vector field $A_k^{\lambda}$ is defined as
\begin{align}
A_{k}^{\lambda}(x) = C_k n_B^{b_k - 2} j_B^{\lambda} ~,
\end{align}
where the baryon current $j_B^{\mu}$ is given by a self-consistent relation,
\begin{align}
j_B^{\mu} \equiv g \int \frac{d^3p}{(2\pi)^3} ~ \frac{p^{\mu} - \sum_{k=1} A_k^{\mu}}{  \sqrt{ m^2 + \bm{\Pi}^2 }} ~ f_{\bm{p}}~,
\end{align}
with $g$ denoting the degeneracy and $f_{\bm{p}}$ the distribution function, and $n_B \equiv \sqrt{j_{B,\mu} j_B^{\mu}}$ is the rest-frame baryon density.
Assuming thermal equilibrium, this leads to a pressure functional of the form
\begin{align}
P_K &= g \int \frac{d^3p}{(2\pi)^3} ~ T ~ \ln \Big[ 1 + e^{-\beta \big(\varepsilon_{\bm{p}} - \mu_B \big)}  \Big] +  \sum_{k=1}^K C_k \left( \frac{b_k - 1}{b_k}\right) n _B^{b_k} ~,
\label{eq:VDF_pressure}
\end{align}
where $T$ is the temperature, $\beta = 1/T$, and $\mu_B$ is the baryon chemical potential, so that the first term can be easily recognized as the pressure of the ideal relativistic Fermi gas. 
This relativistic generalization of the Skyrme potential, through the inclusion of an arbitrary number of potential terms, allows one to consider various possible behaviors of the potential at high baryon densities while retaining the ability to describe the known properties of ordinary nuclear matter around saturation. 
Of note is also the fact that while the form of the potential leading to Eq.~\eqref{eq:VDF_pressure} is relatively simple, polynomials of vector density are the \textit{only} possible form of the interactions if one considers a mean-field model with self-consistently defined baryon current and one demands that the coefficients of the interaction terms are constant~\cite{Sorensen:2021zxd} (the same can be shown to be true for scalar-type interactions~\cite{Sorensen:2021zxd}).

\begin{figure}[t]
    \centering
	\includegraphics[width=0.99\linewidth]{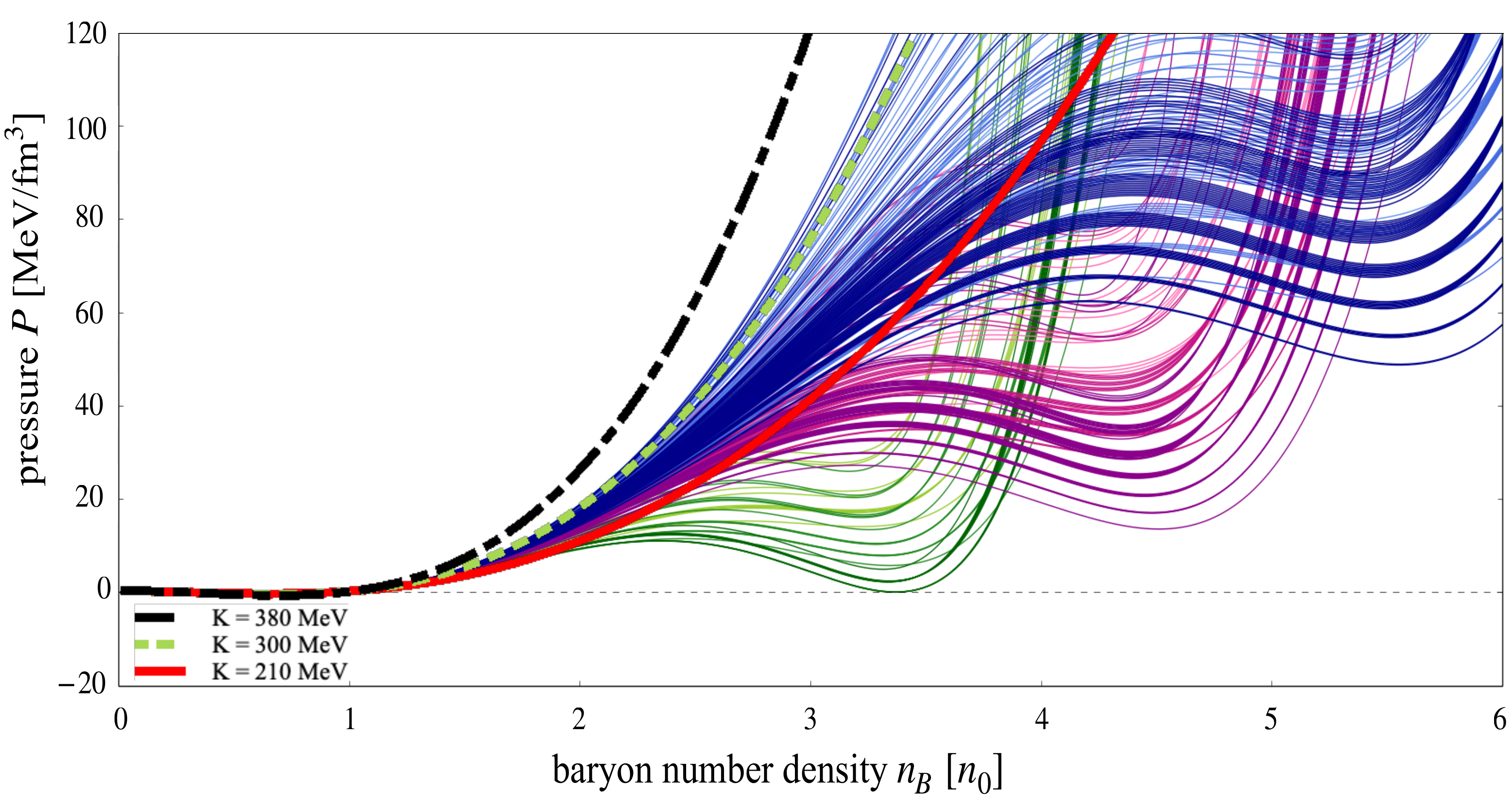}
    \caption{Pressure at $T=0$ as a function of baryon density for 3 EOSs using the Skyrme parametrization and 9 families of VDF EOSs~\cite{Sorensen:2020ygf}. The Skyrme EOSs are marked with thick solid red, short-dashed green, and mixed-dashed black lines, corresponding to incompressibility at saturation of $K_0 = 210$ (soft), $300$ (medium), and $380~\rm{MeV}$ (hard), respectively. Thin green, magenta, and blue curves mark VDF EOSs with a high-density phase transition characterized by a critical density of $n_c^{(Q)} = 3.0n_0$, $4.0n_0$, and $5.0n_0$, respectively, with light, medium dark, and dark variations in the colors of the curves further differentiating between curves with the critical temperature $T_c^{(Q)} = 50$, $100$, and $150~\rm{MeV}$. Within each family, the VDF EOSs further vary with respect to the width of the coexistence region at $T=0$. All of the VDF EOSs describe nuclear matter with $n_0 = 0.160~\rm{fm}^{-3}$, $B_0 = - 16~\rm{MeV}$, and the critical point of the nuclear liquid-gas phase transition at $n_c^{(N)} = 0.06~\rm{fm}^{-3}$ and $T_c^{(N)} = 18~\rm{MeV}$, with incompressibilities averaging at $\langle K_0 \rangle = 273.1 \pm 5.1~\rm{MeV}$. It is easily seen that parametrizations of the VDF EOS can lead to a great variability in the behavior of the EOS at high densities. Figure modified from Ref.~\cite{Sorensen:2021zxd}.
	}
    \label{fig:VDF_2_PTs}
\end{figure}
In the original work~\cite{Sorensen:2020ygf}, the developed formalism has been used to model various critical densities, critical temperatures, and coexistence widths of a possible first-order phase transition at baryon densities $n_B > 2n_0$. 
As shown in Fig.~\ref{fig:VDF_2_PTs}, this leads to a non-trivial dependence of the pressure as a function of baryon density which would be impossible to describe with the standard Skyrme form of the interaction.
One should note that while in this example the VDF model has been employed in a minimal form (utilizing $K=4$ interaction terms) allowing one to describe nuclear matter with a first-order phase transition at high densities, the model can be fit to describe other possible behaviors of the single-particle energy or, equivalently, pressure.
In particular, additional terms can be added to prevent the superluminal behavior of the speed of sound~\cite{Sorensen:2020ygf} at densities above the postulated locations of phase transitions.

The above-described model has been subsequently generalized to enable parametrization of the single-particle potential based on a chosen dependence of the speed of sound squared at zero temperature $c_s^2$ on baryon density~\cite{Oliinychenko:2022uvy}.
Within this approach, the rest-frame single-particle potential, $U(n_B) \equiv A^0(n_B) \big|_{\substack{ \rm{rest}\\\rm{frame} }}$, is given by
\begin{align} 
U(n_B) = \mu_B\big(n_B^{(0)}\big) \exp \Bigg[\int_{n_B^{(0)}}^{n_B} dn' ~   \frac{c_s^2(n')}{n'} \Bigg] -   \mu^*(n_B)~,
\label{eq:pot_cs2}
\end{align}
where $n_B^{(0)}$ is an arbitrarily chosen density at which the corresponding value of the chemical potential $\mu_B \big(n_B^{(0)}\big)$ is known and $\mu^*(n_B) \equiv \big[m^2 + \big(\infrac{6\pi^2 n_B}{g}  \big)^{2/3}\big]^{1/2}$. 
A convenient way of parametrizing such a potential, leading to an analytic expression for $U(n_B)$ (and, consequently, $A^{\mu}(n_B)$), is to consider a piecewise functional form of $c_s^2(n_B, T=0)$ in which the speed of sound squared follows the Skyrme-like parametrization up to some density $n_1$, after which it assumes a constant value in chosen density intervals~\cite{Oliinychenko:2022uvy},
\begin{align} \label{eq:cs2_piecewise}
    c_s^2(n_B) = \left.  \begin{cases}
      c_s^2(\rm{Skyrme}) &~~~ n_B < n_1  \\
      c_1^2 &~~~ n_1 < n_B < n_2 \\
      c_2^2 &~~~ n_2 < n_B < n_3 \\
      \dots & \\
      c_m^2 &~~~ n_m < n_B
    \end{cases} \right.
\end{align}
where $n_1, \dots, n_m$ are chosen values of baryon density, which yields
\begin{align}
U(n_B) = \left.  \begin{cases}
     U_{\mathrm{Sk}}(n_B) & n_B < n_1 \\
     C(n_1) \big(\frac{n_B}{n_1}\big)^{c_1^2} - \mu^*(n_B) &  n_1 < n_B < n_2 \\
     \dots & \\
     C(n_1) \big(\frac{n_B}{n_k}\big)^{c_k^2} \prod\limits^{k}_{i=2} \big(\frac{n_{i}}{n_{i-1}}\big)^{c_{i-1}^2} - \mu^*(n_B) & n_k < n_B < n_{k+1}
     \end{cases} \right. 
     \label{eq:potential_final}
\end{align}
where $U_{\rm{Sk}}(n_B)$ is a Skyrme-like potential and \mbox{$C(n_1) = U_{\mathrm{Sk}}(n_1) + \mu^*(n_1)$} (here, ``Skyrme-like potential'' refers to a relativistic VDF potential with $K=2$ interaction terms parametrized to reproduce chosen values of $n_0$, $B_0$, and $K_0$). Note that Eq.~\eqref{eq:potential_final} makes it evident that this generalization of the VDF potential, straying away from the polynomial form of the EOS, indeed leads to density-dependent coefficients of the interaction terms, $\tilde{C}_{k} \equiv C(n_1) \prod^{k}_{i=2} \big(\frac{n_{i}}{n_{i-1}}\big)^{c_{i-1}^2} $.

\begin{figure}[t]
    \centering
    \includegraphics[width=0.99\linewidth]{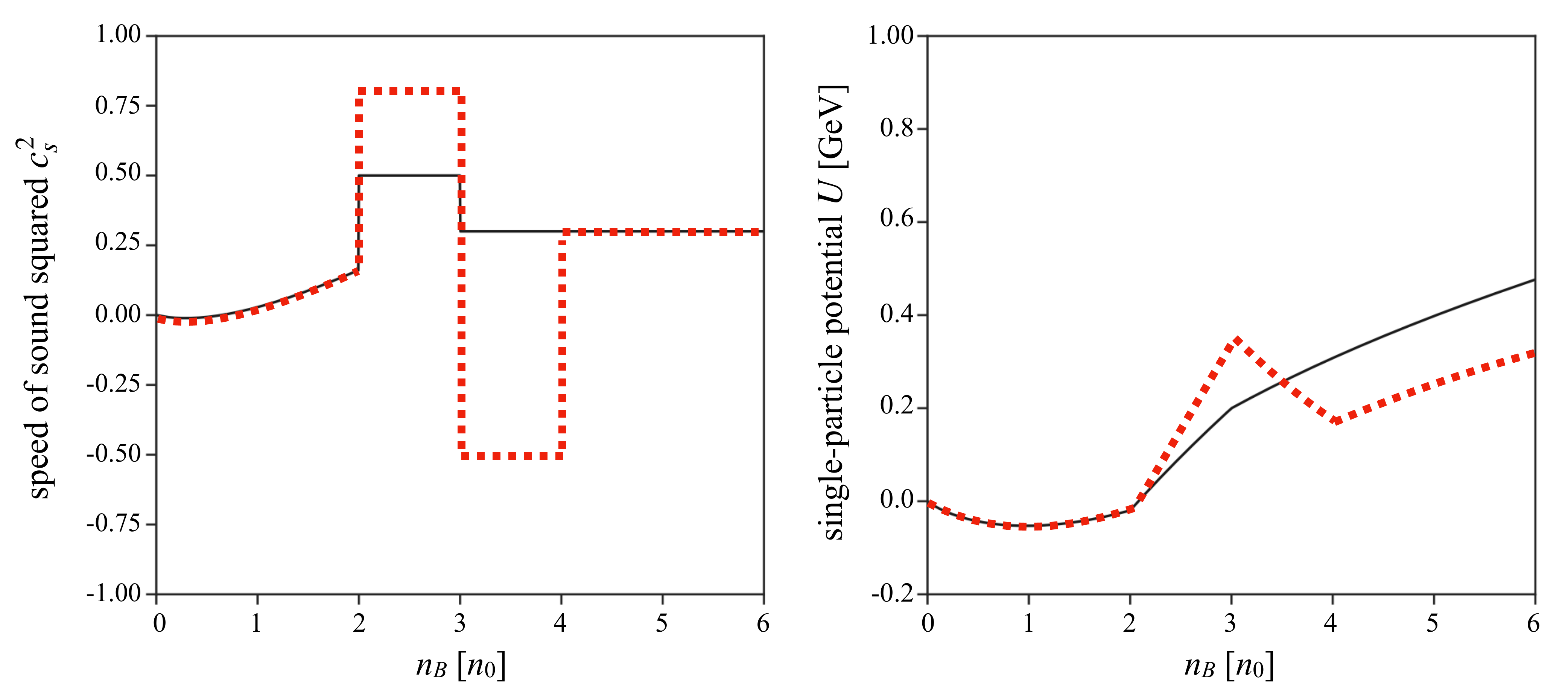}
    \caption{Functional dependence of $c_s^2$ on baryon density for two chosen $c_s^2$ profiles (\textit{left}) and the corresponding rest-frame single-particle potential $U(n_B)$ (\textit{right}) obtained within the $c_s^2$-parametrization of the generalized VDF model~\cite{Oliinychenko:2022uvy}.
	}
    \label{fig:VDF_cs2_parametrization}
\end{figure}

Examples of functional dependence of $c_s^2$ on baryon density and the corresponding single-particle potentials $U(n_B)$ are shown in Fig.~\ref{fig:VDF_cs2_parametrization}.
As in the case of the original VDF model, the $c_s^2$-parametrization supports non-trivial behavior of the EOS or, equivalently, the single-particle potential at high baryon densities.
Moreover, in contrast to the original VDF model where the EOS essentially takes the form of a polynomial, the $c_s^2$-parametrization allows one to vary the behavior of the EOS within the chosen density intervals in an entirely independent way.

The  $c_s^2$-parametrization is not only highly intuitive due to its anchoring in the baryon-density dependence of the speed of sound (which additionally allows one to easily avoid superluminal behavior at any density), but also provides a convenient starting point for performing comprehensive analyses of the corresponding EOSs through simple varying of the parameter sets~$\{c_i^2, n_i\}$.
\begin{figure}[t]
    \centering
    \includegraphics[width=0.94\linewidth]{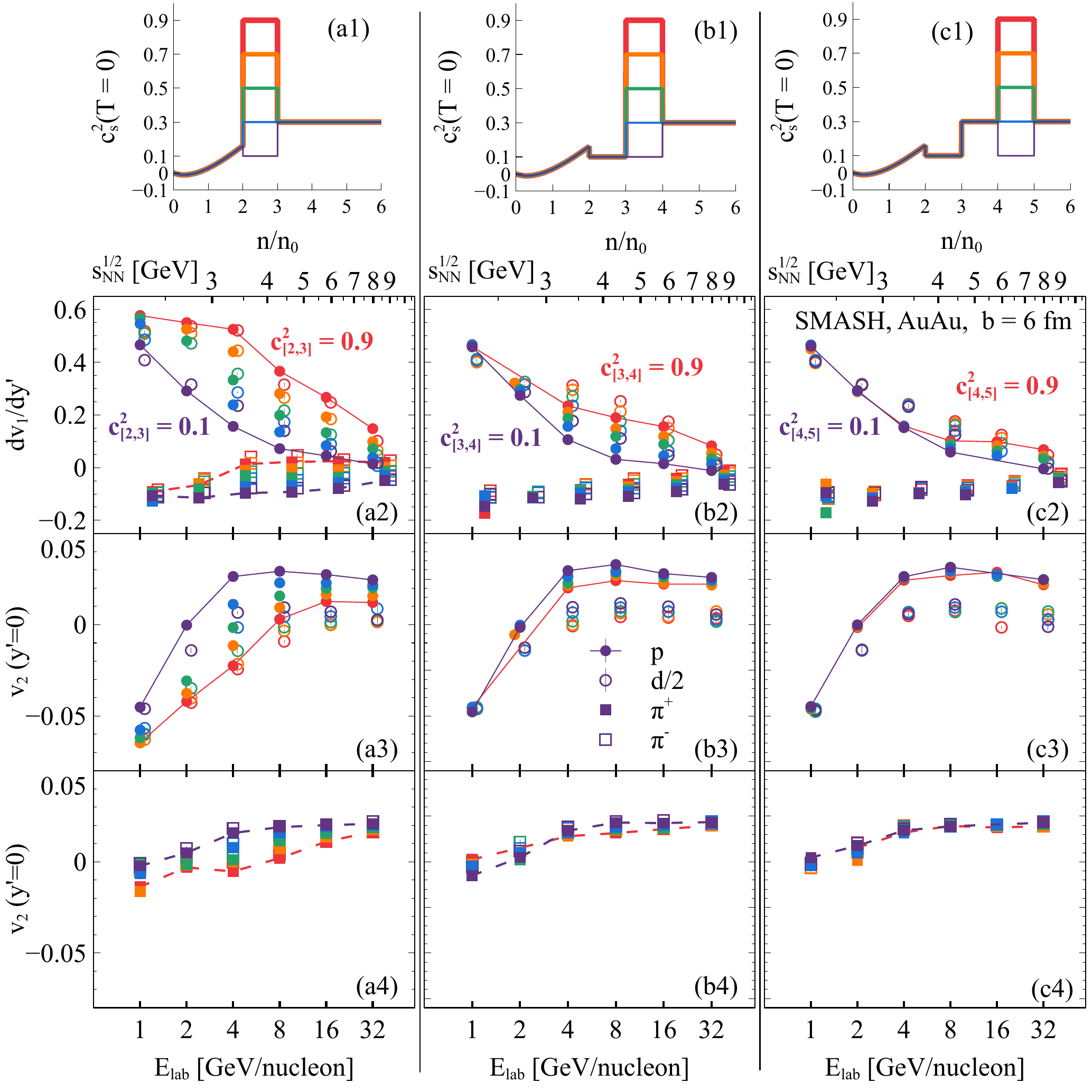}
    \caption{Results of simulations in \texttt{SMASH} using different $c_s^2$-parametrizations of the EOS. For $n_B < 2n_0$, the EOS takes a Skyrme-like form
	. In columns (a), (b), and (c), the EOS is only varied in the density interval $n_B \in (2,3)n_0$, $n_B \in (3,4)n_0$, and $n_B \in (4,5)n_0$, respectively; for densities $n_B > 2n_0$ outside of the varied intervals, $c_s^2$ takes on an arbitrary constant value. The first row shows variations in $c_s^2$ in the above intervals, where it takes representative values of $\{0.1, 0.3, 0.5, 0.7, 0.9\}$ marked with varying colors. These then correspond to results for $dv_1/dy$ (row 2) and $v_2$ (rows 3 and 4) at midrapidity for protons (full circles), deuterons (empty circles), and positive (filled squares) and negative (empty squares) pions; here, the deuteron flow is divided by 2, and the points for deuterons and pions are slightly shifted to the right for clarity. Figure from Ref.~\cite{Oliinychenko:2022uvy}.
	}
    \label{fig:VDF_cs2_sensitivity}
\end{figure}
In particular, one can test the sensitivity of flow observables differentially in chosen density regions~\cite{Oliinychenko:2022uvy} as shown in Fig.~\ref{fig:VDF_cs2_sensitivity}. 
It is evident that changing the EOS for densities $n_B \in (2,3)n_0$ leads to most prominent changes in the proton and pion flow observables within a large range of energies, $\snn \approx 2$--$8~\rm{GeV}$ ($\ekin \approx 1$--$32~\rm{GeV/nucleon}$). 
Changing the EOS for densities $n_B \in (3,4)n_0$ leads to significantly smaller effects, largely constrained to $\snn \approx 3$--$8~\rm{GeV}$ ($\ekin \approx 3$--$32~\rm{GeV/nucleon}$), and changes in the density interval $n_B \in (4,5)n_0$ (and, consequently, likely also for higher densities) lead to only very subtle effects for $\snn \approx 4$--$8~\rm{GeV}$ ($\ekin \approx 8$--$32~\rm{GeV/nucleon}$).
This suggests that the possibilities to constrain the EOS at densities $n_B > 4n_0$ using heavy-ion collisions may be limited, although some sensitivity seems to remain in the deuteron flow (note that the values of the deuteron flow in the figure are divided by 2).

Furthermore, since one can define, without any significant numerical penalty, arbitrarily many intervals characterized by a constant $c_s^2$ and thus, in the limit of an infinite number of intervals, the parametrization can exactly reproduce any $c_s^2(n_B, T=0)$ profile, the approach can be used to explore the consequences of \textit{any} chosen EOS for microscopic transport simulations of heavy-ion collisions as long as the dependence $c_s^2(n_B, T=0)$ is available~\cite{Yao:2023yda}.
Here, however, some caution is warranted: since the approach is based on vector-type interactions, the resulting mean-field potentials do not exhibit any complex temperature dependence (more precisely, the temperature dependence of the EOS comes solely from ideal gas terms).
Since collisions of heavy ions probe temperatures from a few tens to well over a hundred of MeV, this may lead to substantial deviations from the EOS which was used to provide the input $c_s^2(n_B,T=0)$ profile if that EOS has a non-trivial (beyond ideal gas) dependence on temperature.

The two above-discussed parametrizations have been developed with relativistic BUU-type transport in mind and implemented in the transport code \texttt{SMASH} (currently, only the original VDF model is available in the public version of the \texttt{SMASH} code). 
Recently, an implementation of the single-particle potentials allowing one to go beyond the default Skyrme potential, Eq.~\eqref{eq:basic_Skyrme}, has been employed in \texttt{UrQMD}~\cite{Steinheimer:2022gqb}. 
The approach is based on the fact that the density-dependent average potential energy per particle, $\mathcal{V} (n_B)$, can be related to the density-dependent single-particle potential $U(n_B)$ through
\begin{align}
U(n_B) = \parr{\big( n_B  \mathcal{V}(n_B)  \big)}{n_B} ~,
\end{align}
so that $U_i = U\big(n_B (\bm{X}_i)\big)$. 
While the above prescription may appear trivial in the case of Skyrme-like models (for which the single-particle potential is equivalent to the mean-field potential required for the equations of motion), it allows one to extend the treatment of potentials in QMD to more complicated models, including models with varied interactions between multiple species, relativistic fields, and momentum-dependence (or, equivalently, effective masses), as long as the average potential per particle $\mathcal{V}(n_B)$ can be identified by some means.
In particular, this can be achieved by generating a table of the average potential values at a given $n_B$, which is the basis of the implementation of the approach in \texttt{UrQMD}~\cite{Steinheimer:2022gqb}, allowing for studying the influence of almost any EOS on heavy-ion collision observables.
(We note here, however, that similar caution has to be exercised here as in the case of $c_s^2$-parametrized potentials described above, since the approach can capture only the density dependence of a given potential, provided at $T=0$, and any temperature-dependent effects come solely from ideal-gas-like contributions, which may lead to deviations in the behavior of the finite-temperature EOS in the underlying model and in the simulations). 

\begin{figure}[t]
    \centering
    \includegraphics[width=0.99\linewidth]{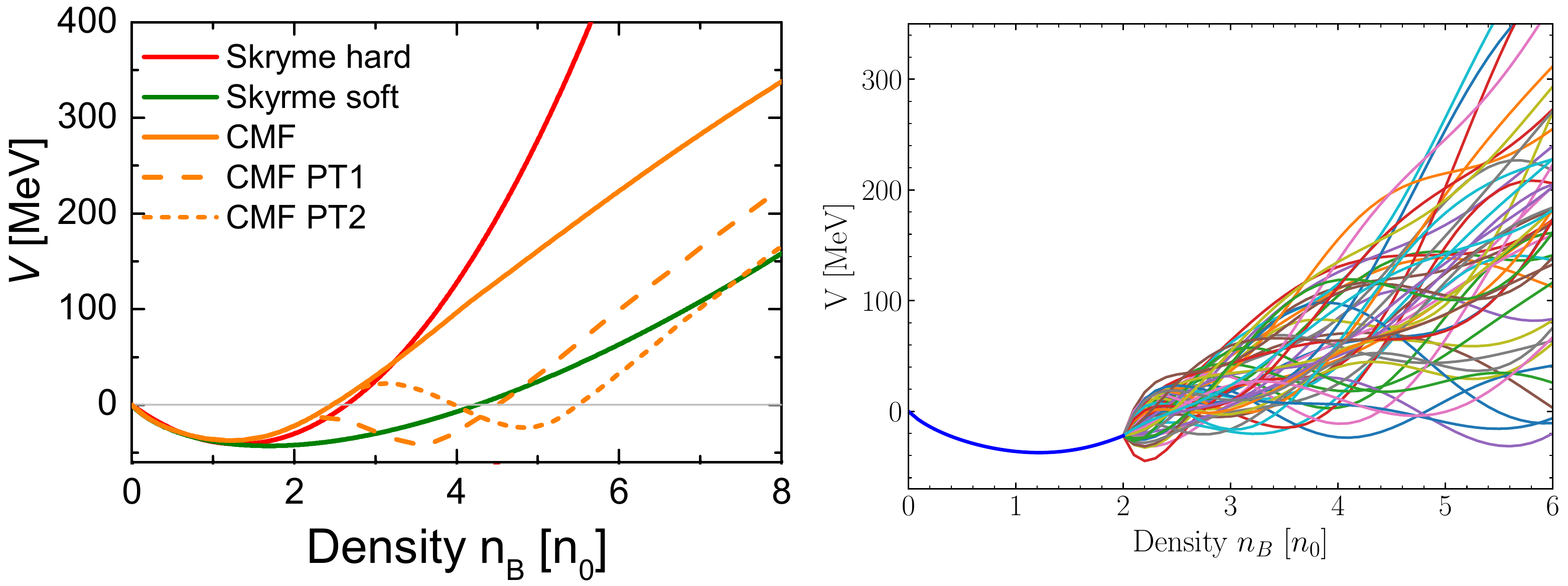}
    \caption{Average potential per particle as a function of baryon number density for 
	an original CMF EOS and 2 CMF EOSs with an additional strong phase transition at high densities (\textit{left})~\cite{Steinheimer:2022gqb} as well as for numerous combinations of the CMF model with a flexible polynomial parametrization of the EOS at high densities (\textit{right})~\cite{OmanaKuttan:2022aml}.
	}
    \label{fig:UrQMD_EOS_parametrization}
\end{figure}
In the initial exploration~\cite{Steinheimer:2022gqb}, this approach has been used to test the effects of multiple EOSs, including in particular EOSs based on the Chiral Mean Field (CMF) model \cite{Papazoglou:1998vr,Steinheimer:2010ib,Motornenko:2019arp} (see the left panel in Fig.~\ref{fig:UrQMD_EOS_parametrization}), which uses an extensive list of degrees of freedom (including a complete list of all known hadrons as well as three light quark flavors and a gluon contribution) and describes the liquid-gas phase transition in ordinary nuclear matter, a transition between quarks and hadronic degrees of freedom, and chiral symmetry restoration introduced through parity doubling in the mean-field approximation~\cite{Motornenko:2020yme}. In a subsequent work~\cite{OmanaKuttan:2022aml}, the same framework has been used to explore effects of variations in the EOS above $n_B=2n_0$ by joining a CMF EOS, describing nuclear matter for $n_B < 2n_0$, with a seventh-degree-polynomial parametrization of the potential energy per baryon $\mathcal{V}$ at $n_B > 2n_0$,
\begin{align}
\mathcal{V}(n_B) = \sum_{i=1}^7 \theta_i \left( \frac{n_B}{n_0} - 1 \right)^i  + C~,
\end{align}
where $\theta_i$ are independently varied parameters and $C$ is a constant ensuring continuity at $n_B = 2n_0$ (see the right panel in Fig.~\ref{fig:UrQMD_EOS_parametrization}).

The above recent developments in modeling the EOS address only its density dependence, while heavy-ion collisions probe dense nuclear matter not only at various densities, but also at varying temperatures and isospin fractions. 
Moreover, as already mentioned above, the nuclear interaction also depends on momentum (or, equivalently, effective mass). 
Comprehensive studies of the EOS through comparisons of simulations with experimental data require development and implementation of potentials able to capture effects due to all of these factors.

The momentum-dependence of the EOS, which is expected to have a large effect on extractions of the EOS at all BES energies
, can be conveniently modeled by including scalar-type interactions, which in the mean-field approximation lead to the dependence of the EOS on scalar density. 
In the past, this has been mostly done by using relativistic mean-field models, such as the Walecka model~\cite{Blaettel:1993uz}, which, however, do not allow flexible parametrizations of the EOS.
A more flexible approach, based on the relativistic Landau Fermi-liquid theory, has been proposed in the past~\cite{Danielewicz:1998pb}.
More recently, following similar ideas, a generalization of the VDF model, based on a vector-scalar density functional (VSDF), has been developed~\cite{Sorensen:2021zxd} to enable maximally flexible parametrizations of the scalar-density-dependence of the EOS, and its implementation in transport simulations is the subject of an ongoing work.
Notably, scalar-type interactions additionally provide a non-trivial (i.e., beyond a thermal contribution of the ideal gas form) temperature dependence. 

The isospin-dependence of the EOS, often quantified in terms of the symmetry energy defined as the isospin-dependent contribution to energy per particle, is expected to be less important for regions of the phase diagram explored in BES. 
It is, however, crucial for experiments colliding unstable isotopes, such as those planned at FRIB~\cite{FRIB400,Sorensen:2023zkk,FRIB-TA_white_paper_motivations}.
Moreover, the symmetry energy is a key component of the EOS connecting nearly-symmetric nuclear matter found on Earth and the highly asymmetric nuclear matter relevant to studies of neutron stars, neutron star mergers, and supernovae~\cite{Sorensen:2023zkk,Lovato:2022vgq}
. 
Finally, while the effects of the symmetry energy may be small for collision energies $\snn \gtrsim 3~\rm{GeV}$ ($\ekin \gtrsim 2.9~\rm{GeV}$), they may still be measurable in the upcoming precision era for heavy-ion physics~\cite{Almaalol:2022xwv}. 
Currently, the prevailing approach to modeling the dependence of the EOS on the isospin fraction is to express the isospin-dependence of the single-particle potential in terms of the expansion coefficients of the symmetry energy around $n_0$~\cite{Baldo:2016jhp}, which essentially gives it a Skyrme-like form.
However, such simple phenomenological approaches may not correctly capture a number of important effects, including the very high-density behavior of the symmetry energy.
Therefore, it is worthwhile to consider ways in which isospin dependence can be flexibly and efficiently modeled in transport simulations.


\subsection{Constraints on the equation of state}
\label{sec:constraints_on_the_EOS}

Recent advances in the use of statistical methods as well as a greater availability of significant high-performance computing resources have brought about a qualitative change in using models of heavy-ion collisions to constrain the EOS. 
While in the past, due to limitations in computational power, comparisons of models with experimental data could only afford testing several different EOSs, nowadays it is possible to perform extensive Bayesian analyses based on simulations that explore tens or even hundreds of different EOSs which then provide input to event emulators.
In view of this, in the following we will largely concentrate on most recent results which take advantage of these new capabilities.

\subsubsection{Bayesian analysis of BES FXT proton directed and elliptic flow data}
\label{Bayesian_cs2_parametrization}

Recent work~\cite{Oliinychenko:2022uvy} performed a Bayesian analysis of proton collective flow data from fixed-target collisions at beam energies $\snn = \{ 3.0, 4.5\}~\rm{GeV}$ ($\ekin = \{2.9, 8.9\}~\rm{GeV/nucleon}$) as measured by the STAR experiment~\cite{STAR:2020dav,STAR:2021yiu}. 
Specifically, the analysis considered the slope of the $p_T$-integrated proton directed flow $dv_1/dy$ and the $p_T$-integrated proton elliptic flow $v_2$, both taken at midrapidity.
Employing a large family of significantly varying EOSs was made possible by using a flexible piecewise parametrization of the EOS~\cite{Oliinychenko:2022uvy}, described in Sec.~\ref{sec:modeling_the_EOS}. 
The family of EOSs was generated by varying independently three EOS parameters: the incompressibility at saturation $K_0$, the value of the speed of sound squared $c_s^2$ between 2 and 3 saturation densities $c_s^2[2,3]n_0$, and the value of $c_s^2$ between 3 and 4 saturation densities $c_s^2[3,4]n_0$; $c_s^2$ above $4n_0$ was taken to be equal $c_s^2[n_B>4n_0] = 0.3$ for all EOSs (an example of the effects of varying $c_s^2[2,3]n_0$ and $c_s^2[3,4]n_0$ is shown in Fig.~\ref{fig:VDF_cs2_parametrization}).

\begin{figure}[t]
    \centering
	\includegraphics[width=0.43\linewidth]{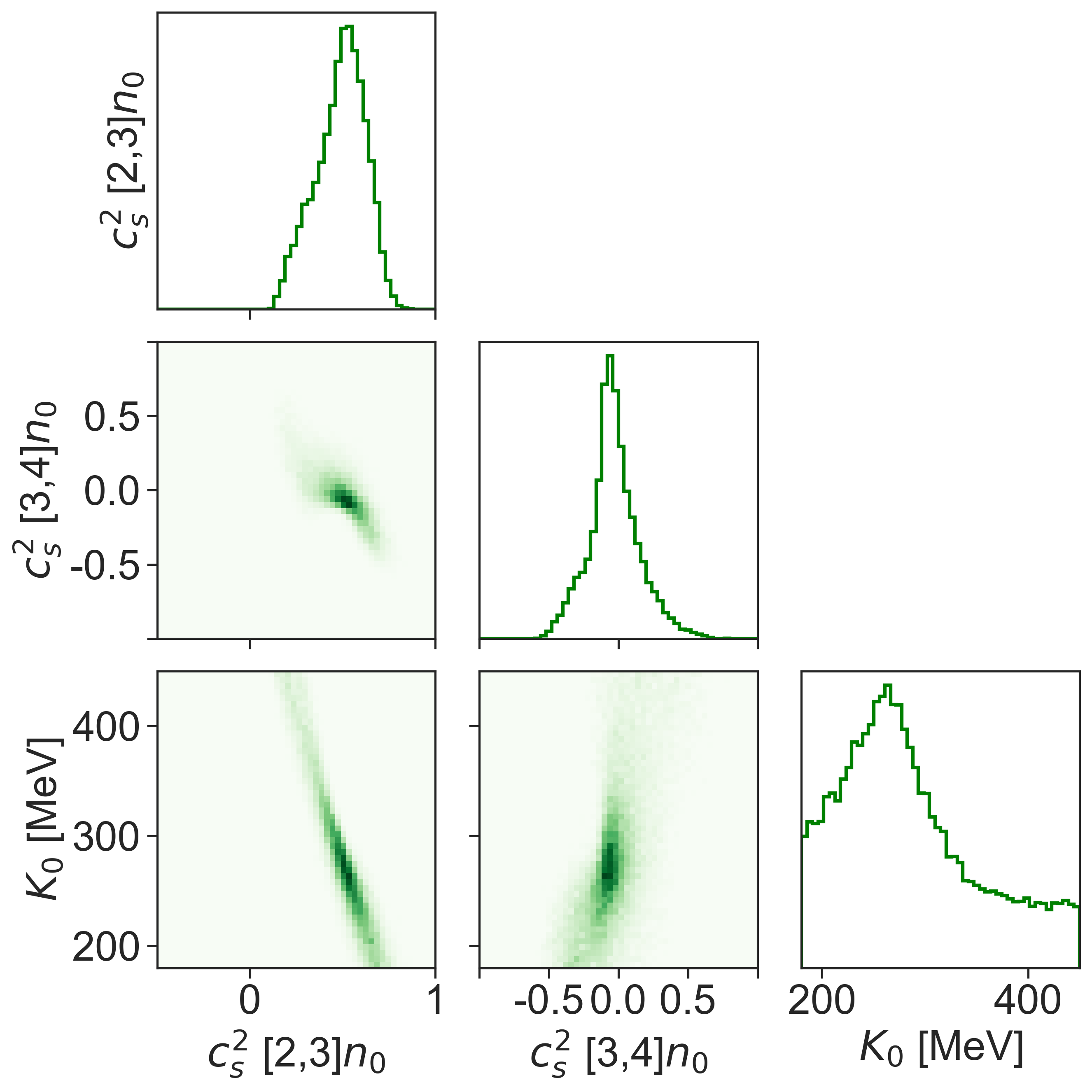}
    \includegraphics[width=0.56\linewidth]{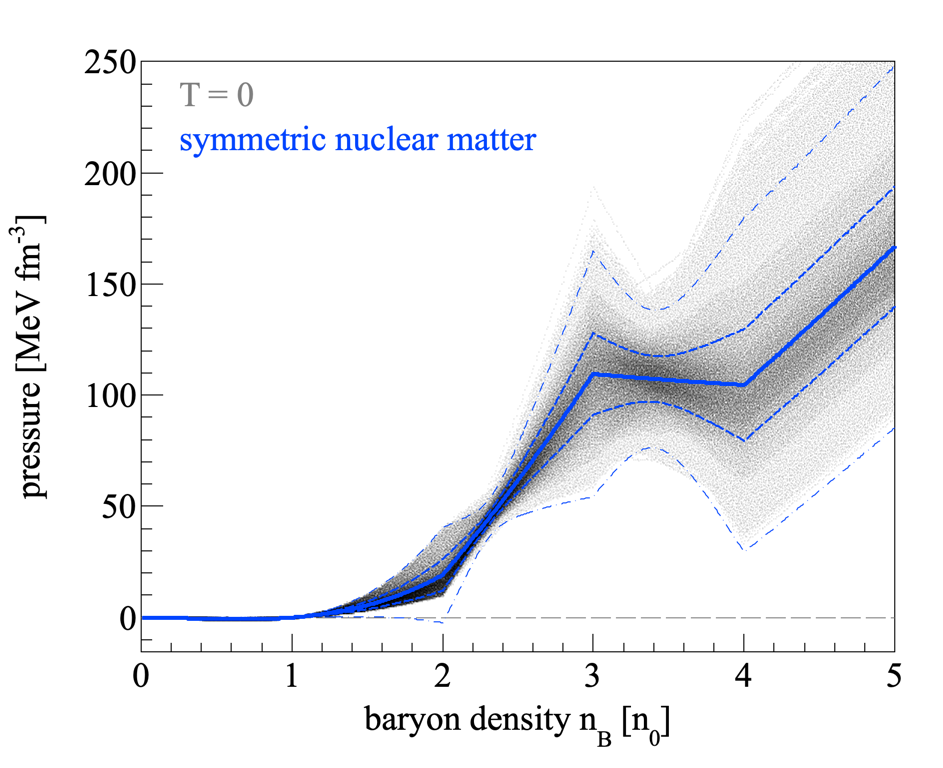}
    \caption{Results from a Bayesian analysis~\cite{Oliinychenko:2022uvy} of STAR experiment proton flow data at $\snn = \{3.0, 4.5 \}~\rm{GeV}$~\cite{STAR:2020dav,STAR:2021yiu}. \textit{Left:} Posterior parameter distribution for incompressibility at saturation $K_0$, the value of the speed of sound squared between 2 and 3 saturation densities $c_s^2[2,3]n_0$, and the value of the speed of sound squared between 3 and 4 saturation densities $c_s^2[3,4]n_0$ obtained using STAR proton flow measurements. The maximum \textit{a posteriori} probability (MAP) parameters are $K_0 = 285 \pm 67~\rm{MeV}$, $c_s^2[2,3]n_0 = 0.49 \pm 0.13$, $c_s^2[3,4]n_0 = -0.03 \pm 0.15$. \textit{Right:} Scatter plot of pressure at zero temperature as a function of baryon number density obtained by sampling the parameters of the EOS from the posterior distributions shown in the left panel. The solid line corresponds to the mean of the pressures, while the thick and thin dashed line correspond to the $\pm1\sigma$ and $\pm3\sigma$ contour around the mean, respectively. Figures from Ref.~\cite{Oliinychenko:2022uvy}.
	}
    \label{fig:VDF_cs2_Bayesian}
\end{figure}

The Bayesian analysis of simulations using the generated family of EOSs and experimental data yielded the following maximum \textit{a posteriori} probability (MAP) parameters: $K_0 = 285 \pm 67~\rm{MeV}$, $c_s^2[2,3]n_0 = 0.49 \pm 0.13$, $c_s^2[3,4]n_0 = -0.03 \pm 0.15$. 
Plots of posterior parameter distributions are shown in the left panel of Fig.~\ref{fig:VDF_cs2_Bayesian}, while a scatter plot of pressure at zero temperature against baryon density, obtained from sampling the EOS parameters from the posterior distribution, is shown in the right panel. 
For the first time, the obtained constraint on the EOS describes both the directed and elliptic proton flow data simultaneously~\cite{Oliinychenko:2022uvy}, a characteristic which was not achieved in the case of the previous state-of-the-art study and led to a comparatively broad constraint~\cite{Danielewicz:2002pu}. 
It is possible, however, that this was not only a result of performing the analysis with a much more flexible EOS, but also a consequence of using new data, given that there appear to be some discrepancies~\cite{Oliinychenko:2022uvy} between flow measurements~\cite{E895:1999ldn,E895:2000maf,E895:2001axb,E895:2001zms} performed by experiments at the Alternating Gradient Synchrotron (AGS) and measurements provided by the STAR experiment.

It is evident both from the values of the MAP parameters and from the resulting behavior of the pressure that the constrained EOS suggests a first-order phase transition for densities between 3 and 4 times the saturation density. 
However, some caution in interpreting the results is advised. 
As elaborated on in Sec.~\ref{sec:modeling_the_EOS}, momentum-dependence of nuclear interactions is a crucial part of modeling, which in particular for the energy range considered in the above study would result in an additional source of repulsion felt by nucleons interacting with a dense medium. 
In the absence of that (physically well-motivated) source of repulsion, describing the experimental data is possible by using an EOS which is relatively \textit{stiffer}, that is characterized by higher values of the speed of sound.
However, as can be seen from the correlation plot between $c_s^2[2,3]n_0$ and $c_s^2[3,4]n_0$ shown in the left panel of Fig.~\ref{fig:VDF_cs2_Bayesian}, a stiffer EOS at densities $n_B \in [2,3]n_0$ prefers a softer EOS for densities $n_B \in [3,4]n_0$. 
In other words, the significant softening of the EOS at high densities, preferred by the results of this Bayesian analysis, may not be a robust prediction given that the used model misses a crucial component of nuclear interactions which likely affects the extracted constraints on the low- and medium-density EOS parameters.
Since the momentum dependence itself likewise needs to be better constrained, especially at high densities, this discussion calls for an improved Bayesian analysis of available data where both the density- and the momentum-dependence of the nuclear EOS are varied.

\subsubsection{Bayesian analysis of proton transverse kinetic energy and elliptic flow data}
\label{Bayesian_Omana_Kuttan}

Another recent study~\cite{OmanaKuttan:2022aml} performed a Bayesian analysis of proton mean transverse kinetic energy $\langle m_T \rangle - m_0$, based on data spanning collision energies $\snn \in [3.83, 8.86]~\rm{GeV}$ ($\ekin \in [5.9, 40.0]~\rm{GeV/nucleon}$)~\cite{E802:1999hit,NA49:2006gaj,STAR:2017sal}, and proton elliptic flow $v_2$ at midrapidity, based on data spanning collision energies $\snn \in [2.24, 4.72]~\rm{GeV}$ ($\ekin \in [0.8, 10.0]~\rm{GeV/nucleon}$)~\cite{E895:1999ldn,CERES:2002eru,FOPI:2004bfz,STAR:2012och,STAR:2020dav,STAR:2021yiu,HADES:2020lob}.
This study used a flexible parametrization of the EOS, described in Sec.~\ref{sec:modeling_the_EOS}, in which a CMF EOS is used for densities below $2n_0$, while the potential energy per baryon at densities $n_B > 2n_0$ is given by a seventh-degree-polynomial with variable parameters.
This led to a sizable family of EOSs with varying behavior at high densities, as shown explicitly in the right panel of Fig.~\ref{fig:UrQMD_EOS_parametrization}.

\begin{figure}[t]
    \centering
	\includegraphics[width=0.99\linewidth]{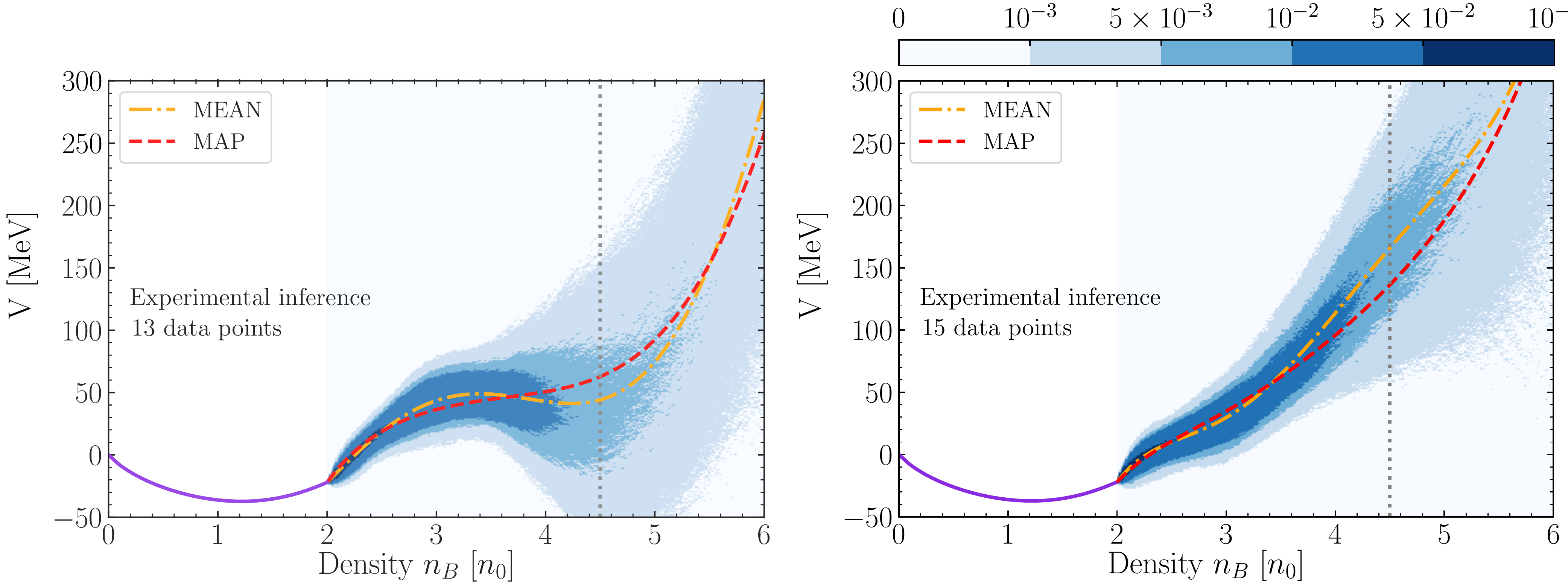}
    \caption{Posterior distributions for the average potential energy per baryon $\mathcal{V}$ obtained from Bayesian analysis~\cite{OmanaKuttan:2022aml} using experimental data for the proton elliptic flow $v_2$ and proton mean transverse kinetic energy $\langle m_T \rangle - m_0$. The purple line shows the CMF EOS, while the red dashed and orange dot-dashed lines show the EOS corresponding to maximum \textit{a posteriori} parameters (MAP) and the mean EOS obtained by averaging the values of sampled potentials at different densities, respectively. The vertical gray line marks the highest average central density achieved in collisions at $\snn = 9 ~\rm{GeV}$ ($\ekin = 41~\rm{GeV/nucleon}$). \textit{Left:} Posterior obtained based on a data set of 13 experimental points excluding values of $\langle m_T \rangle - m_0$ for $\snn = \{3.83, 4.29\}~\rm{GeV}$ ($\ekin = \{5.9, 7.9  \}~\rm{GeV/nucleon}$). \textit{Right:} Posterior obtained based on a data set of all 15 considered experimental points. Figures from Ref.~\cite{OmanaKuttan:2022aml}.
	}
    \label{fig:UrQMD_Bayesian_V_posterior}
\end{figure}
Interestingly, the study explored the influence of either excluding or including constraints based on the mean transverse kinetic energy at the two lowest collision energies, $\snn = \{3.83, 4.29\}~\rm{GeV}$ ($\ekin = \{5.9, 7.9 \}~\rm{GeV/nucleon}$), resulting in two Bayesian analyses constrained by 13 or 15 experimental data points, respectively.
This led to constraints on the potential energy per baryon $\mathcal{V}$ with divergent behavior at large densities, $n_B \gtrsim 3n_0$, as can be seen in Fig.~\ref{fig:UrQMD_Bayesian_V_posterior}.
This result is especially surprising given that sizable differences occur at high densities, that is in a region which should not be significantly affected by the excluded or included data points measured at energies which mostly probe densities $n_B \lesssim 3n_0$ (see Fig.~\ref{fig:VDF_cs2_sensitivity}).

The study then performs an extensive analysis of reasons behind this discrepancy, and suggests either a tension between the $\langle m_T \rangle - m_0$ and $v_2$ data or a shortcoming of the model used for simulations, including a possible need for an explicitly temperature-dependent EOS at higher beam energies where contributions from mesonic degrees of freedom may become dominant.
(Additionally, as in the case of work described in the previous section, this study misses the momentum dependence of the nucleon potential which may lead to temperature-dependent effects.)
Based on the study's closure tests with 13 and 15 data points, one can also surmise that the large discrepancy at high densities may be caused by a small deviation at lower densities (caused by the exclusion or inclusion of the two data points) which is then propagated to higher densities. 
This could be an artifact of constraining the EOS in a polynomial form, which is known for exhibiting uncontrolled behavior away from the region in which the polynomial is fit, and is further supported by the study's conclusion that the constraining power of the chosen data set at densities $n_B > 4n_0$ is weak (agreeing with the study of flow sensitivity to the EOS shown in Fig.~\ref{fig:VDF_cs2_sensitivity}).
Finally, the study compares simulation results with experimental data for the slope of the directed flow $dv_1/dy$ and finds a better agreement for the EOS obtained using 15 data points.

\subsubsection{Prominent constraints from flow observables and discussion}
\label{sec:constraints_and_discussion}

The two  
recent Bayesian analysis studies described above used proton flow observables as the basis for their constraints on the dense nuclear matter EOS.
Historically, flow was likewise often an observable of choice in endeavors aimed at constraining the EOS from heavy-ion collision data.

\begin{figure}[t]
    \centering
	\includegraphics[width=0.8\linewidth]{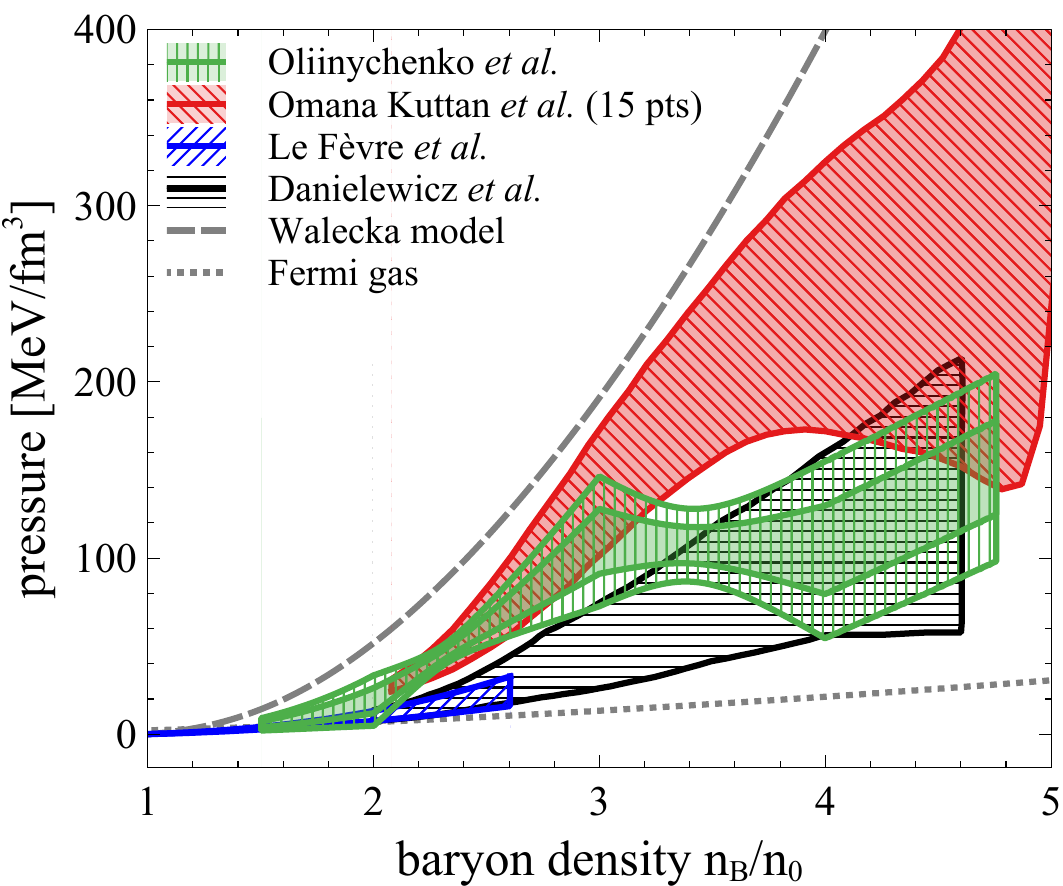}
    \caption{A compilation of selected constraints on the EOS based on comparisons of experimental data on flow observables with microscopic transport simulations carried out using \texttt{pBUU} (region with black horizontal stripes, labeled by ``Danielewicz \textit{et al.}'')~\cite{Danielewicz:2002pu}, \texttt{IQMD} (region with blue backward stripes, labeled by ``Le F\`{e}vre \textit{et al.}'')~\cite{LeFevre:2015paj}, \texttt{SMASH} (region with green vertical stripes, labeled by ``Oliinychenko \textit{et al.}'', where the shaded (unshaded) region corresponds to the 68\% (95\%) confidence interval)~\cite{Oliinychenko:2022uvy}, and \texttt{UrQMD} (region with red forward stripes, labeled with ``Omana Kuttan \textit{et al.}~(15 pts)'', corresponding to the 68\% confidence interval for the analysis using 15 input data points)~\cite{OmanaKuttan:2022aml}.
	}
    \label{fig:flow_constraints_compilation}
\end{figure}
A compilation of well-known constraints on the EOS based on comparisons of experimental data on flow observables with microscopic transport simulations, including the recent studies, is shown in Fig.~\ref{fig:flow_constraints_compilation}.
A prominent constraint on the EOS for baryon densities between $(2$--$4.5) n_0$ was obtained~\cite{Danielewicz:2002pu} by comparing measurements of collective flow from heavy-ion collisions~\cite{Gustafsson:1988cr,EOS:1994kku,E877:1997zjw,E895:2000maf} at center-of-mass energies $\snn = 1.95$--$4.72~\rm{GeV}$ ($\ekin = 0.15$--$10~\rm{GeV/nucleon}$) with results from \texttt{pBUU}~\cite{Danielewicz:1991dh,Danielewicz:1999zn} hadronic transport simulations using Skyrme-type EOSs with different values of the incompressibility at saturation density $K_0$ (see Sec.~\ref{sec:modeling_the_EOS}).
The study results indicate that the symmetric nuclear matter EOS lies between Skyrme-like EOSs characterized by $K_0 =210~\rm{MeV}$ and $K_0 = 300~\rm{MeV}$ (see the region with black horizontal stripes, labeled by ``Danielewicz \textit{et al.}'').  
Subsequently, the elliptic flow data measured at $\snn = 2.07$--$2.52~\rm{GeV}$ ($\ekin = 0.4$--$1.5~\rm{GeV/nucleon}$) by the FOPI collaboration~\cite{FOPI:2011aa} were used together with simulations from Isospin Quantum Molecular Dynamics (\texttt{IQMD})~\cite{Aichelin:1991xy,Hartnack:1997ez} to constrain the incompressibility in a Skyrme-like EOS at $K_0 = 190 \pm 30~\rm{MeV}$~\cite{LeFevre:2015paj}, likewise suggesting a comparatively soft EOS (see the region with blue backward stripes, labeled by ``Le F\`{e}vre \textit{et al.}'').
Also shown are results from the studies described in Sec.~\ref{Bayesian_cs2_parametrization} (see the region with green vertical stripes, labeled by ``Oliinychenko \textit{et al.}'', where the shaded (unshaded) region corresponds to the 68\% (95\%) confidence interval)~\cite{Oliinychenko:2022uvy} and in Sec.~\ref{Bayesian_Omana_Kuttan} (region with red forward stripes, labeled with ``Omana Kuttan \textit{et al.}~(15 pts)'', corresponding to the 68\% confidence interval for the analysis using 15 input data points)~\cite{OmanaKuttan:2022aml}.

The comparison of EOSs constraints presented in Fig.~\ref{fig:flow_constraints_compilation} once again underscores the need to include momentum dependence of the potentials in simulations. 
The two older results suggest a softer EOS at small and moderate densities, compared to the results of the two Bayesian analyses, which can be attributed to the fact that the frameworks used in the former both included momentum dependence. 
On the other hand, it is apparent that the newer studies have the capability to describe a much more varied behavior of the EOS with density.
The task for future studies is to merge these two features within one framework, allowing for a simultaneous comprehensive constraint on density- and momentum-dependence of the EOS, followed by similar extensions to include isospin-dependence, in-medium cross-sections, or cluster production mechanisms.
The need for tighter and more reliable constraints is evident especially at high densities.

\begin{figure}[t]
    \centering
	\includegraphics[width=0.99\linewidth]{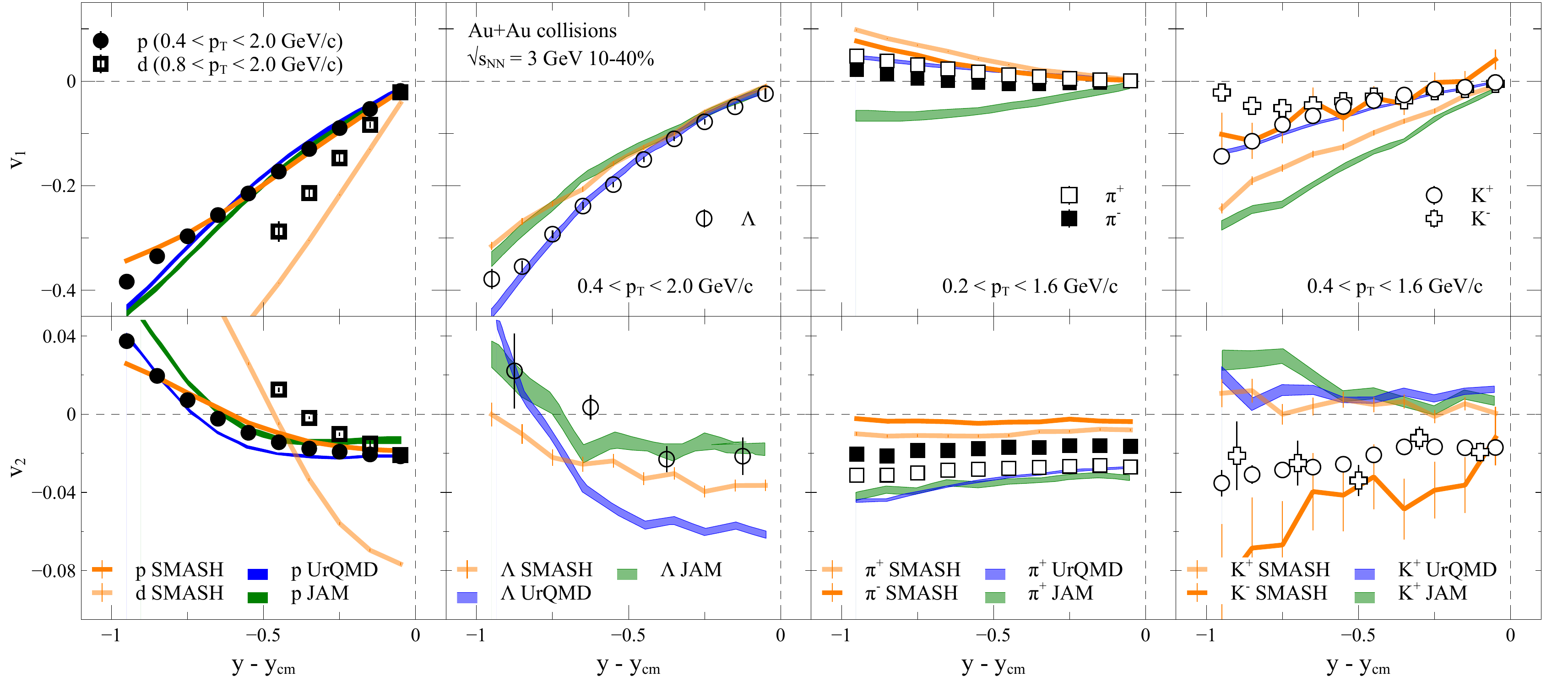}
    \caption{Directed ($v_1$, \textit{top}) and elliptic ($v_2$, \textit{bottom}) flow, integrated over $p_T$, of protons and deuterons (\textit{left panels}), lambda baryons (\textit{center-left panels}), pions (\textit{center-right panels}), and kaons (\textit{right panels}) as a function of rapidity. Experimental data points (symbols) are STAR measurements~\cite{STAR:2021yiu} performed with Au+Au collisions at $\snn=3~\rm{GeV}$ ($\ekin = 2.9~\rm{GeV/nucleon}$) and 10–40\% centrality. Bands represent results from microscopic transport models: \texttt{UrQMD} with an EOS characterized by $K_0 = 380~\rm{MeV}$ (blue bands)~\cite{STAR:2021yiu}, \texttt{JAM} with an EOS characterized by $K_0 = 380~\rm{MeV}$ (green bands)~\cite{STAR:2021yiu}, and \texttt{SMASH} with an EOS characterized by $K_0 = 300~\rm{MeV}$ at moderate densities and a significant softening at high densities~\cite{Oliinychenko:2022uvy}.
	}
    \label{fig:flow_results_many_baryon_species}
\end{figure}
Moreover, given the complex evolution of heavy-ion collisions in this energy regime, a reliable constraint should describe not only proton observables.
To underscore the wildly varying degrees to which simulations can describe different observables, Fig.~\ref{fig:flow_results_many_baryon_species} shows a comparison of flow observables for different species as measured by the STAR experiment~\cite{STAR:2021yiu} and as obtained in three simulation frameworks: \texttt{UrQMD}, \texttt{JAM}, and \texttt{SMASH}.
For results shown in the figure, both \texttt{UrQMD} and \texttt{JAM} simulations use an EOS characterized by $K_0 = 380~\rm{MeV}$, while \texttt{SMASH} simulations use a \mbox{$c_s^2$-parametrized} EOS characterized by $K_0 = 300~\rm{MeV}$, $c_s^2[2,3]n_0 = 0.47$, and $c_s^2[3,4]n_0 = -0.08$, sampled from the posterior distribution obtained~\cite{Oliinychenko:2022uvy} by fitting STAR proton flow data (see Sec.~\ref{Bayesian_cs2_parametrization}), including the data shown in the figure.
Note that while all simulations describe the proton flow reasonably well, with \texttt{SMASH} showing the best agreement (unsurprisingly, since the used EOS has been constrained to describe this data), flow observables for other particle species are consistently rather poorly described by all of the codes.
This suggests that the simulations must either miss or incorrectly treat components necessary to describe the entire system, and calls for further studies of and developing simulation algorithms for, among others, deuteron production, strange interactions, and meson interactions.
Only when multiple observables of importance are well-described, or when deviations from experimental data can be reliably ascribed to elements of the simulations which do not significantly affect the EOS extraction, can constraints on the EOS be accepted at their face value.

%% file: fluctuations.tex
The description of the heavy-ion collisions is inherently
statistical. Therefore, fluctuations are an essential part of that
description. In this section we describe the status and the recent progress in
understanding the physics of fluctuations, both in and out of
equilibrium, with a particular focus on the role fluctuations play in
mapping the QCD phase diagram and the search for the critical point.

\subsection{Fluctuations in thermodynamics, critical fluctuations}

\subsubsection{Relation between equation of state and
  fluctuations}

In thermodynamic equilibrium (by definition) the probability of the
system to be found in a given {\em microscopic}, i.e., quantum, state
depends only on the conserved quantum numbers of this state, such as
energy, momentum, charge, etc. Therefore, the probability of a given
{\em macroscopic} state, characterized by a given set of macroscopic
variables, such the energy, is proportional to the number of
microscopic states with these values of the macroscopic variables.
That probability is equal to the exponential of the entropy, by
definition. Therefore, the dependence of the entropy on energy, etc,
i.e., the equation of state (EOS), determines the probability of the
fluctuations for a system in thermodynamic equilibrium
\cite{landau2013statistical}.

This fundamental relationship between fluctuations and
EOS has been exploited by Einstein \cite{Einstein1910} to
describe the singular behavior of the density fluctuations near
critical points -- the origin of the phenomenon of critical
opalescence.

The entropy in question is the entropy of the open system
which can exchange conserved quantities, such as energy and
charge with the surroundings (thermodynamic bath) at temperature $T$
and chemical potential $\mu$. The corresponding probability
distribution for the values of energy density $\epsilon$ and charge
density $n$ is given by
\begin{equation}
  \label{eq:P=eS}
  \mathcal P(\epsilon,n) \sim \exp\left\{
   V[s(\epsilon,n)-\beta\epsilon + \alpha n]   
      \right\}
\end{equation}
where $\beta=1/T$, $\alpha=\mu/T$ and $V$ is the volume of the
thermodynamic system. In the thermodynamic limit,
i.e., for large $V$, the probability is sharply peaked around the
maximum determined by the familiar conditions:
\begin{equation}
  \label{eq:ds/de=b}
  \left(\frac{\partial s}{\partial\epsilon}\right)_{\!n}-\beta=0,
  \quad
  \left(\frac{\partial s}{\partial n}\right)_{\!\epsilon}+\alpha=0.
\end{equation}
I.e., the entropy $s(\epsilon, n)$ determines the relationship between
conserved densities $\epsilon$, $n$ and the corresponding
thermodynamic quantities $T$ and $\mu$.

The quadratic form of second derivatives of the entropy must be
negative to ensure thermodynamic stability. One can show that this
form is diagonalized using variables $m\equiv s/n$ (specific entropy)
and $p$ (pressure). Probability of small fluctuations in terms of
these variables is given by
\begin{equation}
  \label{eq:Pdmdp}
  \mathcal P \sim \exp\left\{ -
  \frac V2 \left(
    \frac{n^2}{c_p}(\delta m)^2
    +   \frac{\beta}{wc_s^2}(\delta p)^2
    \right)\right\}
\end{equation}

The specific heat $c_p=Tn(\partial m/\partial T)_{p}$ diverges at the
critical point. This corresponds to the probability of fluctuations
developing a ``flat direction'' along $\delta p=0$, where the
fluctuations of the specific entropy
$V\langle(\delta m)^2\rangle=c_p/n^2$ become large.\footnote{Other
  thermodynamic quantities also develop large fluctuations at the
  critical point. For example, $V\langle(\delta n)^2\rangle=\chi_T T$,
  where $\chi_T=(\partial n/\partial\mu)_T$ is also divergent.  The
  specific entropy $m=s/n$ is special from the hydrodynamic point of
  view. Unlike, e.g., baryon density, $m=s/n$ is a normal hydrodynamic
  mode in ideal hydrodynamics, i.e., being a ratio of conserved densities
  specific entropy is a diffusive mode which
  does not mix with propagating sound modes.}
The non-monotonic behavior of fluctuations as the critical point is
being approached and passed during the QCD phase diagram scan has been
proposed as a signature of the QCD critical point \cite{Stephanov:1998dy,Stephanov:1999zu}.

In this section we focus on the fluctuations intrinsic in any system
which affords local thermodynamic, statistical description.  These
fluctuations are determined by the equation of state, as discussed
above.  In the context of heavy-ion collisions we must distinguish
these fluctuations from the fluctuations which are determined by the
initial state of the system. For example, from the fluctuations of the
initial geometry of the system. In experiments, such separation is not
always trivial. Typically it involves selecting collisions with
similar centrality, i.e., similar collision geometry. The effects of
the initial fluctuations are also qualitatively different from those
of thermodynamic fluctuations in that the correlations induced by
initial fluctuations are longer range (in longitudinal rapidity space)
than the thermodynamic fluctuations we discuss.\cite{Kapusta:2011gt}
Most importantly, the $\snn$ dependence of the initial fluctuations does not
reflect the {\em non-monotonicity\/} inherent in the thermodynamic
fluctuations in the vicinity of the critical point.

\subsubsection{Universality and non-gaussianity of critical  fluctuations}

Since the equation of state is universal near critical points, the
fluctuation phenomena are also universal. In this section we shall
describe the universal properties of the fluctuations which are
relevant for the search for the QCD critical point in heavy-ion collisions.

Since the coefficient of the $(\delta m)^2$ term in
Eq.~(\ref{eq:Pdmdp}) vanishes at the critical point, non-gaussian terms
in Eq.~(\ref{eq:Pdmdp}) ($\delta m^3$, etc) become important. This
makes non-gaussianity of fluctuations a telltale signature of the
critical point.

An important, and related, consequence of the vanishing of the
coefficient of $\delta m^2$ is the divergence of the correlation
length $\xi$ of the fluctuations of $m$. In mean-field approximation
$\langle(\delta m)^2\rangle\sim\xi^2$. The divergence of $\xi$ leads
to the breakdown of the mean-field approximation.\footnote{The
  approximation is valid in the thermodynamic limit, i.e.,
  $V/\xi^3\to\infty$. This limit is in conflict with $\xi\to\infty$. } The probability can
no longer be considered as a function of the homogeneous (mean) value
of $\delta m$. It is a functional of the field $\delta m(x)$.
The corresponding scalar field theory, in the limit of the divergent
correlation length, is universal. Any liquid-gas critical point is
described by that theory, as well as other transitions characterized
by a single-component order parameter, such as uniaxial (Ising)
ferromagnets. The full details of the universal theory of  critical
phenomena are beyond the scope of this review, and are covered in many
textbooks and reviews on critical phenomena.\cite{Pelissetto:2000ek}
Here we shall
emphasize only the basic properties which are most relevant for the critical point
search in the beam energy scan experiments.

The theory of the QCD critical point fluctuations is in the
universality class of the Ising model, i.e., $\phi^4$ single-component
scalar field theory in three dimensions. The mapping of variables
between critical region of QCD and the critical region of the Ising
model has been standardized~\cite{Parotto:2018pwx} using 6
parameters: $T_c$, $\mu_{Bc}$, $w$, $\rho$, $\alpha_1$ and
$\alpha_2$. The parameters $T_c$ and $\mu_{Bc}$ set the location of
the QCD critical point, while $\alpha_1$ is the angle of the slope of
the coexistence line (first-order phase transition line) at the
critical point in the $T$ vs $\mu_{B}$ plane. It is also the slope of
$m={\rm const}$ line at the critical point and is obtained by mapping
the zero magnetization line (i.e., zero magnetic field $h=0$) of the
Ising model onto the QCD phase diagram. The angle $\alpha_2$ is the
angle of the line on the QCD phase diagram onto which the constant
temperature line passing through the Ising critical point maps.

The nongaussian cumulants of the order parameter such as $\delta m$
are the same as in the Ising model, but mapped into the QCD phase
diagram. In particular, these cumulants contain singular contributions
which diverge at the critical point
with universal powers of the correlation length, given approximately
by~\cite{Stephanov:2008qz}
\begin{equation}
  \label{eq:cumulants-xi}
  \Delta \langle(\delta m)^3 \rangle\sim \xi^{4.5},\quad
  \Delta \langle(\delta m)^4 \rangle^c \sim \xi^{7}
\end{equation}
where $\Delta$ reminds us that this is a contribution to cumulants,
singular at the critical point. In practice, it does not have to be
dominant. The search for the critical point is aimed at detecting the
non-monotonic dependence of fluctuations measures on the collision
energy $\snn$ as the critical point is approached and passed, i.e.,
as the correlation length increases and then shrinks back to
non-critical background, or baseline, values.

Unlike the quadratic cumulant $\langle(\delta m)^2\rangle$ which,
being a measure of the width of the probability distribution, is
positive, the cubic, and especially, the quartic cumulant could have
either sign,
since non-gaussian cumulants describe the {\em shape} of
the distribution, or, more precisely, its deviation from Gaussian.

\begin{figure}[t]
  \centering
  \includegraphics[height=12em]{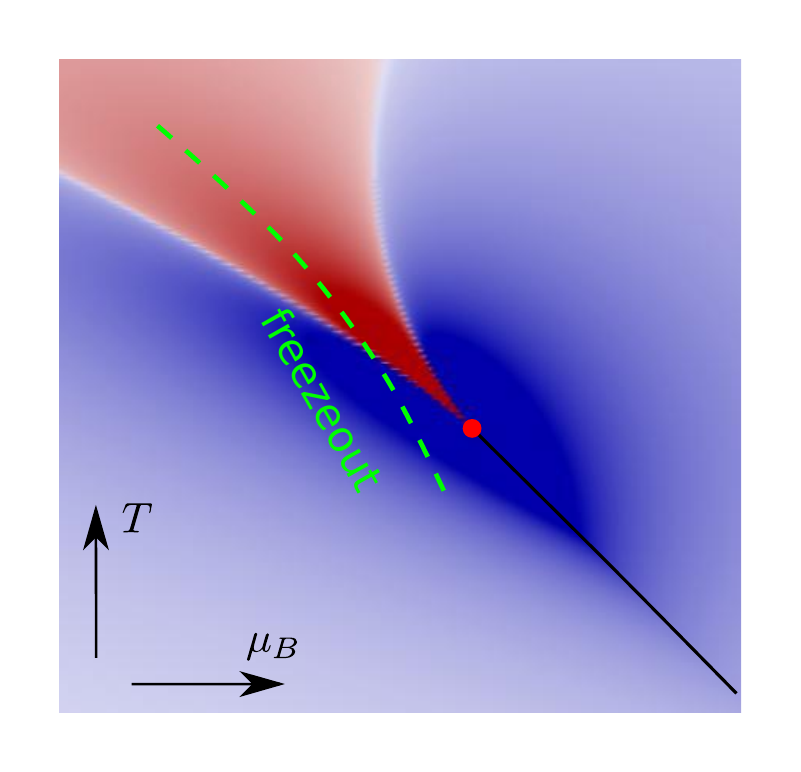}
  \hskip 2em
    \includegraphics[height=12em]{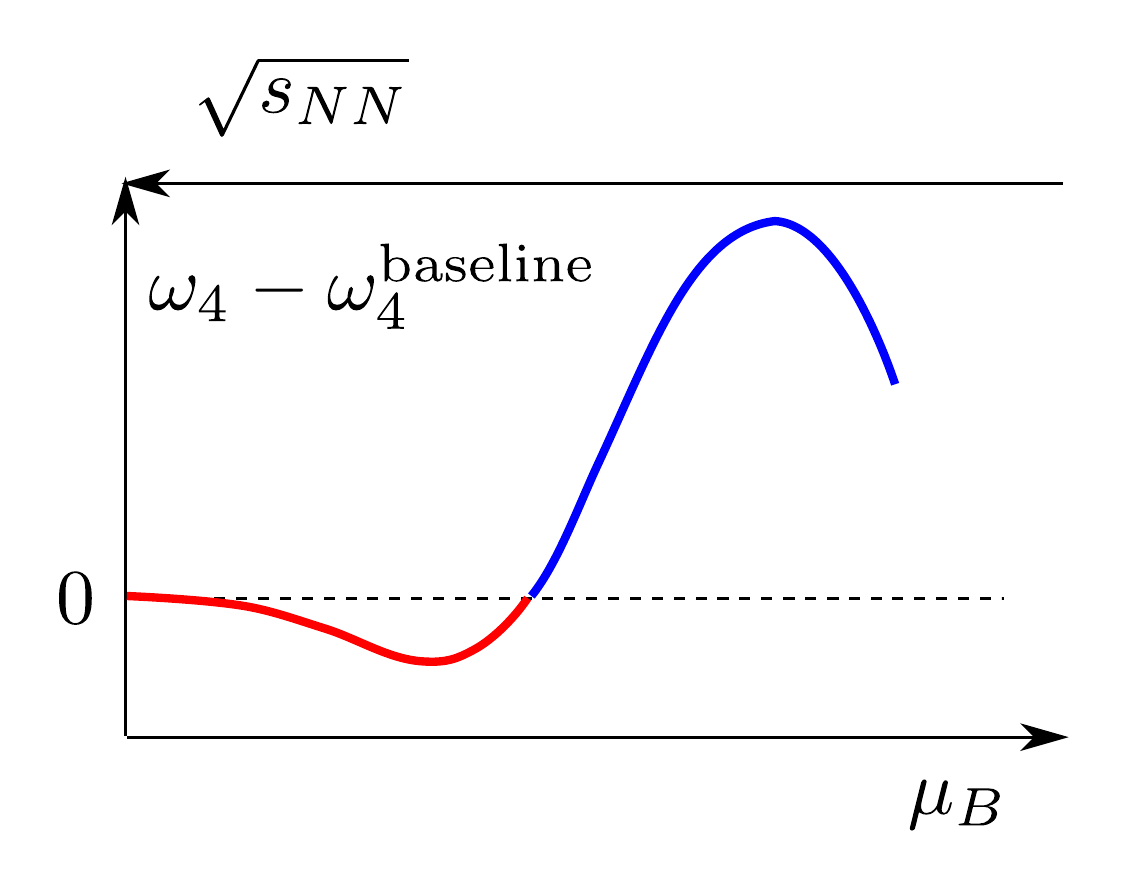}\\
    (a)\hskip 15em (b)
    \caption[]{The equilibrium expectation for the quartic cumulant of
      fluctuations as a function of temperature and baryon chemical
      potential on the QCD phase diagram in the vicinity of the
      critical point. Red and blue colors reflect the sign of the
      cumulant -- negative and positive respectively. The sign changes
      as the QCD phase diagram is scanned by varying $\snn$, and
      thus $T$ and $\mu$, along the freeze-out ``trajectory'' (dashed
      green line).}
    \label{fig:Tmu-kappa4}
\end{figure}

For example, the quartic cumulant around the QCD critical
point is illustrated in
Fig.~\ref{fig:Tmu-kappa4}(a).~\cite{Stephanov:2011pb}
The resulting
dependence on the collision energy, as it is varied in the region where
the freeze-out occurs near the critical point, is illustrated in
Fig.~\ref{fig:Tmu-kappa4}(b). The characteristic non-monotonicity of
this cumulant is one of the signatures of the critical point searched
for in the Beam Energy scan experiments.\cite{Bzdak:2019pkr}

The most recently published experimental measurements of
quartic cumulants by STAR collaboration indicate a non-monotonic
dependence of the type similar to the one shown in
Fig.~\ref{fig:Tmu-kappa4}(b) (see, e.g., Fig.4(2) in Ref.\cite{STAR:2020tga}). While the magnitude of the cumulant
sensitively depends on the EOS, and thus hard to predict (hence the
lack of the vertical scale on Fig.~\ref{fig:Tmu-kappa4}(b)) the
experimental results indicate that monotonic dependence is excluded at
3.1$\sigma$ level.\cite{STAR:2020tga,STAR:2021iop,STAR:2021fge}
While not yet a definitive indication of the
presence of the QCD critical point, this intriguing result motivates
the second phase of the beam energy scan (BES-II)
program which has collected higher
statistics data being processed and analyzed currently.

In this review we shall focus on the recent
progress achieved in connecting these equilibrium fluctuations to the
experimental measurements. There are at least two intermediate steps
between the equilibrium thermodynamic fluctuations and the
experimental observables:

First, like thermodynamic variables
themselves, their fluctuations evolve. This evolution is subject to
conservation laws which may lead to significant memory effects, i.e.,
the lag of the fluctuations behind the equilibrium values.

Second, the thermodynamic/hydrodynamic quantities are not measured by
experiments directly. An important step is connecting the fluctuations
of these quantities to the observables,
such as measures of particle multiplicity fluctuations and correlations.

The
following two sections will review the recent developments in
these two steps connecting theory and experiment.

\subsection{Fluctuations in hydrodynamics}

\subsubsection{Stochastic hydrodynamics}

There has been considerable recent progress in understanding and
describing the {\em dynamics} of local thermodynamic fluctuations in an
evolving medium making up a heavy-ion collision fireball.

There are two complementary approaches, which we shall refer to as
stochastic and deterministic. Both begin with the Landau-Lifshits
theory of hydrodynamic fluctuations\cite{Landau:2013stat2}. In this
theory, generalized to relativistic context in
Ref.\cite{Kapusta:2011gt} , the equations of hydrodynamics are
{\em stochastic}. The evolution of the energy and momentum densities are
described by the conservation law, as usual,
\begin{equation}
  \partial_\mu\breve T^{\mu\nu}=0,
\end{equation}
where the components of the stress-energy tensor $\breve T^{\mu\nu}$
are stochastic variables (as indicated by the 'breve' accent). Since
there are only four independent hydrodynamic variables, corresponding
to conserved energy and momentum densities, the other six components
of $\breve T^{\mu\nu}$ are local functions of these four.\footnote{For simplicity, in this discussion, we do not consider other
hydrodynamic variables, such as baryon density. With an additional
equation for baryon current conservation the extension to five (or
more, if necessary) hydrodynamic variables is straightforward.\cite{An:2019osr} } As usual,
the independent variables are chosen to be the rest energy
$\breve\epsilon$ and four velocity $\breve u$ of the local fluid
element and are related to $\breve T^{\mu\nu}$ by the
condition similar to Eq.~(\ref{eq:flow_landau}):
\begin{equation}
  \label{eq:Teu}
  \breve T^\mu_\nu\breve u^\nu = \breve\epsilon\breve u^\mu.
\end{equation}
The stress tensor in the local rest frame is then related to these
independent variables by a local relationship, containing only spatial
gradients in the local rest frame. This relation is stochastic, and
contains a contribution of the local (point-like correlated on the
hydrodynamic scale) noise $\breve S^{\mu\nu}$:
\begin{equation}
  \label{eq:TTS}
  \breve T^{\mu\nu} = T^{\mu\nu}(\breve\epsilon,\breve u)+\breve S^{\mu\nu}
\end{equation}
where $T^{\mu\nu}(\epsilon,u)$ is the usual constituent relationship between the
components of $T^{\mu\nu}$ and $\epsilon$ and $u$ and their gradients,
usually expanded to first order in gradients: 
\begin{equation}
  \label{eq:Teu}
  T^{\mu\nu}(\epsilon,u) = \epsilon u^\mu u^\nu
  - p(\epsilon) (g^{\mu\nu}-u^\mu u^\nu)
  +  \mbox{viscous gradients}\,.
\end{equation}
 To complete this system
of stochastic equations the correlation function
\begin{equation}
\langle\breve
S^{\mu\nu}(x_1)\breve S^{\alpha\beta}(x_2)\rangle\sim
\delta^{(4)}(x_1-x_2)\label{eq:SS-delta}
\end{equation}
is given by
the fluctuation dissipation theorem, ensuring that the equilibrium
fluctuations and correlations have the correct amplitudes in agreement
with thermodynamics. 

The evolution of fluctuations can be then described by, for example,
directly simulating this system of stochastic equations. This,
however, produces a problem (also known as infinite noise
problem) due to the fact that the noise is singular at $x_1=x_2$ in
Eq.~(\ref{eq:SS-delta}). The resulting solutions are dependent on the
hydrodynamic cutoff, i.e., the finite elementary hydrodynamic cell
size, $b$, complicating the ``continuum limit'' $b\to0$. Some 
solutions to this problem within a numerical simulation
have been proposed and implemented in the
literature \cite{Singh:2018dpk,Chattopadhyay:2023jfm}.

\subsubsection{Deterministic approach to hydrodynamic fluctuations}

The approach which deals with the infinite noise problem {\em before}
the actual numerical simulation is performed has been also developed
recently for Bjorken
flow\cite{Akamatsu:2017,Akamatsu:2018,Martinez:2018}, for arbitrary
relativistic flow\cite{Stephanov:2017ghc,An:2019osr,An:2019csj}, and for non-gaussian fluctuations\cite{An:2020vri,An:2022jgc}.
In this approach one expands in fluctuations around the
average, e.g.,
\begin{equation}
  \label{eq:deltaepsilon}
  \breve \epsilon = \epsilon + \delta\epsilon,\quad
  \breve u = u + \delta u\,,
 \end{equation}
thus obtaining stochastic equations for the
fluctuations $\delta\epsilon$, $\delta u$ on the deterministically
evolving averaged background $\epsilon(x)$ and $u(x)$.
These equations can then be used to derive {\em deterministic}
equations obeyed by the {\em correlation functions} of the fluctuations,
i.e., by averages of the products of the fluctuations, such as $\langle\delta\epsilon(x_1)\delta\epsilon(x_2)\rangle$. 

The short-distance singularity of the noise results in ultraviolet
divergences in the deterministic equations -- the infinite noise
problem. These divergent contributions modify the hydrodynamic
equations by {\em local} terms which can be ``renormalized'' away by
redefining the equation of state and transport coefficients in the
constitutive equations. The resulting system of renormalized
equations for hydrodynamic evolution of the renormalized average
densities as well as the correlation functions is ultraviolet finite,
i.e., cutoff independent, and the continuum limit can be taken.

After the renormalization the averaged hydrodynamic equations keep the
form of usual hydrodynamic equations to first order in gradients with
renormalized, i.e., physical, equation of state and transport
coefficients. However, the finite (i.e., cutoff independent)
fluctuation contributions appear beyond that order. These terms
introduce contributions
non-local in space (effectively being of order 3/2 in gradients) and also
non-local in time, leading to the phenomena known as ``long-time tails''.

The fluctuation correlators are expressed in terms of the
Wigner-transformed equal-time correlation functions. For a fluid at
rest globally, this definition is straightforward:
\begin{equation}
  \label{eq:WAB}
  W_{ab}(t,\bm x;\bm q) = \int {d^3\bm y}
  \left\langle\delta\Psi_a(t,\bm x + \bm y/2)
    \delta\Psi_b(t,\bm x - \bm y/2) \right\rangle
  e^{-i\bm q\cdot\bm y}
\end{equation}
where $\delta\Psi_{a}$ is one of the fluctuation fields, e.g.,
$\delta\epsilon$ or $\delta u$, labeled by index~$a$.

Nongaussianity of fluctuations, important for the critical point
search, is described by {\em connected} correlation functions of $k>2$
fluctuation fields:
\begin{equation}
  \label{eq:Hphi}
  H_{ab\dots}(t,\bm x_a,\bm x_b,\dots)\equiv \left\langle\underbrace{
      \delta\Psi_a(t,\bm x_a)\delta\Psi_b(t,\bm x_b)\dots
    }_{\mbox{$k$ fields}}\right\rangle^\textrm{\!\!connected}
\end{equation}
The corresponding generalization of Wigner transform was introduced in
Ref.\cite{An:2020vri} in terms of the Fourier integral with fixed midpoint $\bm
x\equiv (\bm x_a + \bm x_b +\dots)/k$:
\begin{multline}
  \label{eq:WAB...}
  W_{ab\dots}(t,\bm x;\bm q_a,\bm q_b,\dots)\equiv
  \underbrace{
    \int {d^3\bm y_a}\, e^{-i\bm q_a\cdot\bm y_a}
    \int {d^3\bm y_b}\, e^{-i\bm q_b\cdot\bm y_b}\dots
  }_{\mbox{$k$ integrals}}\\
  \times\delta^{(3)}\left(\frac{\bm y_a+\bm y_b+\dots}{k}\right)
  H_{ab\dots}(t,\bm x+\bm y_a,\bm x+\bm y_b,\dots)
\end{multline}
Due to the delta-function factor in
Eq.~(\ref{eq:WAB...}), the function $W$ does not change if all $\bm q$'s
are shifted by the same vector. This means that one of the $\bm q$
arguments is redundant. In practice, it is sufficient to know the
function $W$ at all values of $\bm q$ which add up to zero. In
particular, the correlator $H_{ab\dots}$ can be obtained via inverse
transformation:
\begin{multline}
  \label{eq:HAB...WAB}
  H_{ab\dots}(t,\bm x_a,\bm x_b,\dots) =
  \underbrace{
    \int \frac{d^3\bm q_a}{(2\pi)^3}\,  e^{i\bm q_a\cdot\bm y_a}
    \int \frac{d^3\bm q_b}{(2\pi)^3}\, e^{i\bm q_b\cdot\bm y_b}\dots
  }_{\mbox{$k$ integrals}}\\
  \times(2\pi)^3\delta^{(3)}\left(\bm q_a+\bm q_b+\dots\right)
  W_{ab\dots}(t,\bm x;\bm q_a,\bm q_b,\dots)\,.
\end{multline}
For $k=2$ the generalized Wigner function
$W_{ab}(t,-\bm q,\bm q)$ coincides with the usual 2-point Wigner function
defined in Eq.~(\ref{eq:WAB}).

The hierarchy of evolution equations was
derived in Ref.\cite{An:2020vri} for fluctuations of density in a diffusion
problem, where the only hydrodynamic variable is the diffusion charge
density $n$ which obeys conservation equation and the Fick's law with
local noise
\begin{equation}
  \label{eq:dtn}
  \partial_t n = - \bm\nabla\cdot\bm N, \quad
  \bm N = -\lambda(n)\bm\nabla\alpha(n) + \mbox{noise}
\end{equation}

\newcommand\qpermii{{\overline{12}}}
\newcommand\qpermiii{{\overline{123}}}
\newcommand\qpermiv{{\overline{1234}}}
\newcommand\tc{\text{c}}

The evolution equations describe the relaxation of the
$k$-point functions $W_k$ of fluctuations of $n$ to
equilibrium values given by thermodynamics in terms of the equation of
state $\alpha(n)$:
\begin{subequations}\label{eq:W_234}
  \begin{equation}\label{eq:W_234-a}
    \partial_tW_2(\bm q)
    =-2\left[\gamma\bm{q}^2W_2(\bm{q})-\lambda\bm{q}^2
       \right]
       \,,
     \end{equation}
     \begin{multline}\label{eq:W_234-b}
       \partial_tW_3(\bm q_1,\bm{q}_2,\bm{q}_3)
       =-3\left[\gamma\bm{q}_1^2W_3(,\bm{q}_2,\bm{q}_3)
         +\gamma'\bm{q}_1^2W_2(\bm{q}_2)W_2(\bm{q}_3)
         \right.\\\left.
         +2\lambda'\bm{q}_1\cdot\bm{q}_2W_2(\bm{q}_3)\right]_\qpermiii
       \,,
     \end{multline}
     \begin{multline}\label{eq:W_234-c}
       \partial_tW_{4}(\bm{q}_1,\bm{q}_2,\bm{q}_3,\bm q_4)
       =-4\left[
         \gamma\bm{q}_1^2W_{4}(,\bm{q}_2,\bm{q}_3,\bm{q}_4)
         \right.\\\left.
           +3\gamma'\bm{q}_1^2W_2(\bm{q}_2)
           W_3(,\bm{q}_3,\bm{q}_4)
     +\gamma''\bm{q}_1^2W_2(\bm{q}_2)W_2(\bm{q}_3)W_2(\bm{q}_4)
        \right.
    \\
    \left.
      +3\lambda'\bm{q}_1\cdot\bm{q}_2W_3(,\bm{q}_3,\bm{q}_4)
      +3\lambda''\bm{q}_1\cdot\bm{q}_2W_2(\bm{q}_3)W_2(\bm{q}_4)
    \right]_\qpermiv
       ,
     \end{multline}
 \end{subequations}
 where $\gamma=\lambda\alpha'$. In Eqs.~(\ref{eq:W_234}) we suppressed
 arguments $t$ and $\bm x$, as they are the same for all functions
 $W_n$. Furthermore, note that all arguments
 of each function $W_n$ add up to zero -- reminiscent of the momentum
 conservation in Feynman diagrams. Therefore, to save space, we
 omitted the first argument in $W_k$ where this argument can be
 inferred from the condition $\bm q_1+\dots+\bm q_k=0$. For example, $
 W_2(\bm q)\equiv W_2(,\bm q)\equiv
 W_2(-\bm q,\bm q)$ or $W_3(,\bm q_3,\bm q_4)\equiv W_3(-\bm q_3-\bm
 q_4,\bm q_3,\bm q_4)$. The symbol
$
   [\dots]_{\overline{1\dots k}}
   $
   denotes the sum over all permutations of $\bm q_1$,\dots,$\bm q_k$\
   divided by $1/k!$, i.e., the average over all
   permutations. Diagrammatic representation of Eqs.~(\ref{eq:W_234})
   is given in Ref.\cite{An:2020vri}.

\subsubsection{Confluent formalism for arbitrary relativistic flow}

For a {\em relativistic\/} fluid with nontrivial velocity gradients
the definition of the equal-time correlator in Eq.~(\ref{eq:Hphi})
and its Wigner transform are insufficient since the rest frame of the fluid is
different at different space-time points, and the concept of ``equal time'' is thus ambiguous. Ref.\cite{An:2019osr}, and for non-gaussian fluctuations,
Ref.~\cite{An:2022jgc}, define a generalization of the Wigner
transform to this case by using local rest frame of the fluid at the
midpoint  $x\equiv(t,\bm x)$ of the correlator. This removes an
arbitrary lab frame from the definition of the correlator and allows
fully Lorentz covariant formulation of hydrodynamics with fluctuations.

The idea is to equip every space-time point $x$ with an orthonormal triad of
basis 4-vectors $e_{\mathring{a}}(x)$, $\mathring a=1,2,3$, or $\bm e(x)$, spatial in the local rest frame
of the fluid at that point, i.e.,
\begin{equation}
\bm e(x)\cdot u(x)=\bm 0.\label{eq:e-dot-u}
\end{equation}
One can then define equal-time $k$-point correlators
as functions of $k$ space-time points expressed as
\begin{equation}
  x_1=x+\bm e(x)\bm\cdot\bm y_1,
\quad\ldots\quad,\quad
x_k=x+\bm e(x)\bm\cdot \bm y_k,\label{eq:x1...xk}
\end{equation}
in terms of the midpoint $x$ and the
separation 3-vectors $\bm y_1,\dots, \bm y_k$, which sum to zero: $\bm
y_1+\ldots+\bm y_k=0$.

Some of the fields in the correlator may not be Lorentz scalars, for
example, fluctuations of velocity, $\delta u$. In this case, one would
like to express fluctuations in a frame associated with the fluid,
such as the local rest frame, rather than an arbitrary lab frame. The
local rest frame of the fluid is, however, different for different
points $x_1,\dots,x_k$. To use the same frame for all points while
making sure that fluctuations represent deviations from fluid at rest
locally, we boost the fluctuation variables (if they are not scalars)
from the rest frame at the point they occur, say, at $x+y_1$, to that
at the midpoint $x$ of the correlator. This operation leads to the
{\em confluent correlator} defined as:
\begin{equation}
  \label{eq:confluent-correlator-def}
   H_{a_1\dots a_k}(x_1,\dots,x_k) \equiv
   \left\langle
     \left[\Lambda(x,x_1)\delta\Psi(x_1)\right]_{a_1}\dots
     \left[\Lambda(x,x_k)\delta\Psi(x_k)\right]_{a_k}
    \right\rangle^{\rm c}
\end{equation}
where $\Lambda(x,x_1)$ performs the boost on the corresponding
fluctuation field $\delta\Psi(x_1)$ such that the 4-velocities in points
$x_1$ and $x$ are related by
\begin{equation}
  \label{eq:Lambda-u}
  \Lambda(x,x_1) u(x_1) = u(x).
\end{equation}
Superscript ``c'' means ``connected'' as in
Eq.~(\ref{eq:Hphi}).

We can now take the multipoint Wigner transform of the confluent
correlator we just defined in Eq.~(\ref{eq:confluent-correlator-def})
with respect to the 3-vectors
$\bm y_1,\dots,\bm y_k$ describing
separation of the points in the local basis at point $x$
(cf. Eq.~(\ref{eq:WAB...}):
\begin{multline}
  \label{eq:W...H}
  W_{a_1\dots a_k}(x;\bm q_1,\dots,\bm q_k)\equiv
    \int {d^3\bm y_1}\, e^{-i\bm q_1\bm\cdot\bm y_1}\dots
    \int {d^3\bm y_k}\, e^{-i\bm q_k\bm\cdot\bm y_k}
    \\
  \times\delta^{(3)}\left(\frac{\bm y_1+\dots+\bm y_k}{k}\right)
   H_{a_1\dots a_k}(x+\bm e(x)\bm\cdot\bm y_1,\dots,x+\bm e(x)\bm \cdot \bm y_k)
 \end{multline}
The inverse is given by (cf. Eq.~(\ref{eq:HAB...WAB}))
\begin{multline}
  \label{eq:H...W}
  H_{a_1\dots a_k}(x+\bm e_1(x)\bm\cdot\bm y_1,\dots,
  x+ \bm e_k(x)\bm\cdot\bm y_k) \\ =
    \int \frac{d^3\bm q_1}{(2\pi)^3}\,  e^{i\bm q_1\bm\cdot\bm y_1}\dots
    \int \frac{d^3\bm q_k}{(2\pi)^3}\, e^{i\bm q_k\bm\cdot\bm y_k}
    \\\times  %
    \delta^{(3)}\left(\frac{\bm q_1+\dots+\bm q_k}{2\pi}\right)
  W_{a_1\dots a_k}(x;\bm q_1,\dots,\bm q_k)\,.
\end{multline}

Care must
be taken also in defining a derivative  with respect to the midpoint
of such a correlation
function in order to maintain the ``equal-time in local rest frame''
condition. The local rest frame is different in points $x$ and
$x+\Delta x$ used to define the derivative.
A derivative which maintains ``equal-time'' condition is introduced in
Ref. \cite{An:2019osr,An:2019csj} and termed
{\em confluent derivative\/}.
It is defined via the $\Delta x\to0$ limit of the following equation:
\begin{multline}
  \label{eq:confluent-nabla-W}
  \Delta x\cdot\bar\nabla  W_{a_1\dots a_k}(x;\bm q_1,\dots,\bm q_k)
  \equiv
   \Lambda(x,x+\Delta x)_{a_1}^{b_1}\dots
   \Lambda(x,x+\Delta x)_{a_k}^{b_k}
   \\\times  W_{b_1\dots b_k}(x;\bm q_1',\dots,\bm q_k')
   - W_{a_1\dots a_k}(x;\bm q_1,\dots,\bm q_k)\,,
\end{multline}
where
\begin{equation}
  \label{eq:q'q}
  \bm
q'=\bm e(x+\Delta x)\cdot[\Lambda(x+\Delta x,x)\bm e(x)\bm\cdot\bm
q]\equiv R(x+\Delta x,x) \bm q\,.
\end{equation}
In Eq.~(\ref{eq:confluent-nabla-W}) the (non-scalar) fluctuation
fields are boosted from point $x+\Delta x$ back to point $x$,
similarly to the definition of the confluent correlator in
Eq.~(\ref{eq:confluent-correlator-def}). In addition, at $x+\Delta x$
we evaluate the function using the set of 3-wavevectors $\bm q_i'$
given by Eq.~(\ref{eq:q'q}), different from the set $\bm q_i$ used at
point $x$. The new set is obtained by representing each vector $\bm q$
as a {\em 4-vector\/} orthogonal to $u(x)$, $\bm e(x)\cdot\bm q$, then
boosting this vector to the rest frame at point $x+\Delta x$ and then
expressing it again as a 3-vector, but now in the basis
$\bm e(x+\Delta x)$ orthogonal to $u(x+\Delta x)$. The resulting
transformation of vector $\bm q$ is a rotation, denoted by $R$ in
Eq.~(\ref{eq:q'q}).

Taking the limit $\Delta x\to0$ one can express confluent derivative as follows:
\begin{equation}
  \label{eq:nabla-W}
  \bar\nabla_\mu W_{a_1\dots a_k} = \partial_\mu W _{a_1\dots a_k}
  + k\left(
    \mathring\omega^{\mathring a}_{\mu\mathring b}q_{1\mathring a}
    \frac{\partial}{\partial q_{1\mathring b}}W_{a_1\dots a_k}
    - \bar\omega^{b}_{\mu a_1}W_{ba_2\dots a_k}
    \right)_{\overline{1\dots k}}\,,
\end{equation}
where the confluent connection $\bar\omega$ is a generator of the
infinitesimal boost~$\Lambda$ and $\mathring\omega$ is a generator of
the infinitesimal rotation $R$:
\begin{align}
  \label{eq:Lambda-omega}
  \Lambda(x+\Delta x,x)^a_{~b} = \delta^a_b - \Delta
  x^\mu\bar\omega^a_{\mu b},
  \\ \label{eq:R-omega}
  R(x+\Delta x,x)^{\mathring a}_{~\mathring b}
  = \delta^{\mathring a}_{\mathring b}
  - \Delta x^\mu\mathring\omega^{\mathring a}_{\mu\mathring b}\,.
\end{align}

The indices $a,b,a_1\dots a_k$ label
fluctuating fields. The confluent connection $\bar\omega^a_{\mu
b}$ is nonzero when indices $a,b$ refer to different components of a
Lorentz vector (such as $\delta u$). In this case the connection
satisfies
\begin{equation}
  \label{eq:nabla-u}
  \bar\nabla_\mu u^\alpha \equiv \partial_\mu u^\alpha
  + \bar\omega^\alpha_{\mu\beta}u^\beta=0,
\end{equation}
(local velocity is ``confluently'' constant), which follows from
Eqs.~(\ref{eq:Lambda-u}) and~(\ref{eq:Lambda-omega}).%
\footnote{For a scalar field (e.g., energy
density, pressure fluctuations, etc) the confluent connection is, of course,
zero.}

The rotation connection $\mathring\omega$ is determined by
Eqs.~(\ref{eq:q'q}) and~(\ref{eq:R-omega}):
\begin{align}
  \mathring\omega^{\mathring b}_{\mu\mathring a}
  =
  e^{\mathring b}_\alpha\left(\partial_\mu e^\alpha_{\mathring a}
  + \bar\omega^\alpha_{\mu\beta}e^\beta_{\mathring a}\right)\,.
\end{align}
Naturally, it satisfies:
\begin{equation}
  \label{eq:nabla-e}
  \bar\nabla_\mu e^\alpha_{\mathring a} \equiv \partial_\mu
  e^\alpha_{\mathring a}
  + \bar\omega^\alpha_{\mu\beta}e^\beta_{\mathring a}
  - \mathring\omega^{\mathring b}_{\mu\mathring a}e^\alpha_{\mathring b}=0\,
\end{equation}
(local basis vectors are confluently constant).

The boost $\Lambda$ is not defined uniquely by Eq.~(\ref{eq:Lambda-u})
-- only up to a rotation keeping $u$ unchanged.  The simplest choice
is the boost without additional rotation, which corresponds to
confluent connection given explicitly by
\begin{equation}
  \label{eq:omega-udu}
  \bar\omega^\alpha_{\mu\beta}=u_\beta\partial_\mu u^\alpha
  - u^\alpha\partial_\mu u_\beta\,.
\end{equation}
For this choice of the confluent connection the rotation connection also
simplifies to
\begin{align}
  \mathring\omega^{\mathring b}_{\mu\mathring a}
  =
  e^{\mathring b}_\alpha
  \partial_\mu e^\alpha_{\mathring a}\,.
\end{align}

\subsubsection{Evolution equations for hydrodynamic fluctuations}

There are five normal hydrodynamics modes, which can be described as two
propagating modes and three diffusive.\footnote{We focus on
  hydrodynamics involving baryon charge. Each additional charge adds one diffusive mode to the count.} The propagating modes
correspond to fluctuations of pressure mixed with the fluctuations of
the longitudinal (with respect to the wave vector $\bm q$)
velocity. The frequency of these modes is $c_s|\bm q|$. The diffusive
modes are the fluctuations of the specific entropy $m\equiv s/n_B$ at
fixed pressure and
transverse velocity. The relaxation rate of these modes is
proportional to the square of their wavenumber $\bm q^2$. The
slowest diffusive mode near the critical point is specific entropy
because its diffusion constant 
vanishes at the critical point.

In this review we shall focus on the slowest diffusive mode $m$ for two
reasons. First, because it is the slowest and therefore the furthest from
equilibrium mode. Second, in equilibrium this mode shows the
largest fluctuations, divergent at the critical point.

The evolution equation for the Wigner-transformed confluent two-point
correlator of the specific entropy fluctuations, $\langle\delta
m\delta m\rangle$, derived in
Ref.\cite{An:2019csj} reads:
\begin{equation}
  \label{eq:LW}
  \mathcal L[W_{mm}(\bm q)]
  = (\partial\cdot u) W_{mm}(\bm q)
  - 2\gamma_{mm}\bm q^2\left[W_{mm}(\bm q)
  - \frac{c_p}{n^2}\right]\,.
\end{equation}
where $\gamma_{mm}\equiv\kappa/c_p$ -- the heat diffusion constant,
and $\mathcal L$ is
the Liouville operator given by
\begin{equation}
  \label{eq:mathcalL}
  \mathcal L [W_{mm}] \equiv \left[
    u\cdot\bar\nabla
    - \partial_\nu u^\mu e_\mu^{\mathring a} e^\nu_{\mathring b}
    q_{\mathring a}\frac{\partial}{\partial q_{\mathring b}}\right]W_{mm}\,.
\end{equation}
The first term is the confluent derivative along the flow of the
fluid, while the second describes stretching and/or rotation of the
vectors $\bm q$ due to the expansion and/or rotation of the fluid. In
the case of expansion one can think of this term as describing the
Hubble-like ``red shift'' of the fluctuation wavevector $\bm q$.
For example, for Bjorken flow the Liouville operator takes the form
\begin{equation}
  \label{eq:L-Bj}
  \mathcal L_{\rm Bj} = \partial_\tau
  - \frac{q_3}{\tau}\frac{\partial}{\partial q_3},
\end{equation}
where the second term describes the ``red shift'', or ``stretching''
of the wave number $q_3$ due to longitudinal
expansion.\footnote{Naturally, we
have chosen the triad of the 4 vectors $\bm e$ in such a way that the
spatial part of the 4-vector $e_3$ points along the direction of the
longitudinal flow, while the $e_1$ and $e_2$ are constant.
In this case the rotation connection ($\mathring\omega$) terms vanish in the
confluent derivative in Eq.~(\ref{eq:nabla-W}). The confluent
connection ($\bar\omega$) terms are absent already for arbitrary flow
because the fluctuating quantity, $m$, is
a scalar. }

The first term on the r.h.s. of Eq.~(\ref{eq:LW}), describes the
scaling of the fluctuation magnitude with the volume of a hydrodynamic
cell, as the cell expands at the rate $\partial\cdot u$. This trivial rescaling
could be absorbed by multiplying $W$ by a conserved density, such as
the baryon density, $n$. The equation for a rescaled function $N_{mm}\equiv
nW_{mm}$ is the same as in Eq.~(\ref{eq:LW}), but without the
$\partial\cdot u$ term.
\begin{equation}
  \label{eq:LN}
  \mathcal L[N_{mm}(\bm q)]
  = 
  - 2\gamma_{mm}\bm q^2\left[N_{mm}(\bm q)
  - \frac{c_p}{n}\right]\,.
\end{equation}

The last term in  Eq.~(\ref{eq:LW}) describes diffusive relaxation of
fluctuations towards equilibrium given by thermodynamic quantity
$c_p/n^2$ (or $c_p/n$ for $N_{mm}$).

It is instructive to compare Eq.~(\ref{eq:LW}) for $W_{mm}$ (or the
corresponding Eq.~(\ref{eq:LN}) for $N_{mm}$) to
Eq.~(\ref{eq:W_234-a}) for the density-density correlator
$\langle\delta n\delta n\rangle$ in the diffusion problem. The main
difference is that the time derivative is replaced by the Liouville
operator, which takes into account the flow of the fluid. The
$(\partial\cdot u)$ term on the r.h.s. (absent when Eq.~(\ref{eq:LW})
is written in terms of $N_{mm}$, Eq.~(\ref{eq:LN})) is also an effect of
the flow -- expansion. The diffusive relaxation terms are different
because the correlated quantities are different, $\delta m\delta m$ in
Eq.~(\ref{eq:LW}) and $\delta n\delta n$ in Eq.~(\ref{eq:W_234-a}). The
coefficients, however, can be mapped onto each other via substitution
\begin{equation}
  \label{eq:nmsubstitution}
  n\to m,\quad \gamma=\lambda\alpha'\to\gamma_{mm}=\frac{\kappa}{c_p},
  \quad \lambda \to \frac{\kappa}{n^2}.  
\end{equation}

The evolution equations for non-gaussian correlators $W_{mmm}$ and
$W_{mmmm}$ are qualitatively similar to those in the diffusion problem
in Eq.~(\ref{eq:W_234-b}) and~(\ref{eq:W_234-c}). However, unlike the
case of $W_{mm}$, where the whole equation~(\ref{eq:LW}) can be
obtained from Eq.~(\ref{eq:W_234-a}) for $W_2$ by the substitution
given by Eq.~(\ref{eq:nmsubstitution}), only the terms containing
leading singularities at the critical point can be obtained from
Eq.~(\ref{eq:W_234-b}) and~(\ref{eq:W_234-c}) by the
substitution~(\ref{eq:nmsubstitution}). There are subleading
singularities, which are due to the nonlinearity of the function
$m(n)$ at constant pressure. These terms are written explicitly in
Ref.\cite{An:2022jgc} and contain factors proportional to second and higher
derivatives of $m$ with respect to $n$ at fixed pressure.

\subsection{Freeze-out of fluctuations and observables}

The previous section was devoted to recent progress in describing the
evolution of fluctuations in hydrodynamics using correlators of
hydrodynamic variables in coordinate space. Experiments do not
measure such densities, or their correlations, directly. Instead the
particle multiplicities and their correlations in {\em momentum} space are
measured. In this section we describe how to connect the theoretical
description in terms of fluctuating hydrodynamics to experimental
observables.

\subsubsection{Event-by-event fluctuations in heavy-ion collisions and their experimental measures}\label{sec:event-event-fluct}

Typical experimental measures are cumulants of the event-by-event
fluctuations or correlations of particle multiplicities.
For example, if $N_p$ is the proton number in an event, its fluctuation
in the event is $\delta N_p\equiv N_p - \langle N_p\rangle$ and
$\langle(\delta N)^2\rangle$ is its quadratic cumulant, or variance, where
$\langle\dots\rangle$ is the event average. Higher-order cumulants,
measuring non-gaussianity of fluctuations, are constructed similarly
(see, e.g., Ref.\cite{Bzdak:2019pkr} for review).

In addition, correlations between particles can be also measured, such
as $\langle\delta N_p\delta N_\pi\rangle$ -- a correlation between
proton and pion multiplicities. Such measures can also include
higher-order correlators\cite{Athanasiou:2010kw}.

Similary to correlation between species one can also consider
correlations between particle with different momenta by counting
particles within small momentum bins, or cells, labeled by the central
momentum of the cell.

The number of particles $dN_i$ in a given momentum cell $dp^3$ is
given by Eq.~(\ref{eq:distr}) in terms of the phase space distribution
function $f_i(x,p)$ integrated over the freeze-out
hypersurface. Therefore the correlations between different momentum
cells and/or between different species, can be expressed in terms of
the correlation functions of fluctuations of $f_i(x,p)$. In this
section we denote such fluctuations $\delta f\equiv f-\langle f\rangle$.\footnote{This should not be
confused with a similar
notation used in Section~\ref{sec:off-eq} to denote deviation of
$\langle f\rangle$ from equilibrium distribution $f_{\rm eq}$.}

We shall also combine the species index (which
includes all discrete quantum numbers, such as mass, spin, isospin,
etc.) together with the coordinate and momentum
into a single composite index $A=\{i,x,p\}$.
Therefore, the general correlator which, upon integration over
freeze-out surface, gives the observable correlation measures has the
form
\begin{equation}
  \label{eq:dfdf}
  \langle \delta f_{i_A}(x_A,p_A) \delta f_{i_B}(x_B,p_B)\dots\rangle \equiv
  \langle \delta f_A \delta f_B\dots\rangle\,.
\end{equation}

The freeze-out of hydrodynamic evolution was already discussed in
Section~\ref{sec:fzout}. The Cooper-Frye prescription determines the
event-by-event {\em averaged} distribution function
$\langle f_A\rangle$ in terms of the hydrodynamic variables, or fields,
$T(x)$ and $\mu(x)$, as in Eq.~(\ref{eq:fk_eq}). In order to convert
hydrodynamic {\em fluctuations} into particle event-by-event
fluctuations we need an analogous freeze-out prescription for {\em
  correlators} in Eq.~(\ref{eq:dfdf}).

\subsubsection{Freeze-out of fluctuations and the maximum entropy method}

The generalization of the Cooper-Frye freeze-out to
fluctuations has been first considered in Ref.\cite{Kapusta:2011gt}.  In the
approach of Ref.\cite{Kapusta:2011gt} the fluctuations of the phase space
distribution function $f(x,p)$ are assumed to be caused by
fluctuations of the hydrodynamic variables/fields $T(x)$ and $\mu(x)$
on which $f(x,p)$ depends. As a result, coordinate space correlations in
$T(x)$ and $\mu(x)$ translate into the phase space correlations in
$f(x,p)$.

This approach has an important flaw, which becomes obvious if
one considers fluctuations in an (almost) ideal gas. In this case,
there are fluctuations of hydrodynamic variables, such as charge
density $n(x)$, but there are no momentum space correlations of
$f(x,p)$, which would, nevertheless, be produced if the approach of
Ref.\cite{Kapusta:2011gt} were to be applied. Instead, the hydrodynamic
fluctuations of $n(x)$ are matched on the particle side by trivial
(Poisson, in the ideal gas case) {\em uncorrelated} fluctuations of the
occupation numbers in each phase-space point.

This problem has been
addressed in Ref.\cite{Ling:2013ksb}  by subtracting this trivial ideal gas
contribution from hydrodynamic fluctuations of $n(x)$ before applying
the procedure of Ref.\cite{Kapusta:2011gt} to the remainder, which is
due to interactions and out-of-equilibrium dynamics.

Generalization of this approach to fluctuations of other hydrodynamic
variables, and to non-gaussian fluctuations, proved
elusive until the principle of maximum entropy for fluctuations was
proposed and implemented in Ref.\cite{Pradeep:2022eil}.
In this approach the matching of conserved hydrodynamic densities such as
$\epsilon(x)$ and $n_q(x)$, defined in Eqs.(\ref{eq-landau}), is done not simply on average, as, e.g., in
Eqs.~(\ref{eq:CF_Nq}) or~(\ref{eq:CF_Tmunu}), but on
an event-by-event basis. I.e.,\footnote{Eq.~(\ref{eq:dndf}) can be
obtained by multiplying {\em fluctuating} equation~(\ref{eq:CF_Nq})
by $u_\mu$.}
\begin{equation}
  \label{eq:dndf}
  \delta n_q (x) = \sum_i \int_{p} q_i \delta f_i(x, p) \equiv
  \int_{\tilde A}q_A \delta f_{\tilde A}(x) \equiv
  \int_A q_A \delta^{(3)} (x-x_A) \delta f_{A}\,,
\end{equation}
where $\int_p$ is a 3-integral over momenta with the
Lorentz-invariant measure and $\sum_i$ is a sum over the species of
particles with id label $i$ (corresponding to mass, spin, isospin,
etc.) carrying charge $q_i$ corresponding to density $n_q$ (e.g., baryon
charge when $n_q=n_B$ is the baryon density). We have also introduced a
convenient shorthand $\int_{\tilde A}$ which denotes the sum and the
momentum space integral together (but no space integration), i.e., the
index $\tilde A=\{i,p\}$, while $A\equiv\{i_A,p_A,x_A\}=\{\tilde A, x_A\}$.
Similarly $f_{\tilde A}(x_A)\equiv
f_A$ (think of each particle species having its own phase space).
Finally, we also introduced $\int_A$ which includes integration over
the whole phase space (momentum $p_A$ {\em and} coordinate $x_A$) of each
particle species. The delta function simply reflects the locality of
freeze-out (i.e., each hydrodynamic cell is converted to particles
located in the hydrodynamic cell at point $x$).
Similarly, matching of the energy-momentum requires
\begin{equation}
  \label{eq:dedf}
  \delta ( \epsilon (x) u^\mu(x)) = 
  \int_{\tilde A}p^\mu \delta f_A(x)\equiv\int_A p^\mu \delta^{(3)} (x-x_A) \delta f_{A}\,.
\end{equation}
It is convenient to organize equations such as Eqs.~(\ref{eq:dndf})
and~(\ref{eq:dedf}) into an indexed array, where lowercase index
$\tilde a$
runs through five hydrodynamic variables $\Psi_{\tilde a}=\{n_q,\epsilon
u^\mu\}$:
\begin{equation}
  \label{eq:dpsidf}
  \delta\Psi_a\equiv\delta\Psi_{\tilde a}(x_a) = \int_A P^A_a \delta f_A\,,
\end{equation}
where
\begin{equation}\label{eq:PAa}
  P^A_a=\{q_A,p_A^\mu\}\delta^{(3)}(x_a-x_A)
\end{equation}
is the array of the contributions of a single particle $A$ at point $x_A$ to
hydrodynamic densities $\Psi_{\tilde a}=\{n,\epsilon u^\mu\}$ in a
cell around point
$x_a$ on the freeze-out surface. Similarly to particle index $A$ it is
convenient to view hydrodynamic field index $a$ as a composite index
labeling both the field and the point on the freeze-out surface where
the value of this field is taken from, i.e.,
$a=\{\tilde a, x_a\}$.

Eq.~(\ref{eq:dpsidf}) for fluctuations imply relationships
between the hydrodynamic correlators in space points
$x_a$, $x_b$, etc.
\begin{equation}
  \label{eq:Hcorr}
  H_{ab\dots}\equiv \langle\delta\Psi_a\delta\Psi_b\dots\rangle
\end{equation}
and particle correlators in phase-space points $A$, $B$, etc.
\begin{equation}
  \label{eq:Gcorr}
  G_{AB\dots}\equiv \langle\delta f_A\delta f_B\dots\rangle
\end{equation}
which have the form:
\begin{equation}
  \label{eq:HG}
  H_{ab\ldots} = \int_{AB\ldots} G_{AB\ldots}P^A_aP^B_b\ldots\,,
\end{equation}

Equations (\ref{eq:HG}) represent constraints on the particle
correlators $G_{AB\dots}$ imposed by conservation laws. These
constraints alone are not enough to completely determine
$G_{AB\dots}$ simply because there are more ``unknowns'' $G_{AB\dots}$
then the constraints.

The situation is similar already for ensemble
(i.e., event) averaged quantities, or one-point functions. In
this case the averaged energy, momentum and baryon density are not
sufficient alone to determine the particle distribution functions
$f_A$. Additional input is needed. In the absence of additional
information, the most natural solution is the one which maximizes the
entropy of the resonance gas into which the hydrodynamically evolved
fireball freezes out. That entropy is given by the well-known
functional of $f_A$:
\begin{equation}
S[f]=\int_A f_A(1-\ln f_A)\,,  \label{eq:Sf}
\end{equation}
(neglecting, for simplicity,
quantum statistics). As observed recently in
Ref.\cite{Everett:2021ulz} ,
maximizing $S[f]$, subject to the
constraints on $f_A$ imposed by conservation laws, produces the
well-known Cooper-Frye freeze-out equation for particle phase-space
distributions, which is widely and successfully used for describing experimental
data for the last fifty or so years (see Section~\ref{sec:fzout}).

In order to apply the maximum entropy approach to fluctuation
freeze-out, on needs the entropy of fluctuations as a functional of
$f_A$  as well as correlators $G_{AB\dots}$. Conceptually, this
entropy, $S[f,G_{AB},\dots]$
represents the (logarithm) of the number of the microscopic
states in the resonance gas ensemble with the given set of
correlators. The single particle entropy $S[f]$ is the value of
$S[f,G_{AB},\dots]$ when all correlators $G_{AB\dots}$ are given by
their values in the equilibrium resonance gas.

The expression for the
functional $S[f,G_{AB},\dots]$ was found in
Ref.\cite{Pradeep:2022eil}. For example, keeping only terms with
out-of-equilibrium two-point correlators it reads
\begin{equation}
  \label{eq:S2}
    S_2[f,G] = S[f]
  + \frac12{\rm Tr\,}\left[
\log(- CG) + CG + 1
\right]\,.
\end{equation}
The last term is always negative except for $G=C^{-1}\equiv\bar G$ which is the
equilibrium value of the correlator $G$, where $C_{AB}=\delta^2
S[f]/(\delta f_A\delta f_B)$. When $G=\bar G$ the last term vanishes
and $S_2$ is maximized with respect to $G$.

However, maximizing the entropy in Eq.~(\ref{eq:S2}) with respect to $G$
under constraints in Eq.~(\ref{eq:HG}) gives
\begin{equation}
  (G^{-1})^{AB} = (\bar G^{-1})^{AB} + (H^{-1}-\bar H^{-1})^{ab}
\, P_{a}^{A}\, P_{b}^{B}\,.
\label{eq:G2}
\end{equation}
In this equation and below the repeated lower case indices $a$, $b$,
etc. imply summation over the set hydrodynamic variables
$\Psi_a\equiv\Psi_{\tilde a}(x_a)$ {\em and} volume integration over
hydrodynamic cells at points $x_a$, $x_b$, etc. Due to the delta
functions in the definition of $P_a^A$ in Eq.~(\ref{eq:PAa}), the
integrals in Eq.~(\ref{eq:G2}) simply set the spatial arguments of
$(G^{-1})^{AB}$ to those of $(H^{-1})^{ab}$, i.e., $x_A=x_a$ and
$x_B=x_b$. When hydrodynamic correlator $H$ equals its equilibrium
value $\bar H$, so does $G$ given by equation~(\ref{eq:G2}).

If deviations of fluctuations from equilibrium resonance gas are
small, the equation~(\ref{eq:G2}) can be linearized in such
deviations. The deviations could be due to non-equilibrium effects,
which have to be small for hydrodynamics to apply, or due to effects
of the critical point, which could be small if we are not too close to
the critical point. Linearized equation relates deviations of the
particle correlators $\Delta G_{AB}= G_{AB}-\bar G_{AB}$ to the
deviations of the hydrodynamics correlators $\Delta H_{ab}=H_{ab}-\bar
H_{ab}$ from the resonance gas values:
\begin{equation}
  \label{eq:DG2}
  \Delta G_{AB}
  =
  \Delta H_{ab}
  (\bar H^{-1}P\bar G)^a_A (\bar H^{-1}P\bar G)^b_B,
\end{equation}

Similarly, the non-gaussian cumulants $G_{ABC\dots}$ of particle
fluctuations can be expressed in terms of the non-gaussian cumulants
of the hydrodynamic variables $H_{abc\dots}$. Such non-linear relations
similar to Eq.~(\ref{eq:G2})
can be derived from the corresponding entropy functional found in
Ref.\cite{Pradeep:2022eil} and we will not reproduce them
here. Linearized relations valid for small deviations from the
equilibrium resonance gas, however, are simple and instructive.
The relationship
is similar to Eq.~(\ref{eq:DG2}), but instead of the ``raw'' deviations
from equilibrium $\Delta G_{AB\dots}$ and $\Delta H_{ab\dots}$, i.e.
correlations relative to equilibrium, the
proportionality relation holds between {\em irreducible}
  relative correlators $\hat\Delta G_{AB\dots}$
defined in Ref.\cite{Pradeep:2022eil}. An irreducible
correlator $\hat\Delta G$ is different from the reducible ``raw''
relative correlator
$\Delta G$ by subtraction of correlations
involving only a smaller subset of the points $AB\dots$. The irreducible $\Delta H$ differs from $\Delta
H$ similarly. The resulting linear relation
generalizes Eq.~(\ref{eq:DG2}):
\begin{equation}
  \label{eq:DhatGDhatH2}
  \widehat\Delta G_{AB\dots}
  =
  \widehat\Delta H_{ab\dots}
  (\bar H^{-1}P\bar G)^a_A (\bar H^{-1}P\bar G)^b_B\ldots,
\end{equation}
where $\widehat\Delta G_{AB\dots}$ and $\widehat\Delta H_{ab\dots}$
denote {\em irreducible} relative correlators for particles and for
hydrodynamic variables, respectively. For two-point (gaussian)
correlators $\hat\Delta G=\Delta G$ and $\hat\Delta H=\Delta H$ and Eq.~(\ref{eq:G2})
reproduces Eq.~(\ref{eq:G2}).

Eq.~(\ref{eq:DG2}) thus solves the problem of translating fluctuations
in hydrodynamics into correlations between particles at freeze-out,
in such a way as to obey the conservation laws on event-by-event
basis. As one can see, it systematically eliminates spurious
``self-correlations'' discussed at the beginning of this section not
only for gaussian, but also for non-gaussian cumulants.

Similarly to the way the maximum entropy approach reproduces and
generalizes Cooper-Frye prescription for event {\em averaged}
observables, the maximum entropy approach to fluctuations reproduces,
justifies, and generalizes prior approaches to freezing out
fluctuations, in particular, of critical fluctuations, as shown in
Ref.\cite{Pradeep:2022eil}. Such a prior approach involving fluctuating
background field $\sigma$ was introduced in
Ref.\cite{Stephanov:1999zu}, generalized to non-gaussian fluctuations
in Refs.\cite{Stephanov:2011pb,Athanasiou:2010kw}, and then further
generalized to {\em non-equilibrium} fluctuations in
Ref.\cite{Pradeep:2022mkf}.

The $\sigma$-field approach\cite{Stephanov:1999zu,Stephanov:2011pb,Athanasiou:2010kw}, however, besides the knowledge of QCD EOS,
requires the knowledge of the properties of the field $\sigma$ such as
its correlation length as well as its coupling to observed particles. These
properties would depend on the nature of this field -- an {\it a priori} unknown
mixture of scalar fields such as chiral condensate, energy, and
baryon number densities. It was also not clear how to deal with 
non-critical fluctuations or contributions of lower point correlations
to higher-point correlators. All these uncertainties are absent in
the maximum entropy approach. The correlations described by
Eq.~(\ref{eq:DhatGDhatH2}) are very similar to the correlations
induced by the $\sigma$ field given by a mixture of hydrodynamic
fields determined by the QCD EOS itself.

The maximum entropy approach thus provides a
direct connection between the fluctuations of the hydrodynamic
quantities and the observed particle multiplicities, with their
fluctuations. This connection is determined by the EOS of QCD in the resonance
gas phase where the freeze-out occurs. The effects of the critical
point and non-equilibrium are encoded in the non-trivial correlations
described quantitatively by Eq.~(\ref{eq:DhatGDhatH2}).


%% file: summary.tex
\section{Summary and outlook}
\label{sec:sum}

Investigations of relativistic heavy-ion collisions across a range of
beam energies, and in particular in the range explored by the BES
program at RHIC which covers the QCD transition region, is the primary
method for systematically mapping the QCD phase diagram in controlled
laboratory experiments.  The significantly increased complexity of
these events, compared to collisions at higher energies, is associated with
non-trivial longitudinal dynamics, evolution of finite conserved
charge densities, and non-equilibrium as well as potential critical phenomena,
pose substantial challenges to theoretical descriptions of this
compelling physics.  Over the last decade, significant efforts have
been dedicated to theoretical modeling and phenomenology aimed at
understanding the interesting physics driving the dynamics of the
created hot and dense QCD systems.

In this chapter, we adopted a pedagogical approach to reviewing theoretical descriptions of heavy-ion collision dynamics, with a particular focus on considerations at finite baryon density. 
These descriptions play a crucial role in model-to-data comparisons which will allow us to extract properties of QCD matter from experimental measurements. 
We delve into recent developments in bulk medium dynamics, covering both multistage hydrodynamic approaches applicable for collisions at \mbox{$\snn \gtrsim 7$ GeV} (Section~\ref{sec:hydro}) and hadronic transport descriptions suitable for collisions at \mbox{$\snn \lesssim 7$ GeV} (Section~\ref{sec:microscopic_transport}). These recent advances in modeling the bulk evolution lay the foundation for studying fluctuations and their dynamics near the QCD critical point, and we provide an overview of recent developments in this intriguing area (Section~\ref{sec:fluctuations}).

Being a review of a rapidly developing research direction, this chapter can only
hope to provide a snapshot of the current state of the art.  Some
questions still require more careful analyses while some theoretical
tools, for example, those dealing with fluctuations, still need to be
incorporated into a fully-fledged description of heavy-ion collisions
before comparison with the experiment. Naturally, much of the future
development of the field will be informed by the
experimental data from the BES-II program at RHIC as well as from
experiments at planned future heavy-ion collision
facilities~\cite{LRPNS:2023}.